\documentclass[11pt, a4paper]{article}
\usepackage{amssymb}
\usepackage{amsmath}
\usepackage{latexsym}
\usepackage{tikz}
\usepackage{sidecap,epstopdf}
\usepackage{subfigure}
\usepackage{color}
 \newtheorem{thm}{Theorem}
\newtheorem{lem}{Lemma}

\newcommand\eq[1] {(\ref{#1})}

\newcommand{\bfm}[1]{\mbox{\boldmath ${#1}$}}

\newcommand{\beqa}{\begin{eqnarray}}
\newcommand{\eeqa}{\end{eqnarray}}
\newcommand{\bequ}{\begin{equation}}
\newcommand{\eequ}[1]{\label{#1}\end{equation}}

\newcommand{\Go}{\omega}

\newcommand{\GO}{\Omega}

\newcommand{\GY}{\Psi}

\newcommand{\BGb}{\bfm\beta}

\newcommand{\BGg}{\bfm\gamma}

\newcommand{\BGx}{\bfm\xi}

\newcommand{\BGG}{\bfm\Gamma}

\newcommand{\CG}{{\cal G}}
\newcommand{\CH}{{\cal H}}

\newcommand{\CM}{{\cal M}}
\newcommand{\CN}{{\cal N}}

\newcommand{\CP}{{\cal P}}

\newcommand{\CS}{{\cal S}}
\newcommand{\CT}{{\cal T}}

\newcommand{\CV}{{\cal V}}

\newcommand{\BCT}{{\bfm{\cal T}}}

\def\Be{{\bf e}}

\def\Bx{{\bf x}}
\def\By{{\bf y}}
\def\Bz{{\bf z}}

\def\BB{{\bf B}}
\def\BC{{\bf C}}
\def\BD{{\bf D}}
\def\BE{{\bf E}}

\def\BG{{\bf G}}
\def\BH{{\bf H}}
\def\BI{{\bf I}}

\def\BO{{\bf O}}
\def\BP{{\bf P}}

\def\BR{{\bf R}}

\def\BV{{\bf V}}
\def\BW{{\bf W}}
\def\BX{{\bf X}}
\def\BY{{\bf Y}}
\def\BZ{{\bf Z}}

\newcommand{\beq}{\begin{equation}}
\newcommand{\eeq}{\end{equation}}
\newcommand{\overliner}{\begin{eqnarray}}
\newcommand{\earr}{\end{eqnarray}}
\newcommand{\beqn}{\begin{equation*}}
\newcommand{\eeqn}{\end{equation*}}
\newcommand{\overlinern}{\begin{eqnarray*}}
\newcommand{\earrn}{\end{eqnarray*}}

\title{Eigenvalue problem in a solid with many inclusions: asymptotic analysis }
\date{}
\author{V.G.  Maz'ya\footnote{Department of Mathematics, Link\"oping University, SE-581 83 Link\"oping, Sweden.}, A.B. Movchan\footnote{Department of Mathematical Sciences, University of Liverpool, Liverpool L69 3BX, U.K.}\,\, and M.J. Nieves\footnote{Mechanical Engineering and Materials Research Centre, Liverpool John Moores University, James Parsons Building, Byrom Street, Liverpool L3 3AF, U.K.}}
\begin{document}
\maketitle
\begin{abstract}
We construct the asymptotic approximation to the first eigenvalue  and corresponding  eigensolution of Laplace's operator inside a domain containing a cloud of small rigid inclusions. The separation of the small inclusions is characterised by a small parameter which is much larger compared with the nominal size of  inclusions.
Remainder estimates for the approximations to the first eigenvalue and associated eigenfield  are presented. 
Numerical illustrations are given to demonstrate the efficiency of the asymptotic approach compared to conventional numerical techniques, such as the finite element method, for three-dimensional solids containing
 clusters of small inclusions.
\end{abstract}
\section{Introduction and highlights of results}\label{intro}

The method of uniform asymptotic approximations for solids with large clusters of small defects has been developed in the series of papers \cite{MM_meso_1}, \cite{Maz_MMS}, \cite{MMN_Mesoelast}, \cite{MMN_Mesoelast_voids} and the book \cite{MMN_book}. The singular perturbation approach is applicable to the cases of clouds  containing large numbers of inclusions/voids with different boundary conditions on their surfaces. 

While the relative size of the inclusions is small, their overall number may be large, and the homogenisation algorithms for such mesoscale type domains are challenging, as discussed in \cite{MarKhrus} and \cite{Murat}.

In particular, the change of  eigenvalues due to a singular perturbation of the domain is an interesting and challenging problem, which is discussed in detail in \cite{OPTH1} for domains containing finite number of small inclusions.

Moreover, uniformity of asymptotic approximations for the eigenfunctions is a serious challenge, which is not addressed in the existing literature for eigenfunctions corresponding to large clusters of small inclusions. 

\subsection{Background and previous results}   

The method of compound asymptotic approximations is systematically presented in \cite{OPTH1, OPTH2} for solutions to a range of boundary value problems with small holes and irregular boundary points. This method can lead to  asymptotic expansions for integral characteristics of several quantities such as energy, stress-intensity factors and eigenvalues associated with such problems.
The method is versatile and has been used in the monograph \cite{KMMII} to treat problems concerning multi-structures  commonly found in civil engineering and many other applications in physics and applied mathematics.

When periodicity is prevalent within a  multi-structure or composite, powerful homogenisation based approaches are used to model these situations using the notion of an average medium \cite{Bakhvalov, SP}. This is a very effective tool in characterising the behaviour of the microstructure of composites, such as those  re-enforced by periodically placed fibers \cite{PC1, PC2} that are subjected to different loads. Averaging procedures have been adopted in \cite{W_PC}, to model the overall behaviour of materials with regions containing randomly distributed  inclusions.

The method proposed in \cite{OPTH1, OPTH2} was  important in  the recent development in the asymptotic treatment of solutions to boundary value problems with non-smooth loading terms
in singularly and regularly perturbed domains \cite{CRM}. 

Uniform approximations of singular solutions  in  a domain with a small rigid perforation have been presented in \cite{JCOM}. 
Uniform  approximations of fields in solids with impurities supplied with different boundary conditions have appeared, for instance, in \cite{Sob_vol} for traction free boundaries or in  \cite{AMS_tran} for  transmission conditions.  
The asymptotic scheme uses model problems in the domain without defects and boundary layers posed in the exterior of a single defect. Different boundary conditions require different boundary layers. In the case of rigid boundaries, corresponding to the Dirichlet boundary conditions, we invoke the the notion of  capacity associated with the inclusion \cite{OPTH1}. When the Neumann conditions are supplied on small voids, the asymptotic algorithm must be modified and  dipole characteristics for the impurity should be used to construct correction terms in the approximation \cite{MMP}.
Approximate  Green's functions in thin or long rods have appeared in \cite{MMAS}.  
Uniform asymptotics in multiply perforated bodies for problems of vector elasticity were constructed in   \cite{RAN, AA}. Uniform asymptotic approximations of Green's kernels have been used to study a Hele-Shaw flow containing several obstacles in \cite{Pecketal}.

\subsubsection*{Compound asymptotic expansions}
In \cite{OPTH1}, the method of compound asymptotic expansions is used to develop asymptotic formulae for a variety of eigenvalue problems for Laplace's operator in two- or three-dimensional  domains with small rigid inclusions or voids. Approximations of this type allow one to determine the behaviour of the effect of the perturbation to the first eigenvalues when these defects are introduced. In contrast with what is analysed here, these approximations are built on the assumption that the small defects are  separated by a finite distance and are not situated near the external boundary.
Extension of the results to the vector case of elasticity is demonstrated. In addition, asymptotics of the first eigenvalue and eigenfunction are constructed for the case of a Riemannian manifold with a small rigid inclusion. 

For the body  a single rigid inclusion and with zero external forces on the exterior, a complete asymptotic series is constructed for the first eigenvalue and the corresponding eigenfunction. For this mixed problem, to leading order, the approximation to the first eigenvalue does not contain information about the position of the inclusion inside the body. It does rely on knowing the capacity of the small inclusion and the volume of the set without inclusions, which   depend on the shape and size of the inclusion or body, respectively. The leading order approximation to the eigenfunction uses the capacitary potential for the exterior  of the inclusion, see \cite{OPTH1, MMN_book}, which decays sufficiently fast at infinity, along with model problems for the domain without the defect (which includes Green's function for this domain). A similar approach can be used to tackle the equivalent problem but for a  domain containing an arrangement of a finite number of inclusions. In this case, the size of the first eigenvalue is shown to grow as the number of inclusions increases. 
For the two-dimensional case, the functions used to construct static boundary layers in  the exterior of small holes
 have a logarithmic growth at infinity.

Other asymptotic approximations for the first eigenvalue and the associated eigenfunction of the Laplacian, that rely on the use of the capacity of an inclusion, include those for Dirichlet's problem in a 3-dimensional domain with a small inclusion \cite{OPTH1}.  

When rigid inclusions are introduced into the body, one can expect the first eigenvalue to increase, which is a feature predicted by the  asymptotic approximations. As mentioned above, asymptotic representations of the type found in \cite{OPTH1} are useful in determining how the geometry of the perforated domain influences the change in the first eigenvalue when a void is introduced. Here, boundary layers are constructed using dipole fields for the void, which decay quicker than those in the case of a rigid inclusion.
As a result, the asymptotic approach demonstrates that the perturbation to the first eigenvalue of Laplace's operator is smaller than  the case when a rigid inclusion is situated in this domain. 
In addition, introducing a void into the domain does not necessarily increase the first eigenvalue as with the case of a Dirichlet type inclusion.
 One can find cases where this quantity decreases or increases and this change depends on the position of the hole or properties of the first eigenfunction for the domain without holes.
 
In \cite{OPTH1}, asymptotics of eigenvalues and eigenfunctions are presented for Dirichlet's problem on a Riemannian manifold with a small hole. In particular, here the leading order term of the first eigenvalue depends on the logarithmic capacity of the small inclusion. Examples of this approximation have been demonstrated for the  surface of the sphere with a small rigid inclusion.

The compound asymptotic approximations mentioned above provide a framework for the the extension of the theory to more complicated systems, such as that found in vector elasticity. In \cite{OPTH1},  approximations for first eigenvalues and associated  eigenfunctions  for elastic bodies containing small soft inclusions in three-dimensional and planar bodies with cavities are presented.

 \subsubsection*{Homogenisation approximations}
 
 Initial boundary value problems for diffusion phenomena  in heavily perforated solids  have been considered in \cite{MarKhrus}, using homogenisation based techniques.  As the overall number of perforations becomes large  the convergence of the considered problem  to a limit problem is studied and the authors show the appearance of additional terms in the governing equations. 
For the Dirichlet problem, such a term is proportional to the limit problem's solution  and  its coefficient depends on the capacity of the perforations.
In the scenario when Neumann conditions are imposed on the voids, such additional terms include those which show that during the diffusion process in the perforated medium, this medium has a memory.
For the diffusion problem, if one considers an asymptotic approximation inside such a medium, the boundary layers for small holes or cavities decay exponential fast away from the defects.
It should be noted that for the problems treated in  \cite{MarKhrus} explicit asymptotic representations of the fields inside the perforated domains is not given, whereas results of this type based on the method of compound asymptotic expansions appear in, for example,  \cite{OPTH1, MM_meso_1, Maz_MMS}.

 The methods developed in \cite{MarKhrus}, assume the defect size and the minimum separation between neighbouring defects satisfy a constraint similar to that imposed here in (\ref{epstod}). This constraint is unavoidable in the analysis as it governs the solvability of the system (\ref{sys_C_thm}) as shown in section \ref{algsys}. 
 The homogenisation approach of \cite{MarKhrus} also depends on the microstructure of the perforated medium satisfying some periodicity constraints or that is governed by some probability law. In this paper, the analysis relies on no such assumptions on the position of the defects.
 
 The eigenvalue problem for the Laplacian inside a heavily perforated $n$-dimensional  solid ($n\ge 2$) containing voids, corresponding to the Neumann conditions, have also been treated in \cite{MarKhrus}. Again, to treat this problem asymptotically, one should invoke the dipole characteristics of individual voids that enjoy a greater decay than those in the case of rigid inclusions if one considers the far-field behaviour. There, in addition to understand the convergence to the limit problem, the authors also analyse the spectrum in the limit and how this arises as the number of voids grows. Again, explicit asymptotic representations are not given for both eigenvalues and corresponding eigenfunctions. 
 
 Compared with \cite{MarKhrus}, we analyse the eigenvalue problem for the Laplacian inside a
  domain with a  densely perforated region containing rigid inclusions, with the Dirichlet boundary conditions. This approach leads to an explicit asymptotic structure for both the first eigenvalue and corresponding eigenfunction for this problem (see Theorems \ref{th1} and \ref{th1a}). In addition, the asymptotic approximation of the eigenfunction is uniform throughout the strongly  perforated solid. 
  

The  approximations for cluster configurations work well when the holes are few and are separated far from each other and remote from  the exterior boundary. 
In particular, interesting effects on the governing equations  can be observed when the number of obstacles  in a region increase,  while their nominal size decreases. This has been studied in \cite{MarKhrus}, where an equation representing the effective properties of a heavily perforated medium appears in this limit. 
The analysis of a collection of many randomly distributed obstacles has been considered in \cite{Fig_1} for the Dirichlet problem  and \cite{Fig_2} for a mixed problem of the Laplacian. There, the convergence of the governing equation to the limit operator was studied.

Here, we seek a different type of approximation suitable for the case when the small inclusions can be close to one another and their number is large. Such approximations, are known as mesoscale asymptotic approximations, which do not require any assumptions on the periodicity of the cluster of defects. They serve the   intermediate case between a finite number of voids and a cluster of defects. Mesoscale approximations originated in  \cite{MM_meso_1}, concerning the Dirichlet boundary value problem for the Laplacian in a densely  perforated domain. Mixed boundary value problems for a domain with many small voids were treated in \cite{Maz_MMS}. 
Extension of the mesoscale approach to vector elasticity has been carried out for a solid with a large number of small rigid defects \cite{MMN_Mesoelast} and voids \cite{MMN_Mesoelast_voids}.
A collection of approximations of Green's kernels and solutions to boundary value problems 
in domains with  finite collections or mesoscale configurations of perforations, respectively, can be found in the monograph \cite{MMN_book}.
Applications of the mesoscale approach have also appeared in \cite{ChallaSini2, ChallaSini3} where the remote scattered field produced by a  cluster in an infinite medium has been studied.

\subsection{Highlights of the results}

\begin{figure}%
  \centering
      %
\centering
        \includegraphics[width=0.9\textwidth]{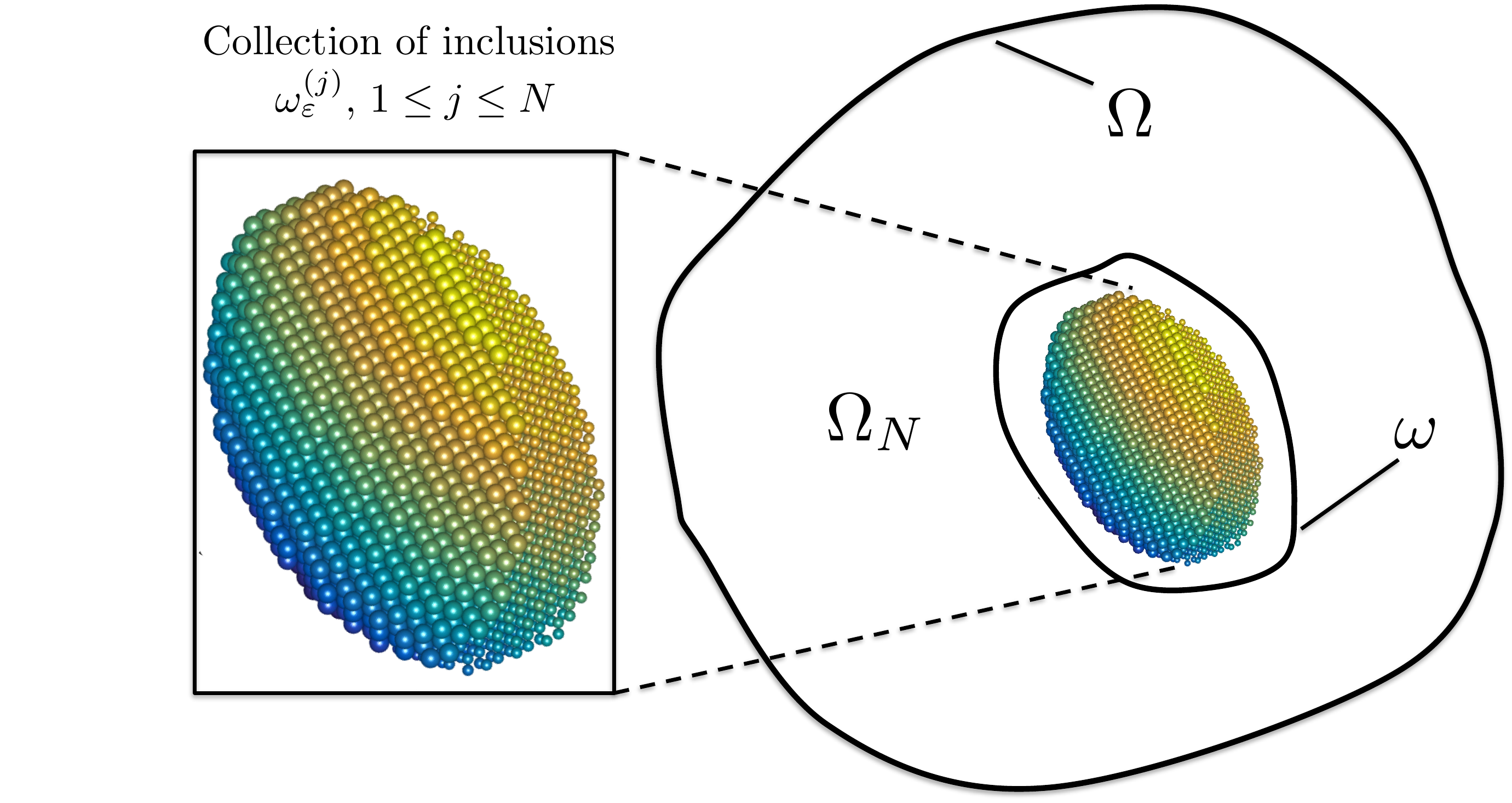}
        \caption{A non-periodic  cluster of inclusions $\omega^{(j)}_\varepsilon$, $1\le j \le N$, contained inside the set $\omega$, which is a subset of $\Omega_N:=\Omega\backslash \cup_{j=1}^N \overline{\omega_\varepsilon^{(j)}}$. }
         \label{fig:exmeso-0_intro}
        \end{figure}

In the present paper, we extend the analysis of eigenvalues and eigenfunctions in solids with a finite number of holes, in \cite{OPTH1}, to the case  of large clusters of small inclusions, as shown in Fig. \ref{fig:exmeso-0_intro}.  
The asymptotic approximation for the first eigenvalue and corresponding eigenfunction of the Laplacian for various boundary value problems in  domains with a single small hole, can be found in \cite{OPTH1}. The case of elasticity is also considered there, along with the extension to  the scalar case with multiple defects.
Asymptotic analysis of the spectral problem for elasticity in an  anisotropic and inhomogeneous body has been carried out in \cite{Nazarov}.
The spectral problem for the plate containing a single small clamped  hole and corresponding asymptotics of the first eigenvalue and corresponding eigenfunction can be found in \cite{Campbell}. 
For Dirichlet problems, asymptotics of spectra for $-\Delta$ inside $n$-dimensional domains with a single small ball has been treated in \cite{Ozawab, Ozawaa,  Ozawac}. For mixed problems, asymptotics of eigenfunctions and eigenvalues for the Laplacian in a 2-dimensional domain containing a small circular hole with the Neumann or Robin condition were constructed in \cite{Ozawa2, Ozawa3}.  A similar analysis of spectra has been carried out for domains in  $\mathbb{R}^n$ containing a spherical void \cite{Ozawa4}. Homogenisation based techniques have also been developed in \cite{Craster} to tackle problems when periodic lattices are subjected to high-frequency vibrations.

We consider an eigenvalue problem in a three-dimensional domain $\Omega_N$ containing a cluster of $N$ small inclusions $\omega^{(j)}_\varepsilon$, $1\le j \le N$, with homogeneous Dirichlet boundary conditions on their surfaces, and the  Neumann boundary condition on the exterior boundary $\partial \Omega$.  Here $\Omega$ is the set without any inclusions and  $\Omega_N:=\Omega\backslash \cup_{j=1}^N\overline{\omega_{\varepsilon}^{(j)}}$. Each inclusion $\omega^{(j)}_\varepsilon$ has smooth boundary, a diameter characterised by a small parameter $\varepsilon$ and contains an interior point $\BO^{(j)}$, $1\le j \le N$. We assume the minimum separation between any pair of such points within the cloud is characterised by $d$, defined by
\[d=2^{-1}\min_{\substack{k\ne j \\ 1\le j, k\le N}}|\BO^{(k)}-\BO^{(j)}|\;.\]
In addition to the above sets, we assume there exists a set $\omega\subset \Omega_N$ such that
\begin{equation}\label{assumptions}
\cup_{j=1}^N \omega^{(j)}_\varepsilon \subset \omega\;,\quad \quad \text{dist}(\cup_{j=1}^N \omega^{(j)}_\varepsilon, \partial \omega)=2d\quad \text{ and } \quad \text{dist}(\omega, \partial \Omega)=1\;.
\end{equation}
For $D\subset \mathbb{R}^3$ we denote by $|D|$ the three-dimensional measure of this set.

 We construct a high-order approximation for the first eigenvalue $\lambda_N$, and develop a uniform asymptotic approximation of the corresponding eigenfunction $u_N$, which is 
a solution of:
\begin{equation}\label{uN1}
\Delta_{\Bx} u_N(\Bx)+\lambda_N u_N(\Bx)=0\;, \quad \Bx \in \Omega_N:=\Omega\backslash \cup_{j=1}^N\overline{\omega^{(j)}_\varepsilon}\;,
\end{equation}
\begin{equation}\label{uN2}
\frac{\partial u_N}{\partial n_\Bx}(\Bx)=0\;, \quad \Bx \in\partial  \Omega\;,
\end{equation}
\begin{equation}\label{uN3}
{u_N}(\Bx)=0\;, \quad \Bx \in\partial  \omega_\varepsilon^{(j)}, \quad 1\le j\le N\;,
\end{equation}
where $N$ is  considered to be large. 

Our approximations rely on model problems in $\Omega$ and the exterior of $\omega_\varepsilon^{(j)}$, $1\le j \le N$. In particular, the approximation is formed using 
\begin{enumerate}
\item the regular part $\CH$ of Neumann's function $\CG$ in $\Omega$,
\item the capacitary potential $P^{(j)}_\varepsilon$ of $\omega^{(j)}_\varepsilon$, 
\item  quantities  such as the capacity $\text{cap}(\omega_\varepsilon^{(j)})$ of the set $\omega_\varepsilon^{(j)}$ and 
\begin{equation}\label{eqGAMOM}
\Gamma_\Omega^{(j)}=\frac{1}{|\Omega|}\int_{\Omega}\frac{d\Bz}{4\pi|\Bz-\BO^{(j)}|}\;.
\end{equation}
\end{enumerate}
Here we present the following theorem concerning the first eigenfunction for Laplace's operator in $\Omega_N$:

\begin{thm}\label{th1} Let 
\begin{equation}\label{epstod}
\varepsilon< c\, d^3
\end{equation}
where $c$ is a sufficiently small constant.
Then the asymptotic approximation of the eigenfunction $u_N$, which is a solution of $(\ref{uN1})$--$(\ref{uN3})$ in $\Omega_N$, is given by
\begin{eqnarray}
u_N(\Bx)&=&1+\sum_{j=1}^N
{ C_j \Gamma^{(j)}_\Omega\text{\emph{cap}}(\omega^{(j)}_\varepsilon)}
\nonumber \\
&&+\sum^N_{j=1} C_j \{P^{(j)}_\varepsilon(\Bx)-\text{\emph{cap}}(\omega^{(j)}_\varepsilon)\CH(\Bx, \BO^{(j)})\}+R_N(\Bx)\;,
\label{uN_app}
\end{eqnarray}
where $R_N$ is the remainder term, 
and the coefficients $C_k$, $1\le k \le N$, satisfy the solvable algebraic system
\begin{eqnarray}
&&1+C_k(1-\text{\emph{cap}}(\omega^{(k)}_\varepsilon)\{\CH(\BO^{(k)},\BO^{(k)})-\Gamma_\Omega^{(k)} \})\nonumber \\ \nonumber\\
&&+\sum_{\substack{j\ne k\\ 1\le j\le N}} {C_j\text{\emph{cap}}(\omega_\varepsilon^{(j)})}\Big\{\CG(\BO^{(k)}, \BO^{(j)})+\Gamma^{(j)}_\Omega
\Big\}=0\;,\quad 1\le k \le N\;.\label{sys_C_thm}
\end{eqnarray}
 Here $R_N$ satisfies the estimate
\begin{equation}\label{remainderest1}
\|R_N\|_{L_2(\Omega_N)}\le \text{\emph{Const} } \varepsilon^{2} d^{-6} \;.
\end{equation}

\end{thm}

We also present the next theorem, for the corresponding first eigenvalue:
\begin{thm}\label{th1a} Let the small parameters $\varepsilon$ and $d$ satisfy $(\ref{epstod})$
 Then the first eigenvalue  $\lambda_N$ corresponding to the eigenfunction $u_N$ admits the approximation
\begin{equation}\label{lambda_N_app}
\lambda_N=-\frac{1}{|\Omega|}\sum^N_{j=1} {C_j\text{\emph{cap}}(\omega^{(j)}_\varepsilon)}+
O(\varepsilon^{2} d^{-6} )
\;.
\end{equation}
\end{thm}

\begin{figure}\centering
     \subfigure[][]{
       \label{fig:exmeso-0}
\centering
        \includegraphics[width=0.6\textwidth]{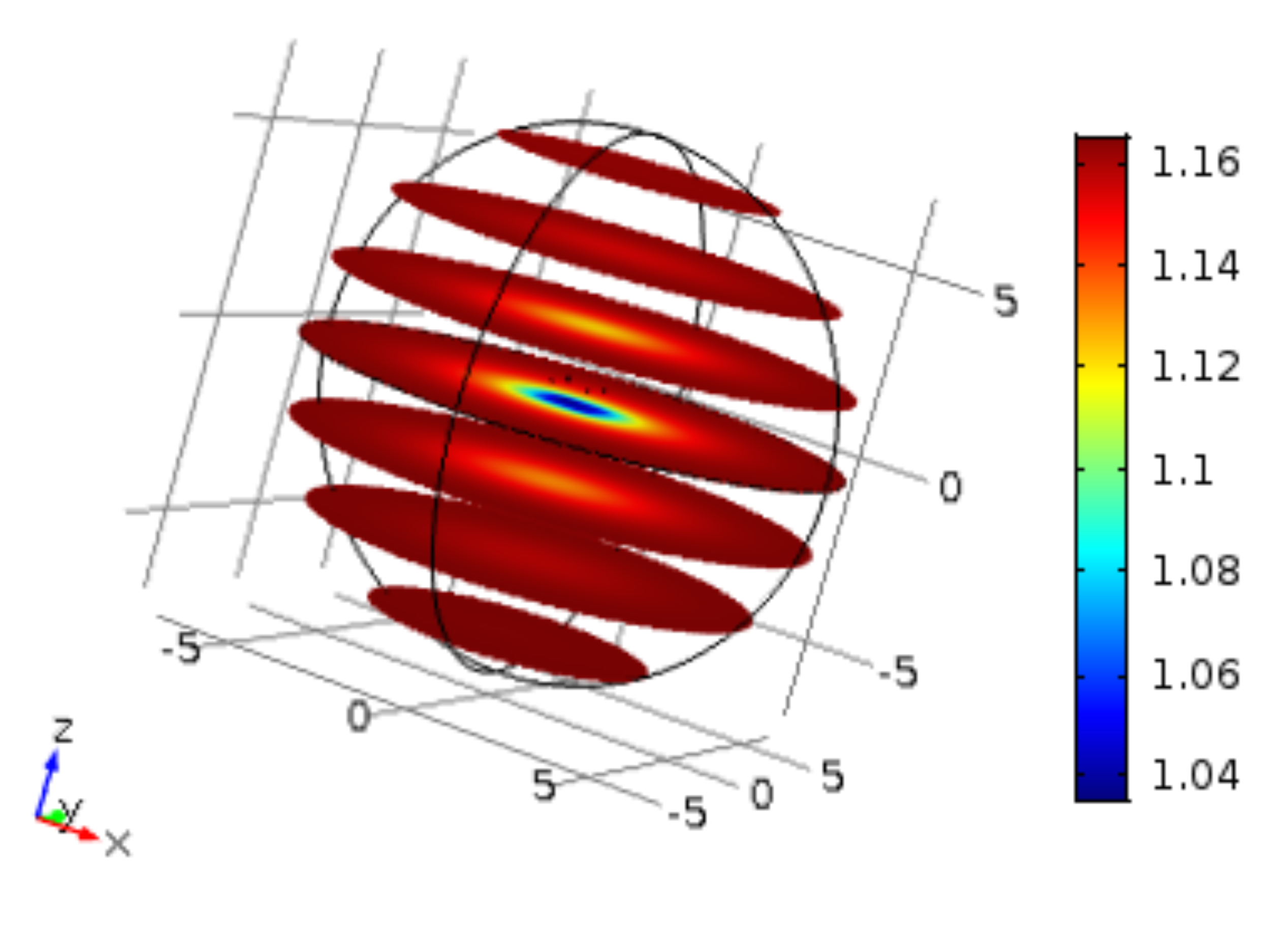}}\\
         \centering
     \subfigure[][]{
       \label{fig:exmeso-a}
\centering
        \includegraphics[width=0.49\textwidth]{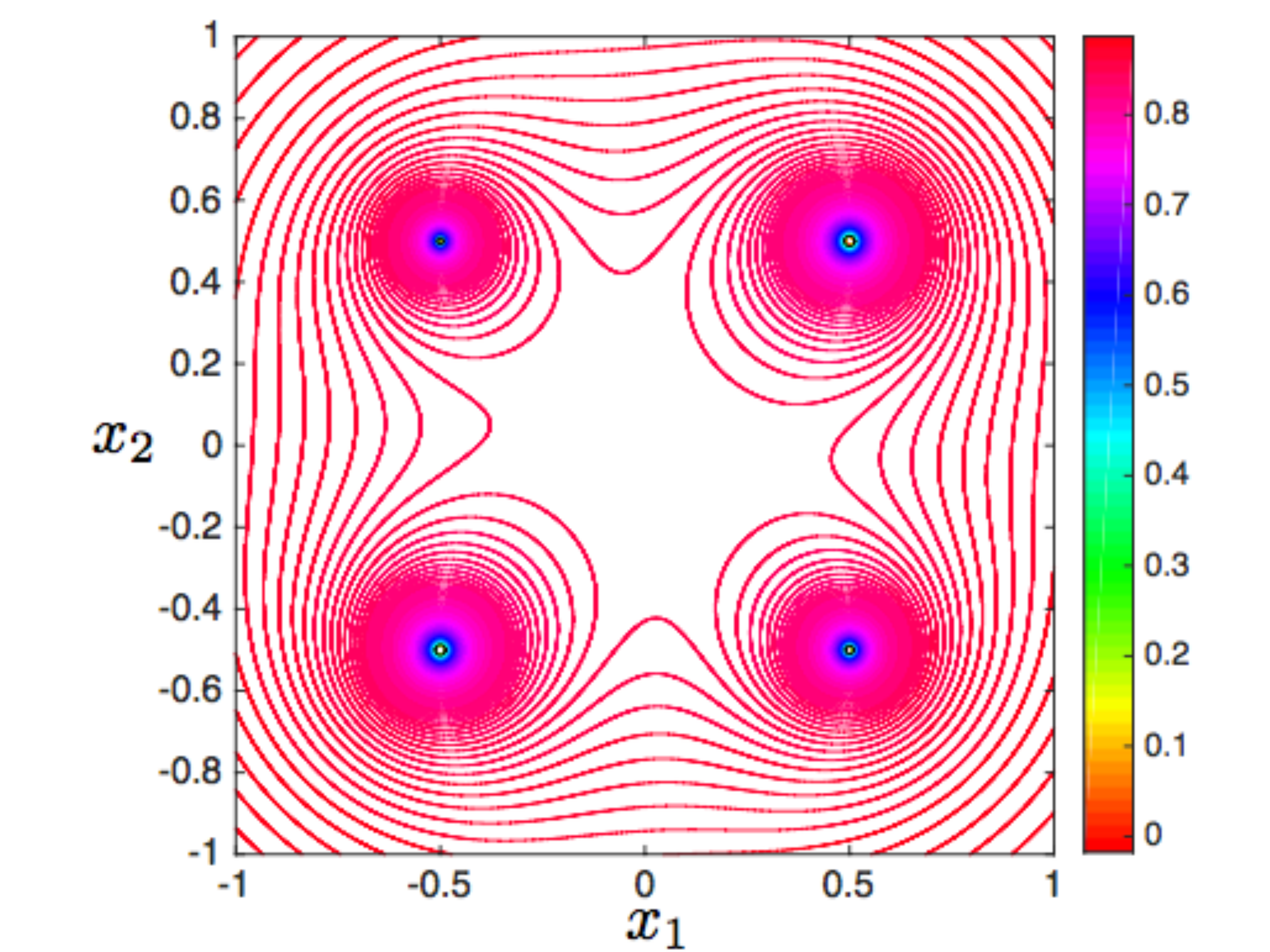}}
         \subfigure[][]{%
       \label{fig:exmeso-b}%
\centering
        \includegraphics[width=0.49\textwidth]{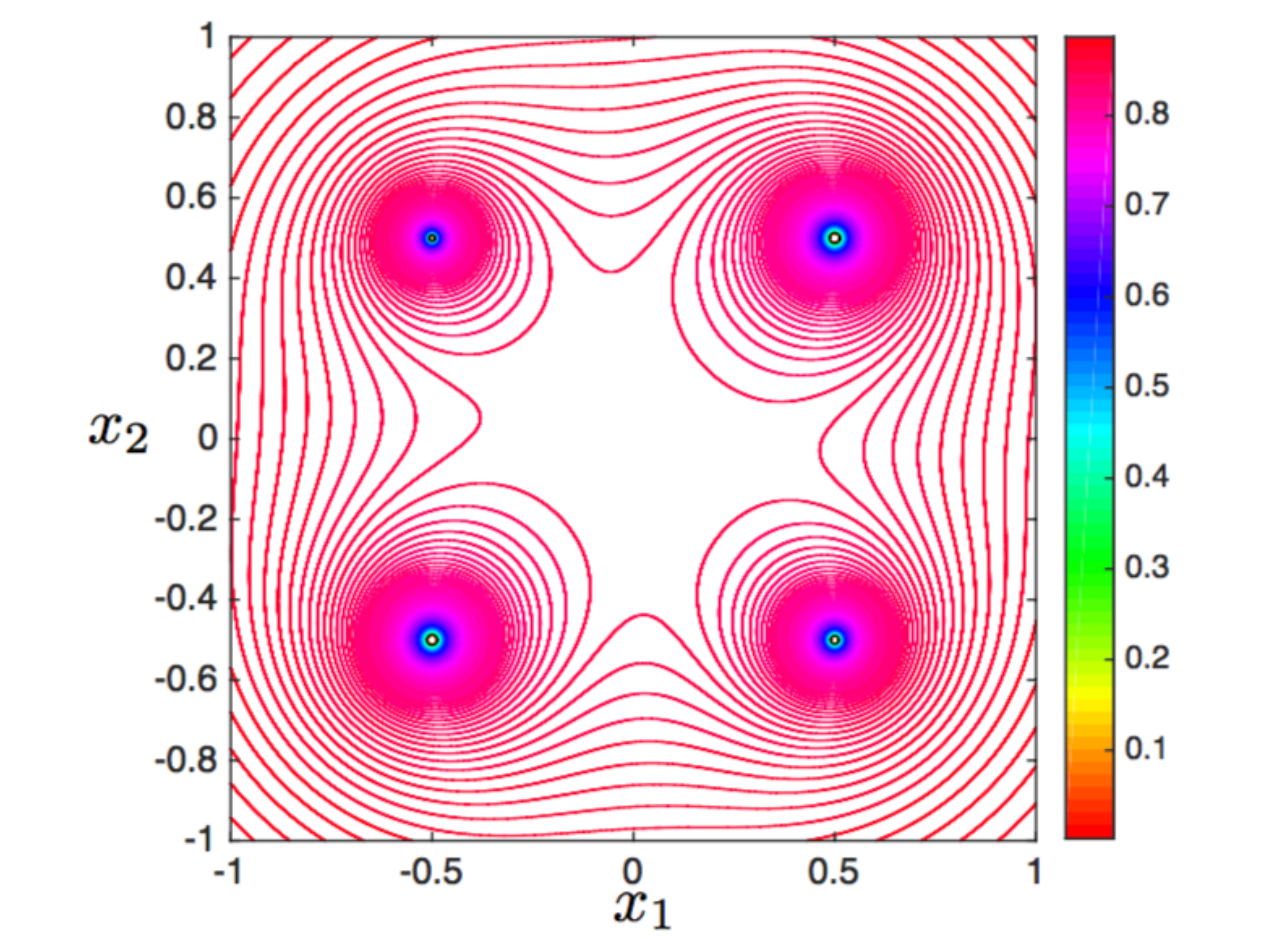}}
         \subfigure[][]{
       \label{fig:exmeso_c}%
\centering
        \includegraphics[width=0.49\textwidth]{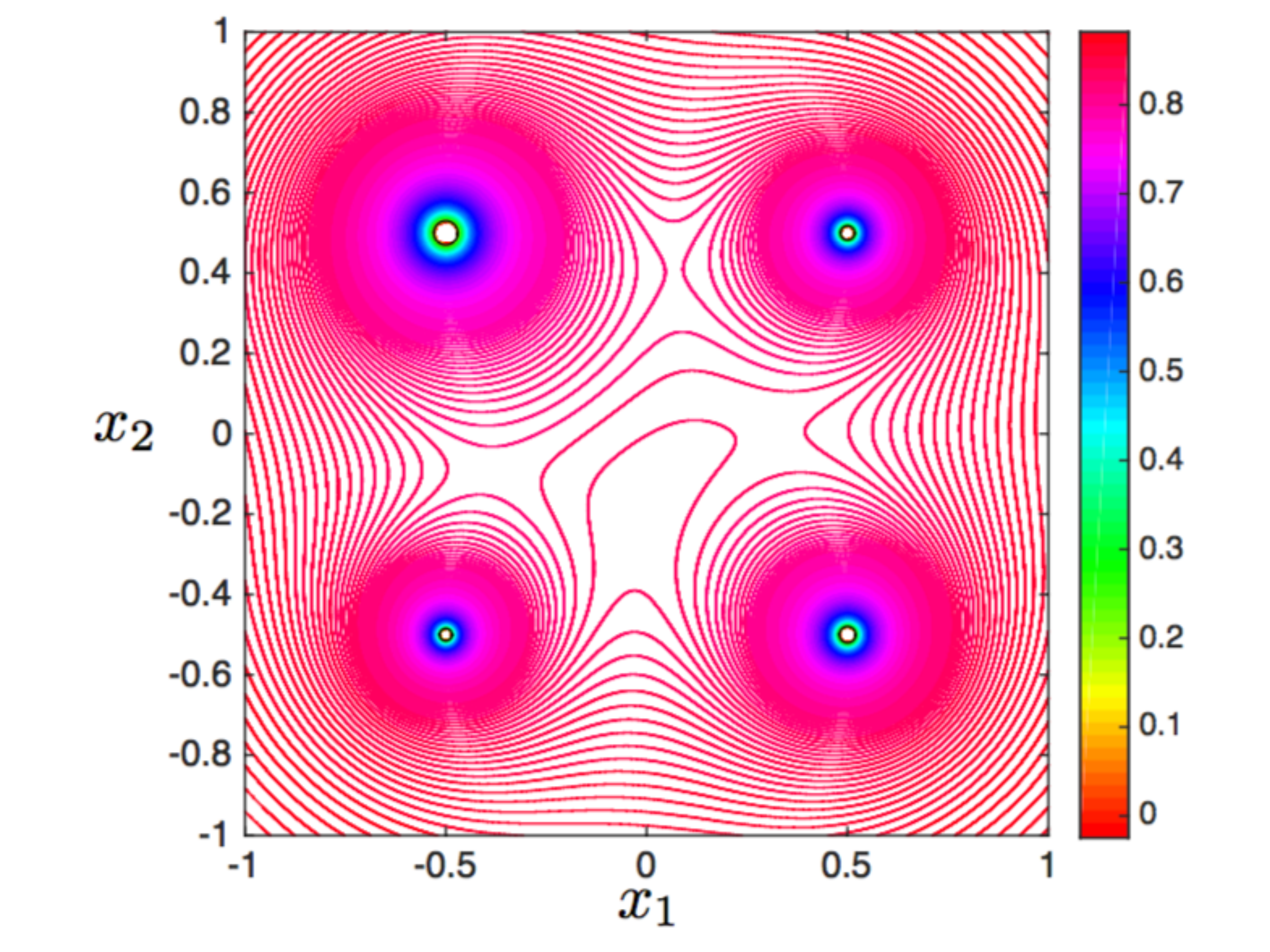}}
           \subfigure[][]{
       \label{fig:exmeso_d}%
\centering
        \includegraphics[width=0.49\textwidth]{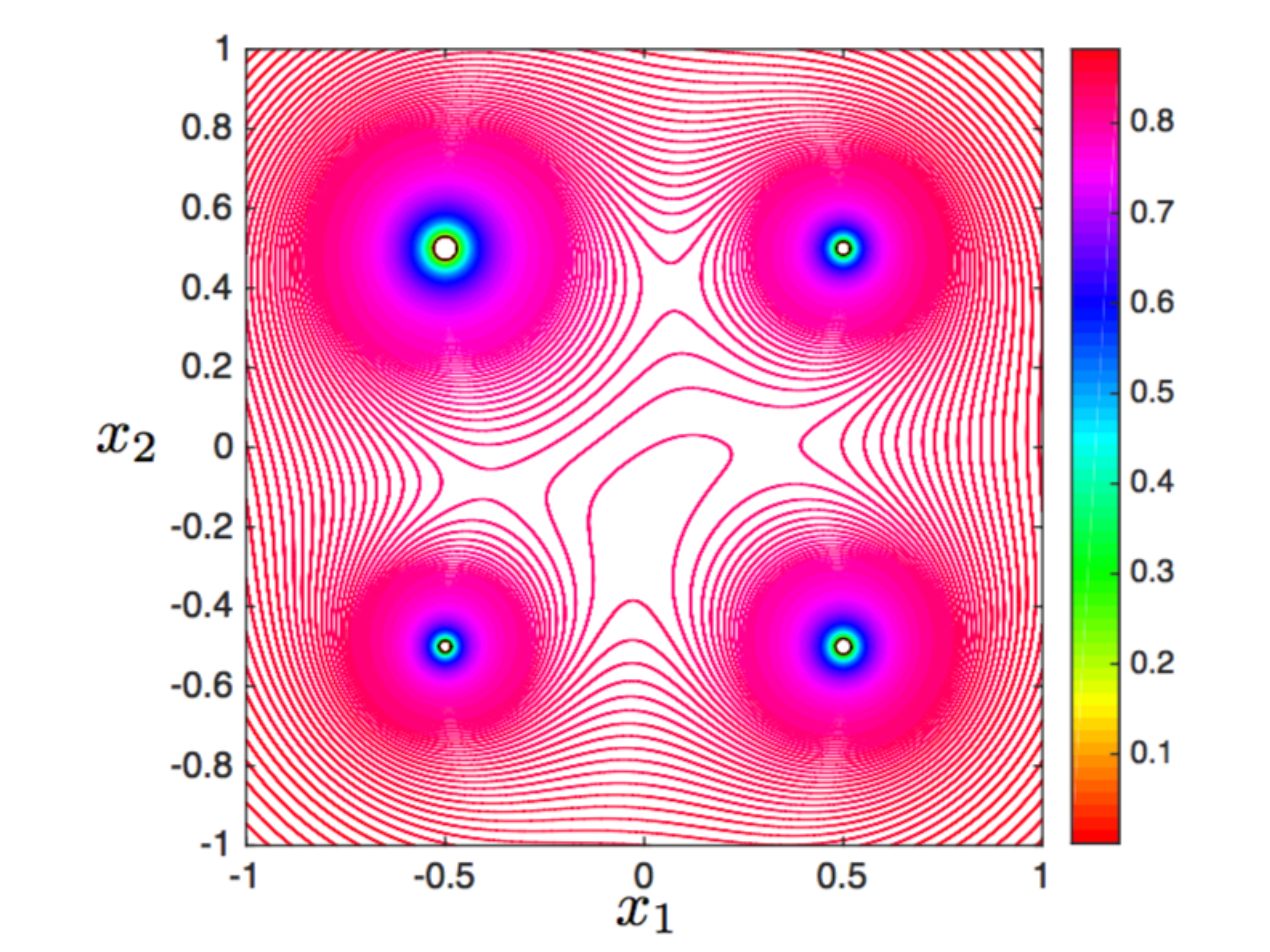}}

\caption[]{ (a) A slice plot of the eigenfield inside a sphere, containing 8 small spherical inclusions, computed using the method of finite elements in COMSOL on a mesh with 1477957 elements. Contour plot of the eigenfield along the planes  (b) $x_3=-0.5$  and (d) $x_3=0.5$ based on the computations from COMSOL. The contour plot of the eigenfield on the planes (c) $x_3=-0.5$ and (e)  $x_3=0.5$ computed using the asymptotic approximation (\ref{uN_app}).  The average absolute error between the computations in (b) and (c) is $2.1 \times 10^{-3}$, whereas  between (d) and (e) it is $3.3 \times 10^{-3}$.
}%
\label{fig:exmeso}%
\end{figure}

\begin{figure}
     \centering
     \subfigure[][]{
       \label{fig:exmeso-ab}
\centering
        \includegraphics[width=0.7\textwidth]{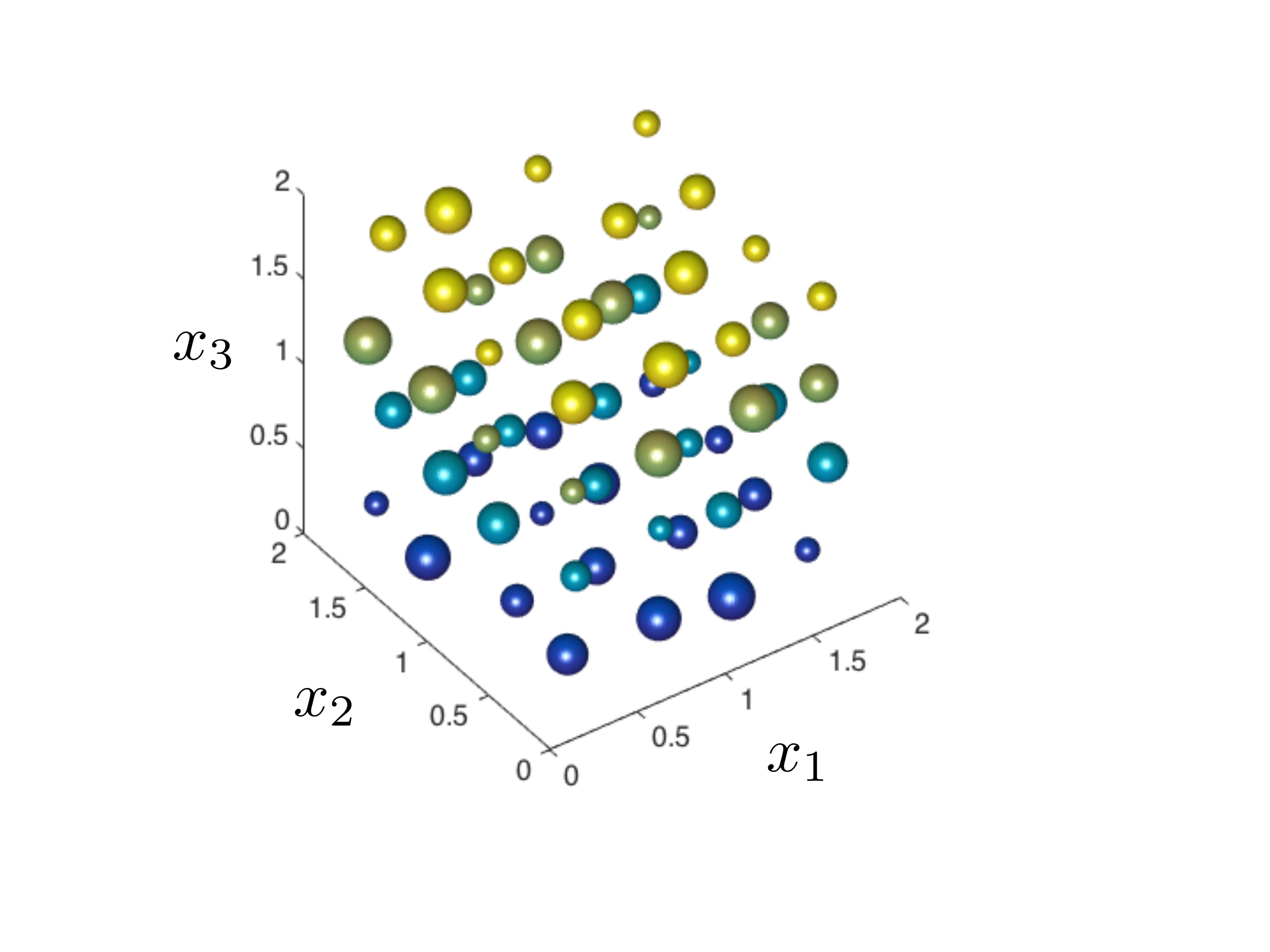}}
         \subfigure[][]{
       \label{fig:exmesoa-a}
\centering
        \includegraphics[width=0.49\textwidth]{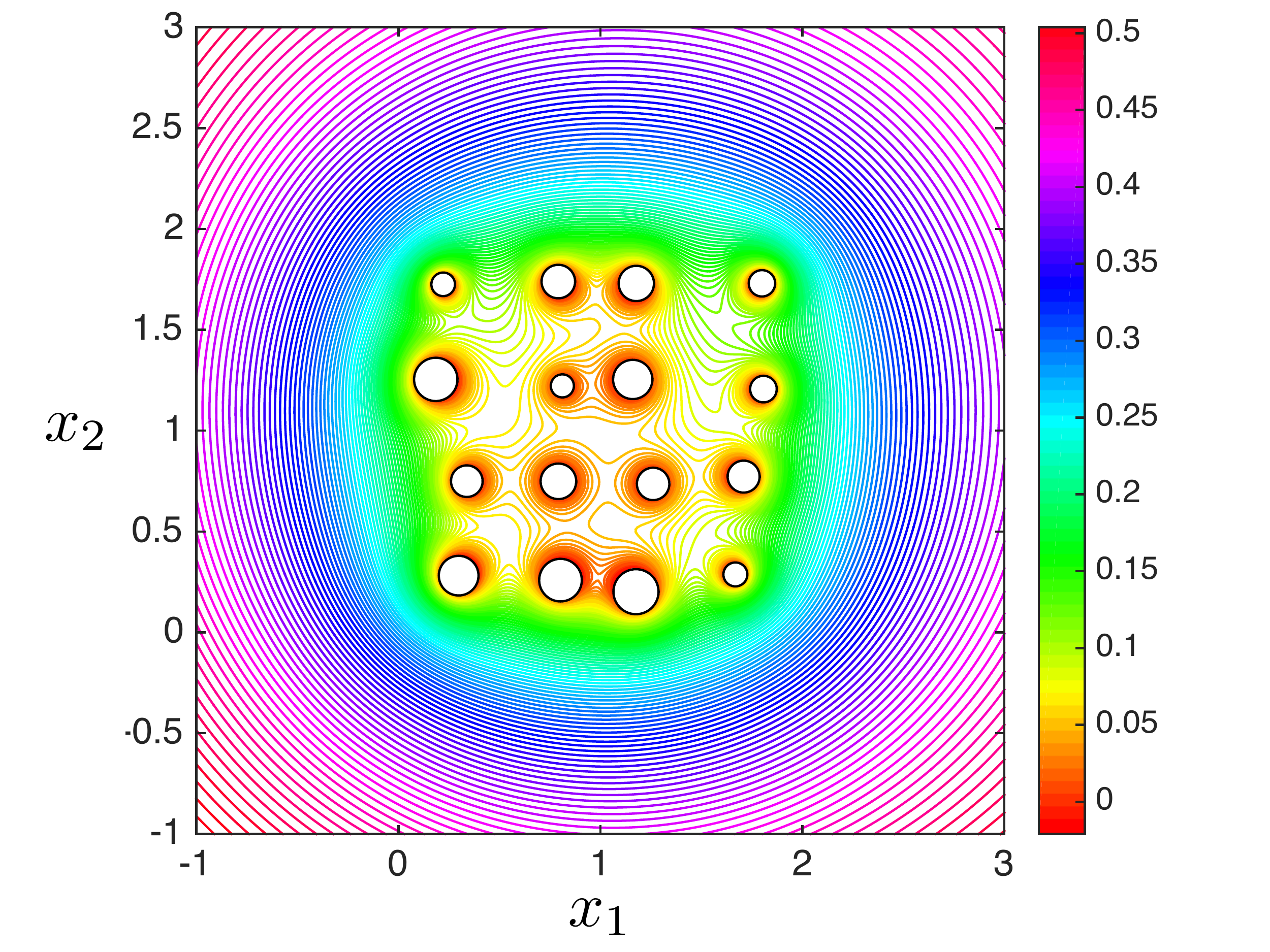}}
         \subfigure[][]{%
       \label{fig:exmesoa-b}%
\centering
        \includegraphics[width=0.49\textwidth]{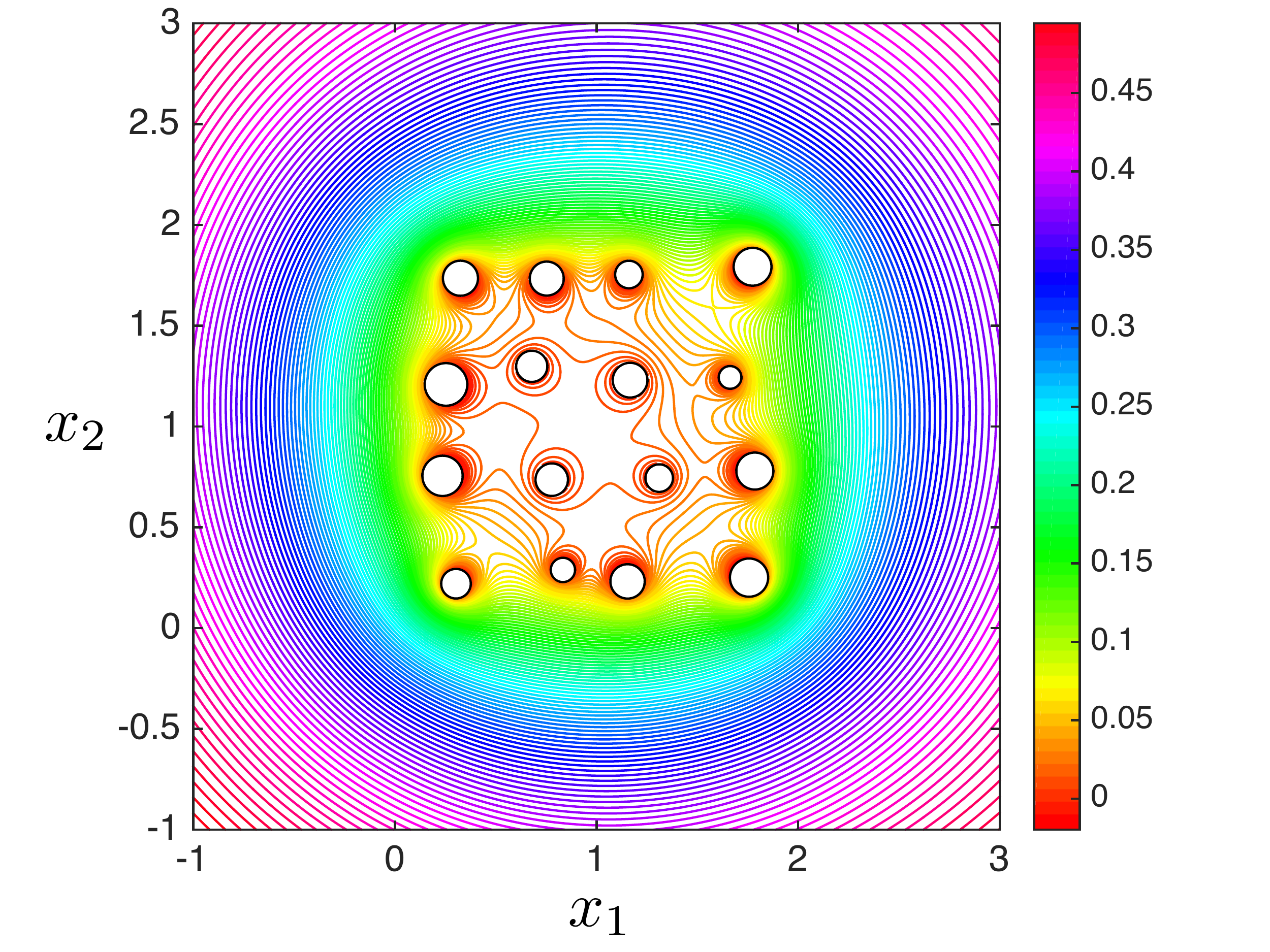}}
         \subfigure[][]{
       \label{fig:exmesoa_c}%
\centering
        \includegraphics[width=0.49\textwidth]{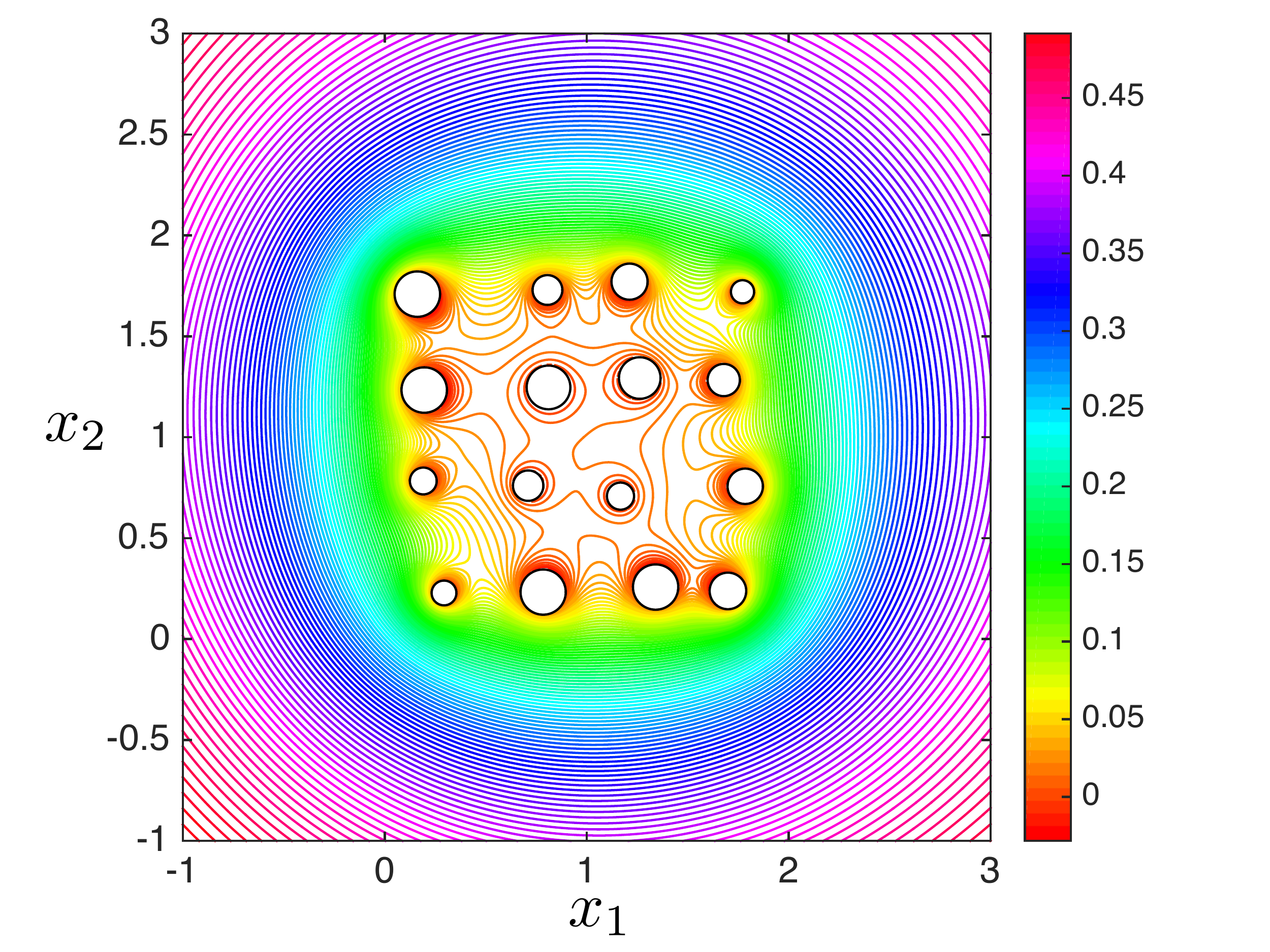}}
           \subfigure[][]{
       \label{fig:exmesoa_d}%
\centering
        \includegraphics[width=0.49\textwidth]{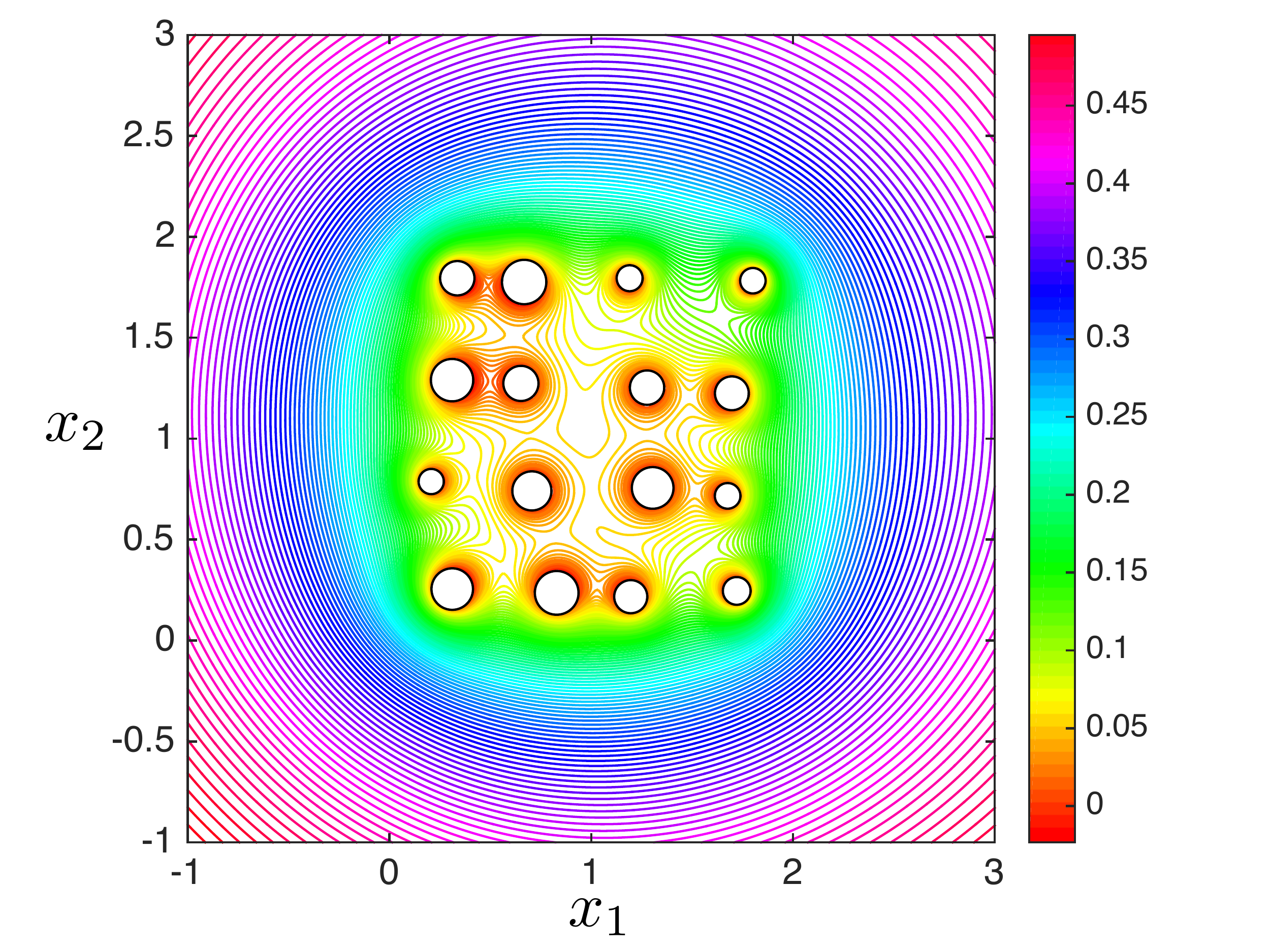}}

\caption[]{(a) The cloud of 64 small inclusions contained in the cube $(0,2)^3$.  (b)--(e) The asymptotic approximation for  eigenfield corresponding to the first eigenvalue in the ball of radius 7, centred at the origin, and  containing the cloud of inclusions. We show the cross-sectional plots on the planes  (b) $x_3=0.25$, (c) $x_3=0.75$, (d) $x_3=1.25$ and (e) $x_3=1.75$.}%
\label{fig:exmeso_aa}%
\end{figure}

For the purpose of illustration, in Fig. \ref{fig:exmeso} we show the analytical asymptotic approximation versus the finite element simulation produced for a cluster of 8 Dirichlet-type inclusions on several cross-sections. The first eigenfunction in the overall three-dimensional domain with the cluster of inclusions is shown on Fig. \ref{fig:exmeso-0}. 

The amount of memory required to run finite element computations increases substantially when the number $N$ of inclusions becomes large. For example, in 3-dimensions, with $N=64$ inclusions in a cluster COMSOL fails due to lack of memory on a standard 16GB workstation. On the other hand, the proposed asymptotic algorithm remains robust and efficient with the results shown in Figure \ref{fig:exmeso_aa}. The positions and the radii of inclusions are arbitrary, subject to constraints outlined earlier. In addition to the 3-dimensional illustration in Figure \ref{fig:exmeso-ab}, we also show several cross-sectional plots in Figures \ref{fig:exmesoa-a}--\ref{fig:exmesoa_d}. The asymptotic approximations are uniform and take into account mutual interaction between the inclusions with the cluster.

The structure of the article is as follows. In section \ref{modprob} we formally introduce model problems necessary to compute the approximations (\ref{uN_app}) and (\ref{lambda_N_app}). Formal asymptotic derivations of (\ref{uN_app}) and (\ref{lambda_N_app}) are then given in section \ref{formalalg}. Solvability of the system (\ref{sys_C_thm}) is proven in section \ref{algsys}. We provide the steps used to attain the  remainder estimates (\ref{remainderest1}) and  (\ref{lambda_N_app}) in section \ref{remainderestsec}. 
and going further in section  \ref{secHOA}, the higher-order approximation for the first eigenvalue and corresponding eigenfunction are given along with the completion of the proof of Theorems \ref{th1} and \ref{th1a}. A comparison of the approximations of Theorems \ref{th1} and \ref{th1a} are compared with those produced by the method of compound asymptotic expansions \cite{OPTH1} in section \ref{finiteVlarge}.
 In section \ref{numsec}, we further demonstrate the effectiveness of the approach presented here, by comparing (\ref{lambda_N_app}) with numerical computations of eigenvalues for  solids containing non-periodic clusters produced in COMSOL. In section \ref{hsolsec}, we discuss the homogenised problem obtained from the algebraic system (\ref{sys_C_thm}) in the limit as the number of inclusions within the cluster grow. 
  Finally in the Appendix, we present technical steps of the derivation to a higher order approximation of the first eigenvalue and corresponding eigenfunction given in section \ref{secHOA}. 

\section{Model problems}\label{modprob}
We now introduce solutions to model problems that are necessary in constructing the  asymptotic approximations for $\lambda_N$ and $u_N$.
\begin{enumerate}
\item {\bf The Neumann function in $\Omega$. } Here, $\CG$ denotes the Neumann function  in $\Omega$, which is a solution of 
\begin{equation}\label{N1}
\Delta_{\Bx} \CG(\Bx, \By)+\delta(\Bx-\By)-\frac{1}{|\Omega|}=0\;, \quad \Bx \in \Omega\;,
\end{equation}
\begin{equation}\label{N2}
\frac{\partial \CG}{\partial n_\Bx}(\Bx, \By)=0\;, \quad \Bx \in\partial  \Omega\;.
\end{equation}
This definition of $\CG$ is also supplied with the orthogonality condition
\[\int_{\Omega} \CG(\Bx, \By) d\Bx=0\;,\]
which implies the symmetry of $\CG$:
\[\CG(\Bx, \By)=\CG(\By, \Bx), \quad \Bx, \By\in \Omega\;.\]
We also introduce the regular part $\CH$ of the Neumann function as
\[\CH(\Bx,\By)=\frac{1}{4\pi |\Bx-\By|}-\CG(\Bx, \By)\;.\]
\item {\bf Capacitary potential for the inclusion $\omega_\varepsilon^{(j)}$. } The capacitary potentials $P^{(j)}_\varepsilon$, $1\le j \le N$, are used to construct boundary layers in the exterior  of the small inclusions. 
The function $P^{(j)}_\varepsilon$ solves 
\begin{equation*}\label{P1}
\Delta P_\varepsilon^{(j)}(\Bx)=0\;, \quad \Bx \in \mathbb{R}^3\backslash \overline{\omega^{(j)}_\varepsilon}\;,
\end{equation*}
\begin{equation*}\label{P2}
 P_\varepsilon^{(j)}(\Bx)=1\;, \quad \Bx \in \partial \omega_\varepsilon^{(j)}\;,
\end{equation*}
\begin{equation*}\label{P3}
 P_\varepsilon^{(j)}(\Bx)\to 0 \;, \quad \text{ as } |\Bx|\to \infty\;.
 \end{equation*}
 The behaviour of the capacitary potential far from the inclusion $\omega^{(j)}_\varepsilon$ is characterised by the  capacity of this set, defined as
 \begin{equation*}\label{P4}
 \text{cap}(\omega^{(j)}_\varepsilon)=\int_{\mathbb{R}^3\backslash \overline{\omega^{(j)}_\varepsilon}}|\nabla P_\varepsilon^{(j)}(\Bx)|^2d\Bx\;.
 \end{equation*}
 \begin{lem} $($\text{\emph{see }} \emph{\cite{MM_meso_1}}$)$
 For $|\Bx-\BO^{(j)}|>2 \varepsilon$, the capacitary potential admits the asymptotic representation
 \begin{equation*}\label{P5}
 P^{(j)}_\varepsilon(\Bx)=\frac{ \text{\emph{cap}}(\omega^{(j)}_\varepsilon)}{4\pi |\Bx-\BO^{(j)}|}+O\Big(\frac{\varepsilon^2}{|\Bx-\BO^{(j)}|^2}\Big)\;.
 \end{equation*}
  \label{lem1}
 \end{lem}

\end{enumerate}
\section{Formal asymptotic algorithm}\label{formalalg}

We now derive formal asymptotics for the first eigenvalue $\lambda_N$ and corresponding  eigenfunction $u_N$. 


First we state the asymptotic approximation for the first eigenvalue of the Laplacian in $\Omega_N$:
\begin{lem}\label{lemform1}
The formal approximation to the first eigenvalue of $\Delta$ in $\Omega_N$ is given by
\begin{eqnarray}\label{lamn_2ord}
\lambda_N=\Lambda_N+\lambda_{R,N}\;,
\end{eqnarray}
where 
\begin{equation}\label{WN61}
\Lambda_N=-\frac{1}{|\Omega|}\sum^N_{j=1} {C_j\text{\emph{cap}}(\omega^{(j)}_\varepsilon)}\;,
\end{equation}
 $C_j$, $1\le j \le N$, satisfy the algebraic system
\begin{eqnarray}
0&=&1+C_k(1-\text{\emph{cap}}(\omega^{(k)}_\varepsilon)\{\CH(\BO^{(k)},\BO^{(k)})-\Gamma_\Omega^{(k)} \})\nonumber\\ \nonumber\\&&+\sum_{\substack{j\ne k\\1\le j\le N}} {C_j \text{\emph{cap}}(\omega^{(j)}_\varepsilon)}\{\CG(\BO^{(k)}, \BO^{(j)})+\Gamma_\Omega^{(j)}\}
\;, \quad 1\le k \le N\;,
\label{form_sys}
\end{eqnarray}
 and $\lambda_{R,N}$ is the remainder of the approximation.
\end{lem}

We present the formal scheme leading to the preceding approximation and of the first eigenfunction $u_N$  contained in the next lemma.
\begin{lem}\label{lemform}
The formal approximation of the eigenfunction $u_N$ of problem $(\ref{uN1})$--$(\ref{uN3})$ has the form
\begin{eqnarray}
u_N(\Bx)=U(\Bx)+R_N(\Bx)\;,\label{un_2ord}
\end{eqnarray}
where
\begin{eqnarray}
U(\Bx)&=& 1+\sum_{j=1}^N
{ C_j \Gamma^{(j)}_\Omega\text{\emph{cap}}(\omega^{(j)}_\varepsilon)}
\nonumber \\
&&+\sum^N_{j=1} C_j \{P^{(j)}_\varepsilon(\Bx)-\text{\emph{cap}}(\omega^{(j)}_\varepsilon)\CH(\Bx, \BO^{(j)})\}
\;,\label{uN=U+RN}
\end{eqnarray}
the coefficients $C_j$ satisfy the linear algebraic system {$(\ref{form_sys})$}
and the function $U$,
defined according to $(\ref{uN=U+RN})$, satisfies the problem
\begin{eqnarray*}
&&\Delta U(\Bx)+\Lambda_N U(\Bx)=f_N(\Bx)\;, \quad \Bx\in \Omega_N\;,\\
&&\frac{\partial U(\Bx)}{\partial n}=\psi(\Bx)\;, \quad \Bx \in \partial \Omega\;,\\
&&U(\Bx)=\phi_k(\Bx)\;, \quad \Bx \in \partial \omega_\varepsilon^{(k)}, 1\le k \le N\;,
\end{eqnarray*}
where 
\begin{eqnarray*}
&&|f_N(\Bx)| =O\Big(\varepsilon^2 d^{-3}\Big( d^{-3}+\sum_{j=1}^N \frac{|C_j|}{|\Bx-\BO^{(j)}|}\Big)\Big)\;, \quad \Bx \in \Omega_N\;,\\
&&|\psi(\Bx)|=O\Big(\sum_{j=1}^N\frac{\varepsilon^2|C_j|}{|\Bx-\BO^{(j)}|^3}\Big)\;,\quad \Bx \in \partial \Omega\;,\\
&&|\phi_k(\Bx)|=O\Big(\varepsilon^2 \Big(d^{-3}
+\sum_{\substack{j\ne k\\1\le j\le N}} \frac{ |C_j|}{|\BO^{(k)}-\BO^{(j)}|^2}\Big)\Big)\;, \Bx\in \partial \omega_\varepsilon^{(j)}\;, 1\le k \le N\;.\nonumber\\
\end{eqnarray*}
\end{lem}

\emph{Proof of Lemmas  \ref{lemform1}  and \ref{lemform}. }Let
\begin{eqnarray}\label{Ufn}
U(\Bx)&=&1+\sum_{j=1}^N C_j P_\varepsilon^{(j)}(\Bx)+u_1(\Bx)\;.
\end{eqnarray}
 It is assumed the remainders $R_N(\Bx)$ and $\lambda_{R,N}$ in (\ref{un_2ord}) and (\ref{lamn_2ord}) are of the order  $O(\varepsilon^2 d^{-6})$.
In addition, we will show
\begin{eqnarray*}
u_1(\Bx)=O(\varepsilon d^{-3})\quad\text{ and } \quad  \Lambda_N=O(\varepsilon d^{-3})\;.\nonumber \\
\label{ansatz} 
\end{eqnarray*}

{\bf\emph{The governing equation in $\Omega_N$. }\rm}According to  (\ref{Ufn}), it holds that
\begin{eqnarray*}
&&0=\Delta U(\Bx)+\Lambda_N U(\Bx)\nonumber\\
&=&\Delta \Big(1+\sum_{j=1}^N C_j P_\varepsilon^{(j)}(\Bx)+u_1(\Bx) \Big)+ \Lambda_N\Big(1+\sum_{j=1}^N C_j P_\varepsilon^{(j)}(\Bx)+u_1(\Bx)\Big)\nonumber\\
&&\text{ for }\Bx\in \Omega_N \;.
\end{eqnarray*}
Since the capacitary potentials are harmonic, this implies in $\Omega_N$ that
\begin{eqnarray}
&&\Delta U(\Bx)+\Lambda_N U(\Bx)=
\Delta u_1(\Bx)+ \Lambda_N \Big(1+\sum_{j=1}^N C_j P_\varepsilon^{(j)}(\Bx)+u_1(\Bx) \Big)\;.\label{eqmethod3}
\end{eqnarray}
For $\Bx\in \Omega_N$, one can write
\begin{equation*}
P^{(j)}_\varepsilon(\Bx)=\frac{\text{cap}(\omega_\varepsilon^{(j)})}{4\pi |\Bx-\BO^{(j)}|} 
+O\Big(\frac{\varepsilon^2}{|\Bx-\BO^{(j)}|^2}\Big)\;,\label{eqlemP2}
\end{equation*}
%
(see \cite{OPTH1,OPTH2}).
As a result, returning to (\ref{eqmethod3}) we then have
\begin{eqnarray}
&&\Delta U(\Bx)+\Lambda_N U(\Bx)\nonumber\\
&&=\Delta u_1(\Bx) +\Lambda_N + O(\varepsilon^2 d^{-6})+O\Big(\varepsilon^2 d^{-3} \sum_{j=1}^N \frac{|C_j|}{|\Bx-\BO^{(j)}|}\Big)
\;.\label{eqmethod4}
\end{eqnarray}

{\bf \emph{Exterior boundary condition.}\rm } 
Next we consider the normal derivative of $U(\Bx)$ on $\partial \Omega$. We have 
\begin{equation*}
\frac{\partial U(\Bx)}{\partial n}=\frac{\partial }{\partial n}\Big\{1+\sum_{j=1}^N C_j P_\varepsilon^{(j)}(\Bx)+u_1(\Bx)
 \Big\}\quad\;, \Bx \in \partial \Omega\;.
\end{equation*}
Using Lemma \ref{lem1}, this can be updated to 
\begin{eqnarray}
&&\frac{\partial U(\Bx)}{\partial n}=\frac{\partial }{\partial n}\Big\{\sum_{j=1}^N \frac{C_j \text{cap}(\omega_\varepsilon^{(j)})}{4\pi |\Bx-\BO^{(j)}|} +u_1(\Bx) \Big\}
+O\Big(\sum_{j=1}^N\frac{\varepsilon^2|C_j|}{|\Bx-\BO^{(j)}|^3}\Big)\;,\label{ext_bc_1_a}
\end{eqnarray}
for $\Bx \in \partial \Omega$.

\vspace{0.1in}{\bf \emph{The terms $u_1$ and $\Lambda_N$. }\rm}
 Consulting (\ref{eqmethod4}) and (\ref{ext_bc_1_a}), we set
\begin{equation}\label{WN21}
\Delta u_1(\Bx)=-\Lambda_N\;, \quad \Bx\in \Omega\;,
\end{equation}
\begin{equation}\label{WN31}
\frac{\partial u_1(\Bx)}{\partial n}=-\frac{\partial }{\partial n}
\Big\{\sum^N_{j=1} \frac{C_j\text{cap}(\omega^{(j)}_\varepsilon)}{4\pi |\Bx-\BO^{(j)}|}\Big\}
\;, \quad \Bx\in \partial \Omega\;,
\end{equation}
and we prescribe that
\begin{equation}\label{WN41}
\int_\Omega u_1(\Bx)d\Bx=0\;.
\end{equation}
Note that according to this problem, the term $\Lambda_1$ can be computed using Green's identity in $\Omega$ to give
\begin{eqnarray*}
-|\Omega|\Lambda_N&=&\int_\Omega \Delta u_N(\Bx)d\Bx=\int_{\partial \Omega}\frac{\partial u_N}{\partial n}(\Bx)dS_{\Bx}\nonumber \\
&=& -\int_{\partial \Omega}\frac{\partial }{\partial n_{\Bx}}
\Big\{\sum^N_{j=1} \frac{C_j\text{cap}(\omega^{(j)}_\varepsilon)}{4\pi |\Bx-\BO^{(j)}|}\Big\}dS_{\Bx}\nonumber\\
&=&\sum^N_{j=1} {C_j\text{cap}(\omega^{(j)}_\varepsilon)}\;.\label{WN51}
\end{eqnarray*}
Thus, from this we prove (\ref{WN61})
of Lemma \ref{lemform1}.

In addition, $u_1$ can be constructed in the form
\begin{equation}\label{WN1}
u_1(\Bx)=-\sum_{j=1}^N C_j\text{cap}(\omega^{(j)}_\varepsilon) \Big\{\CH(\Bx, \BO^{(j)})-
\Gamma_\Omega^{(j)}
\Big\}\;,
\end{equation}
with $\Gamma_\Omega^{(k)}$ specified in (\ref{eqGAMOM}).
It can be checked this satisfies (\ref{WN21})--(\ref{WN41}).

\vspace{0.1in}{\bf \emph{Interior boundary conditions on small inclusions. }\rm} Taking the trace of $U(\Bx)$ on the boundary of $\partial \omega^{(k)}_\varepsilon$, $1 \le k \le N$, and using the definition of the capacitary potentials gives
\begin{eqnarray*}
U(\Bx)&=&1+C_k+\sum_{\substack{j\ne k\\1\le j\le N}} C_j P_\varepsilon^{(j)}(\Bx)+u_1(\Bx) 
\;.
\end{eqnarray*}
Next, Taylor's expansion about $\Bx=\BO^{(k)}$ and Lemma \ref{lem1} can be employed in the above condition to obtain
\begin{eqnarray*}
U(\Bx)&=&1+C_k+\sum_{\substack{j\ne k\\1\le j\le N}} \frac{C_j \text{cap}(\omega^{(j)}_\varepsilon)}{4\pi |\BO^{(k)}-\BO^{(j)}|} +u_1(\BO^{(k)})\nonumber
\nonumber \\&&+O(\varepsilon^2 d^{-3})+O\Big(\sum_{\substack{j\ne k\\1\le j\le N}} \frac{\varepsilon^2 |C_j| }{|\BO^{(k)}-\BO^{(j)}|^2}\Big)\label{intbc_11a}
\end{eqnarray*}
for $\Bx \in \partial \omega^{(k)}_\varepsilon$, $1\le k \le N$. 
According to (\ref{WN1}), this is equivalent to 
\begin{eqnarray}
U(\Bx)&=&1+C_k(1-\text{cap}(\omega_\varepsilon^{(k)})\{\CH(\BO^{(k)},\BO^{(k)})-\Gamma_\Omega^{(k)} \})\nonumber\\ \nonumber \\
&&+\sum_{\substack{j\ne k\\1\le j\le N}} {C_j \text{cap}(\omega^{(j)}_\varepsilon)}\{\CG(\BO^{(k)}, \BO^{(j)})+\Gamma_\Omega^{(j)}\}
\nonumber \\&&+O(\varepsilon^2 d^{-3})+O\Big(\sum_{\substack{j\ne k\\1\le j\le N}} \frac{\varepsilon^2 |C_j| }{|\BO^{(k)}-\BO^{(j)}|^2}\Big)\;.\label{intbc_1}
\end{eqnarray}
We then set up a system of algebraic equations with respect to $C_k$, $1\le k\le N$, as\begin{eqnarray}
0&=&1+C_k(1-\text{{cap}}(\omega^{(k)}_\varepsilon)\{\CH(\BO^{(k)},\BO^{(k)})-\Gamma_\Omega^{(k)} \})\nonumber\\ \nonumber\\&&+\sum_{\substack{j\ne k\\1\le j\le N}} {C_j \text{cap}(\omega^{(j)}_\varepsilon)}\{\CG(\BO^{(k)}, \BO^{(j)})+\Gamma_\Omega^{(j)}\}
\;, 
\label{eqC0}
\end{eqnarray} 
to remove the leading order term in (\ref{intbc_1}). The preceding together with (\ref{lamn_2ord}) and (\ref{WN61}) prove Lemma \ref{lemform1}.

\vspace{0.1in}{\bf \emph{The problem for $U$. }\rm}
As a result of equations (\ref{eqmethod4}), (\ref{WN21}), 
we have that $U$ satisfies
\begin{eqnarray}
&&\Delta U(\Bx)+\Lambda_N U(\Bx)= O(\varepsilon^2 d^{-6})+O\Big(\varepsilon^2 d^{-3} \sum_{j=1}^N \frac{|C_j|}{|\Bx-\BO^{(j)}|}\Big)\;,
\quad \Bx \in \Omega_N\;.\nonumber\\
\label{eqU}
\end{eqnarray}
On the exterior boundary, owing to (\ref{ext_bc_1_a}) and  (\ref{WN31}) we obtain
\begin{eqnarray}
&&\frac{\partial U(\Bx)}{\partial n}=O\Big(\sum_{j=1}^N\frac{\varepsilon^2|C_j|}{|\Bx-\BO^{(j)}|^3}\Big)\;, \quad \Bx \in \partial \Omega\;.\label{ext_bc_1}
\end{eqnarray}
The algebraic system  (\ref{eqC0}) 
with (\ref{intbc_1}) provide on the interior boundaries
\begin{eqnarray}
U(\Bx)&=&
O(\varepsilon^2 d^{-3})
+O\Big(\sum_{\substack{j\ne k\\1\le j\le N}} \frac{\varepsilon^2 |C_j|}{|\BO^{(k)}-\BO^{(j)}|^2}\Big)\label{intbc_1x}
\end{eqnarray}
for $\Bx\in \partial \omega_\varepsilon^{(k)}$, $1\le k \le N$.


By combining (\ref{un_2ord}), (\ref{Ufn}),  (\ref{WN1}),  (\ref{eqC0}) and (\ref{eqU})--(\ref{intbc_1x}), we arrive at the proof of Lemma \ref{lemform}. \hfill $\Box$



\section{The algebraic system and its solvability}\label{algsys}
In this section, it will be shown that the algebraic system (\ref{eqC0})  identified in the previous sections is solvable.
Here we rewrite the system (\ref{form_sys}) 
as
\begin{eqnarray*}
0&=&1+C_k(1-\text{cap}(\omega_\varepsilon^{(k)})\CH(\BO^{(k)},\BO^{(k)}))\nonumber\\ \nonumber\\&&+\sum_{\substack{j\ne k\\1\le j\le N}} {C_j \text{{cap}}(\omega^{(j)}_\varepsilon)}g(\BO^{(k)}, \BO^{(j)})-\Gamma_\Omega^{(k)}\sum_{j=1}^N C_j\text{cap}(\omega_\varepsilon^{(j)})\;,
\label{form_sys_2}
\end{eqnarray*}
where
\begin{equation}\label{defg}
g(\Bx, \By)=\CG(\Bx, \By)+\Gamma_\Omega(\By)+\Gamma_\Omega(\Bx)\;,
\end{equation}
and 
\[\Gamma_\Omega(\Bx)=\frac{1}{|\Omega|}\int_\Omega \frac{d\Bz}{4\pi|\Bz-\Bx|}\;.\]
This system can then be written in matrix form as:
\begin{equation}
\label{sys1}
-\BE=(\BI-\BH\BD+\BG\BD-\BGG\BD)\BC\;,
\end{equation}
where $\BI$ is the $N \times N$ identity matrix, 
\[\BC=(C_1,\dots, C_N)^T\;, \quad 
\BE=\sum_{j=1}^N\Be^{(N)}_j\;,
\]
and $\Be^{(N)}_i=[\delta_{ij}]_{j=1}^N$.
In addition  $\BG=[G_{ij}]_{i,j=1}^N$  with
\[G_{ij}=\left\{\begin{array}{ll}
\displaystyle{g(\BO^{(i)}, \BO^{(j)})}\;,&\quad \text{ for }i\ne j\;,
\\ \\
0\;, & \quad \text{otherwise,}
\end{array}\right.  \]
and 
\[\BH=\text{diag}_{1\le j \le N}\{\CH(\BO^{(j)}, \BO^{(j)})\}\;,\]
\[\BGG=[\Gamma_\Omega^{(j)}]_{i,j=1}^N\;, \quad \BD=\text{diag}_{1\le j \le N}\{\text{cap}(\omega^{(j)}_\varepsilon)\}\;.\]
\subsection*{Solvability of the algebraic systems}
We consider the system (\ref{sys1}),
whose rows 
can be written as in 
(\ref{eqC0}), and here we show the invertibility of the $N\times N$ matrix $\BI+(\BG-\BH-\BGG)\BD$. 

Taking the scalar product of (\ref{sys1}) with $\BD\BC$ one obtains
\begin{eqnarray}
-\langle \BD \BC, \BE \rangle&=&\langle \BD\BC, \BC\rangle+\langle \BD\BC, \BG\BD\BC\rangle \nonumber\\
&&-\langle  \BD\BC, \BH \BD\BC \rangle-\langle  \BD\BC, \BGG \BD\BC \rangle\;.\label{NE_sys}
\end{eqnarray}

In proving the solvability of (\ref{eqC0}), we need the following estimates:

\begin{lem}\label{ineq_sys_est}
The  estimates
\begin{eqnarray}
 |\langle  \BD\BC, \BH \BD\BC \rangle| &\le& \text{\emph{Const} }\varepsilon\, \langle \BC, \BD\BC\rangle\;, \label{CHC}\\
 | \langle  \BD\BC, \BGG \BD\BC\rangle| &\le &\text{\emph{Const} }\varepsilon\,  d^{-3}\langle \BC, \BD \BC\rangle\label{CGC}
\end{eqnarray}
and 
\begin{equation}\label{import_ineq}
\langle \BD\BC, \BG\BD\BC\rangle\ge -\text{\emph{Const} } d^{-1}\langle \BD \BC,  \BD \BC\rangle\;.
\end{equation}
hold.
\end{lem}

{\bf \emph{Proof of $(\ref{CHC})$ and $(\ref{CGC})$. }\rm} Since the regular part $\CH$ is bounded in $\omega$, one has that
 \begin{eqnarray*}
 |\langle  \BD\BC, \BH \BD\BC \rangle|&=&\Big|\sum_{k=1}^N(C_k\text{cap}(\omega_\varepsilon^{(k)}))^2 \CH(\BO^{(k)}, \BO^{(k)})\Big| \nonumber\\
 &\le &\text{Const }\varepsilon\, \langle \BC, \BD\BC\rangle\;,
 \end{eqnarray*}
 which is (\ref{CHC}).
 In addition, using {(\ref{eqGAMOM})}  gives
 \begin{eqnarray}
| \langle  \BD\BC, \BGG \BD\BC \rangle|&\le& \sum_{k=1}^N  \sum_{j=1}^N |C_kC_j\text{cap}(\omega_\varepsilon^{(k)})
\text{cap}(\omega_\varepsilon^{(j)})\Gamma_\Omega^{(k)}|\nonumber\\
&\le &\text{Const }   \sum_{k=1}^N\sum_{j=1}^N |C_kC_j\text{cap}(\omega_\varepsilon^{(k)})\text{cap}(\omega_\varepsilon^{(j)})|\nonumber\;.
\end{eqnarray}
The Cauchy inequality then implies
\begin{eqnarray*}
| \langle  \BD\BC, \BGG \BD\BC\rangle|&\le& \text{Const } \langle\BD \BC, \BC\rangle \sum_{k=1}^N  \text{cap}(\omega_\varepsilon^{(k)})\nonumber \\
&\le &\text{Const }\varepsilon d^{-3}\langle\BD \BC, \BC\rangle\;,
 \end{eqnarray*}
 proving (\ref{CGC}).
 
\vspace{0.1in}{\bf\emph{Proof of $(\ref{import_ineq})$.\rm}  }The term 
\begin{equation}\label{term1}
\langle \BD\BC, \BG\BD\BC\rangle= \sum_{k=1}^N C_k\text{cap}(\omega^{(k)}_\varepsilon)\sum_{\substack{j\ne k\\1\le j\le N}}{g(\BO^{(k)}, \BO^{(j)})} {C_j \text{cap}(\omega^{(j)}_\varepsilon)}\;. 
\end{equation}
According to (\ref{N1}) and (\ref{N2}) the function $g$ defined in (\ref{defg}) satisfies
\begin{equation}
\label{eq1g}
\Delta_{\BX} g(\BX,\BY)+\delta(\BX- \BY)=0\;, \quad  \BX, \BY\in \Omega\;,
\end{equation}
\[\frac{\partial g}{\partial n_\BX}(\BX,\BY)=\frac{\partial \Gamma_\Omega}{\partial n_\BX}(\BX)\;, \quad \BX\in \partial \Omega\;, \BY\in \Omega\;.\]
It is also true  from (\ref{defg}) that 
\[g(\BX,\BY)=g(\BY,\BX), \quad \BX\ne \BY\;.\]
As a result, application of Green's formula to $g(\BZ, \BX)$ and $g(\BZ, \BY)$ shows that this function satisfies the orthogonality condition:
\begin{equation}
\label{orthogg}
\int_{\partial \Omega}g(\BZ,\BY)\frac{\partial \Gamma_\Omega}{\partial n_\BX}(\BZ) dS_{\BZ}=0\;.
\end{equation}
Here, (\ref{eq1g}) shows that $g$ is harmonic if $\BX \ne \BY$. Using this, (\ref{term1}) can be rewritten with the mean value theorem inside disjoint balls to give
\begin{eqnarray}\label{term2}
&&\langle \BD\BC, \BG\BD\BC\rangle\nonumber\\
&=&\frac{48^2}{\pi^2 d^6}\sum_{k=1}^N \sum_{j=1}^N \int_{B^{(k)}}\int_{B^{(j)}} C_k\text{cap}(\omega^{(k)}_\varepsilon){g(\BX, \BY)} {C_j \text{cap}(\omega^{(j)}_\varepsilon)}d\BX d\BY\nonumber\\
&&-\frac{48}{\pi d^3}\sum_{k=1}^N(C_k \text{cap}(\omega_\varepsilon^{(k)}))^2\int_{B^{(k)}}  g(\BX, \BO^{(k)})d \BX\;.
\end{eqnarray}
where $B^{(j)}=\{\BX: |\BX-\BO^{(j)}|<d/4\}$. 

Next the fact $g(\Bx, \BO^{(k)})=O(|\BX-\BO^{(k)}|^{-1})$ allows for the estimate
\begin{equation}\label{intg1}
\int_{B^{(k)}}  g(\BX, \BO^{(k)})d \BX\le \text{Const } d^2\;.
\end{equation}
The function
\begin{equation*}\label{est1term}
\Theta(\Bx)=\left\{\begin{array}{ll}
C_k\text{cap}(\omega^{(k)}_\varepsilon)\;, &\quad \Bx \in B^{(k)}\\
0&\quad \text{otherwise},
\end{array}\right.
\end{equation*}
can be employed to the double sum in (\ref{term2}) to yield:
\begin{eqnarray}\label{term3}
&&\sum_{k=1}^N \sum_{j=1}^N \int_{B^{(k)}}\int_{B^{(j)}} C_k\text{cap}(\omega^{(k)}_\varepsilon){g(\BX, \BY)} {C_j \text{cap}(\omega^{(j)}_\varepsilon)}d\BX d\BY\nonumber\\
&=&\int_\Omega \int_\Omega \Theta(\BX)g(\BX,\BY)\Theta(\BY)d\BX d \BY\;.
\end{eqnarray}
Next set 
\begin{equation*}\label{eqh}
h(\BX)=\int_\Omega g(\BX, \BY)\Theta(\BY)d\BY\;.
\end{equation*}
This function satisfies
\begin{equation}\label{Dh}
\Delta_\Bx h(\BX)=-\Theta(\BX)\;, \quad\BX\in \Omega\;,
\end{equation}
\[\frac{\partial h}{\partial n_\BX}(\BX)=\frac{\partial \Gamma_\Omega}{\partial n_\Bx}(\BX)\int_\Omega \Theta(\BY)d\BY\;.\]
Note that owing to (\ref{orthogg}):
\[\int_{\partial \Omega} h(\BX)\frac{\partial h}{\partial n_\BX}(\BX)dS_\BX=\int_{\partial \Omega} h(\BX)\frac{\partial \Gamma_\Omega}{\partial n_\BX}(\BX)dS_\BX \int_\Omega \Theta(\BY)d\BY=0\;.\]
Thus, after integration by parts, one can show using this and (\ref{Dh}) that
\[\int_\Omega \int_\Omega \Theta(\BX)g(\BX,\BY)\Theta(\BY)d\BX d \BY=\int_\Omega |\nabla h(\Bx)|^2d\Bx \ge 0\;.\] 
Then, this estimate, (\ref{term1}), (\ref{term2}), (\ref{intg1}) and (\ref{term3})
prove (\ref{import_ineq}), completing the proof.
\hfill $\Box$

\vspace{0.1in}\begin{lem}\label{lemcoeffs}
Let the small parameters $\varepsilon$ and $d$ satisfy the inequality
\begin{equation}\label{epsvsd1}
\varepsilon < c\, d^3
\end{equation}
where $c$ is a sufficiently small constant.
Then the system $(\ref{sys1})$ is solvable and the estimate
\begin{equation}\label{C0jest}
\sum_{j=1}^N C_j^2\le \text{\emph{Const }}  d^{-3}\;,
\end{equation}
holds.
\end{lem}

 
\emph{Proof. } We start from (\ref{NE_sys}), and use the 
Cauchy inequality to obtain
\begin{eqnarray*}
\langle \BE,\BD\BE \rangle^{1/2}\langle\BC, \BD\BC\rangle^{1/2}&\ge&  \langle  \BC, \BD\BC\rangle+\langle \BD\BC, \BG\BD\BC\rangle \nonumber \\
&&-\langle  \BD\BC, \BH \BD\BC \rangle-\langle  \BD\BC, \BGG \BD\BC\rangle
\label{eqproof1}
 \end{eqnarray*}
 
Now from Lemma \ref{ineq_sys_est},  we have
\begin{eqnarray*}  \langle \BE,\BD\BE \rangle^{1/2}& \ge& \langle  \BC, \BD\BC\rangle^{1/2}\Big(1-\text{Const } \Big(d^{-1}\frac{\langle \BD\BC, \BD\BC\rangle}{ \langle  \BC, \BD\BC\rangle}+\varepsilon +\varepsilon d^{-3}\Big)\Big) \nonumber\\\nonumber
\\
&\ge & \langle  \BC, \BD\BC\rangle^{1/2}\Big(1-\text{Const } ( d^{-1}\max_k\{\text{cap}(\omega^{(k)}_\varepsilon)\}+\varepsilon +\varepsilon d^{-3})\Big)\;.\nonumber\\
\label{eqproof2}
\end{eqnarray*}
Since $\text{cap}(\omega^{(k)}_\varepsilon)=O(\varepsilon)$, $1\le k \le N$, the preceding inequality shows that the system is solvable for $\varepsilon$ and $d$ satisfying (\ref{epsvsd1}). The estimate (\ref{C0jest}) then follows immediately.  The proof is complete. 
\section{Remainder estimates}\label{remainderestsec}
In this section we present the remainder estimate for approximations associated with the first eigenvalue $\lambda_N$ and the corresponding eigenfunction $u_N$ required for the proof of Theorems \ref{th1} and \ref{th1a}.

We begin 
by introducing auxiliary functions that enable the estimates for the remainders of our formal approximations to be carried out via integrals over domains local to the boundaries of $\Omega_N$.

\vspace{0.1in}{\bf \emph{Auxiliary functions. }\rm}
Let 
\begin{equation*}
\Psi_0(\Bx)=\sum_{j=1}^N C_j \Big\{P_\varepsilon^{(j)}(\Bx)-\frac{\text{cap}(\omega_\varepsilon^{(j)})}{4\pi |\Bx-\BO^{(j)}|}\Big\}
\end{equation*}
and for $k=1,\dots, N$,
\begin{eqnarray}\label{1psikh}
\Psi_k(\Bx)&=&-C_k \text{cap}(\omega_\varepsilon^{(k)})(\CH(\Bx, \BO^{(k)})-\CH(\BO^{(k)},\BO^{(k)}))\nonumber \\ \nonumber \\
&&-\sum_{\substack{j\ne k\\1\le j\le N}} {C_j \text{{cap}}(\omega^{(j)}_\varepsilon)}\CG(\BO^{(k)}, \BO^{(j)})
\nonumber\\
&&+\sum_{\substack{j \ne k\\ 1\le j \le N}} C_j \{P_\varepsilon^{(j)}(\Bx)-\text{cap}(\omega_\varepsilon^{(j)})\CH(\Bx, \BO^{(j)})
\}\;.
\end{eqnarray}
It can be verified that 
\begin{equation}\label{eqDNRN}
\frac{\partial U}{\partial n}=\frac{\partial \Psi_0}{\partial n}\;, \quad \Bx \in \partial \Omega\;,
\end{equation}
\begin{equation}\label{eqDNRN_2}
{ U_N}={ \Psi_k}\;, \quad \Bx \in \partial \omega^{(k)}_\varepsilon\;, 1\le k \le N\;,
\end{equation}
and  
\begin{eqnarray}
&&\Delta \Psi_0(\Bx)=0\;, \quad \Bx \in \Omega_N\;,\nonumber \\
&&\Delta \Psi_k(\Bx)+\Lambda_N=0\;, \quad \Bx \in \Omega_N\;, \quad 1\le k \le N\;.\label{Eqn_PsiN}
\end{eqnarray}
 
 Let $B_r^{(j)}=\{\Bx:|\Bx-\BO^{(j)}|<r\}$.
 In addition, let $\chi_\varepsilon^{(j)}\in C^\infty_0(B^{(j)}_{3\varepsilon})$, which is equal to 1 on $B_{2\varepsilon}^{(j)}$. These cut-off functions will be used to reduce certain integrals over $\Omega_N$ to integrals in the vicinity of the small inclusions. 

The same approach  will be applied to integrals in the vicinity of $\partial \Omega$, which are obtained using the cut-off function $\chi_0\in C^\infty_0$. This function is chosen to be equal to one in $\{\Bx: \text{dist}(\Bx, \partial \Omega)\le 1/6, \Bx \in \Omega\}$ and zero inside the set $\{\Bx: \text{dist}(\Bx, \partial \Omega)\ge 1/2, \Bx \in \Omega\}$. In what follows, $\CV:=\{\Bx: 0<\text{dist}(\Bx, \partial \Omega)\le 1/2, \Bx \in \Omega\}$.

\subsection*{The function $\sigma_N$}
 Now we use the auxiliary functions to construct
\begin{equation}\label{Sign}
\sigma_N=A \,\{U-\chi_0 \Psi_0 -\sum^N_{j=1} \chi^{(j)}_\varepsilon \Psi_j\}\;,
\end{equation}
where the constant $A$ is chosen to enable
\[\|\sigma_N\|_{L_2(\Omega_N)}=1\;.\]
According to (\ref{eqDNRN})--(\ref{Eqn_PsiN}), 
\begin{eqnarray}\label{eqsig1}
&&\Delta \sigma_N+\Lambda_N\sigma_N=F_N\;, \quad \Bx \in \Omega_N\;,\\
&& \frac{\partial \sigma_N}{\partial n}=0\;, \quad \Bx \in \partial \Omega\;,\\
 &&{\sigma_N}=0\;, \quad \Bx \in \partial \omega_\varepsilon^{(k)}, 1\le k \le N\;,\label{eqsig3}
\end{eqnarray}
where 
\begin{eqnarray}
F_N&=&A\{\Delta U+\Lambda_NU\}\nonumber\\
&&-A\{\Delta( \chi_0 \Psi_0)+\Lambda_N\chi_0\GY_0\}\nonumber\\
&&-\sum^N_{j=1}A\{\Delta( \chi^{(k)}_\varepsilon \Psi_k)+\Lambda_N\chi^{(k)}_\varepsilon\GY_k\}\;, \quad \Bx \in \Omega_N\;.
\label{FNrep}
\end{eqnarray}

In the next section we prove the following Lemma.
\begin{lem}\label{lemremest}
Let 
\[\varepsilon <c\, d^3\;,\]
where $c$ is a sufficiently small constant. Then the estimates 
\begin{equation}\label{remainderest}
\|\sigma_N-u_N\|_{L_2(\Omega_N)}\le \text{\emph{Const} } {\varepsilon^{3/2} d^{-9/2}} \;,
\end{equation}
and
\begin{equation}\label{lambda_remainderest}
|\lambda_N-\Lambda_N|\le \text{\emph{Const} }{\varepsilon^{3/2} d^{-9/2} }\;,
\end{equation}
hold.
\end{lem}

\subsection*{Estimate of $F_N$}    

\vspace{0.1in}
We first consider an estimate for $F_N$ in (\ref{eqsig1}) and (\ref{FNrep}) in $L_2(\Omega_N)$.
Here we show 
\begin{equation}\label{FNestL2}
\|F_N\|_{L_2(\Omega_N)}\le \text{Const } \varepsilon^{3/2} d^{-9/2}\;.
\end{equation}
Terms appearing in $F_N$ can be further expanded to give
\begin{eqnarray}
F_N&=&A\{\Delta U+\Lambda_NU\}-A\{2\nabla \chi_0 \cdot \nabla\Psi_0+\Psi_0\Delta \chi_0\}\nonumber \\
&&-A\sum^N_{j=1}\{2\nabla \chi^{(k)}_\varepsilon \cdot \nabla\Psi_k+\Psi_k\Delta \chi^{(k)}_\varepsilon-\chi_\varepsilon^{(k)} \Lambda_N\}\nonumber\\
&&-A\Lambda_N\Big\{\chi_0\GY_0+\sum_{k=1}^N\chi^{(k)}_\varepsilon\GY_k\Big\}, \quad \Bx \in \Omega_N\;.
\label{FNrep2}
\end{eqnarray}
 This provides
\begin{eqnarray}&&\|F_N\|_{L_2(\Omega_N)}^2\le \text{Const }\Big\{\|\Delta U+\Lambda_N U\|_{L_2(\Omega_N)}^2+\|\nabla \Psi_0\|^2_{L_2(\CV)}+\|\Psi_0\|^2_{L_2(\CV)}\nonumber\\\nonumber
\\&&+\varepsilon^{-2}\sum^N_{k=1}\Big[\|\nabla \Psi_k\|^2_{L_2(B^{(k)}_{3\varepsilon} \backslash {\omega_\varepsilon^{(k)}})}+\varepsilon^{-2}\|\Psi_k\|^2_{L_2(B^{(k)}_{3\varepsilon} \backslash {\omega_\varepsilon^{(k)}})}\Big]+\CP+\CS
\Big\}\label{FNest}
\end{eqnarray}
where 
\begin{eqnarray}
\CP&=&\Lambda_N^2
\sum_{k=1}^N\|\chi_\varepsilon^{(k)} \|^2_{L_2(B^{(k)}_{3\varepsilon}\backslash{\omega_\varepsilon^{(k)}})}\label{Pterm1st}\\
\nonumber
\\
\CS&=&\Lambda_N^2\Big\|\chi_0\Psi_0+\sum_{k=1}^N\chi^{(k)}_\varepsilon \Psi_k\Big\|^2_{L_2(\Omega_N)}\label{Sterm1st}\;.
\end{eqnarray}
Thus (\ref{FNestL2}) can be achieved if the right-hand side of (\ref{FNest}) is estimated.

\subsection*{Inequalities associated with $\Psi_0$ }
Here, as a result of Lemma \ref{lem1} and the Cauchy inequality, we have the estimate
\begin{eqnarray}
\|\Psi_0\|^2_{L_2(\CV)}&\le& \text{Const }\varepsilon^4 \int_{\CV}\Big|\sum_{j=1}^N\frac{|C_j|}{|\Bx-\BO^{(j)}|^2}\Big|^2\, d\Bx\nonumber\\
&\le&\text{Const }\varepsilon^4 \sum^N_{m=1}|C_m|^2\sum_{j=1}^N\int_\CV \frac{d\Bx}{|\Bx-\BO^{(j)}|^4}\;.\nonumber
\end{eqnarray} 
Since $\text{dist}(\omega, \partial \Omega)=O(1)$, using Lemma \ref{lemcoeffs}, we arrive at 
\begin{equation}\label{estpsi_0}
\|\Psi_0\|^2_{L_2(\CV)}\le \text{Const }\varepsilon^4 d^{-6}\;.
\end{equation}
Using  similar approach to the estimate (\ref{estpsi_0}), one can show that
\begin{equation}
\label{estT1}
\|\nabla\Psi_0\|_{L_2(\CV)}^2\le \text{Const } \varepsilon^4 d^{-6}\;.
\end{equation}

\subsection*{Inequalities associated with $\Psi_k$, $1\le k \le N$}
Now we prove that 
\begin{eqnarray}
&& \sum_{k=1}^N\|\Psi_k\|_{L_2(B^{(k)}_{3\varepsilon}\backslash{\omega_\varepsilon^{(k)}})}^2 \le \text{Const }  \varepsilon^7 d^{-9}\label{PsikestL21}\\
&& \sum^N_{k=1}\|\nabla \Psi_k\|^2_{L_2(B^{(k)}_{3\varepsilon} \backslash {\omega_\varepsilon^{(k)}})}\le \text{Const }\varepsilon^3d^{-9}\;. \label{DPsikestL21}
\end{eqnarray}
\vspace{0.1in}{\bf \emph{
{Proof of inequality $(\ref{PsikestL21})$}.}}
The terms $\Psi_k$ are estimated in $L_2(B_{3\varepsilon}^{(k)}\backslash {\omega_\varepsilon^{(k)}})$ as follows. The Taylor expansion about 
$\Bx=\BO^{(k)}$ gives
\begin{eqnarray}
&& \int_{B^{(k)}_{3\varepsilon}\backslash {\omega^{(k)}_\varepsilon}}\Big|C_k \text{cap}(\omega_\varepsilon^{(k)})(\CH(\Bx, \BO^{(k)})-\CH(\BO^{(k)},\BO^{(k)}))\Big|^2d\Bx \le \text{Const }\varepsilon^7|C_k|^2\;.
\nonumber\\
\label{ineq_psi_1}
\end{eqnarray}
We note that using Taylor's expansion about $\Bx=\BO^{(k)}$
\begin{eqnarray}
&&\int_{B^{(k)}_{3\varepsilon}\backslash {\omega^{(k)}_\varepsilon}}\Big| \sum_{\substack{j\ne k\\1\le j\le N}} {C_j \text{{cap}}(\omega^{(j)}_\varepsilon)}\CG(\BO^{(k)}, \BO^{(j)})\nonumber
\\&&-\sum_{\substack{j \ne k\\ 1\le j \le N}} C_j \{P_\varepsilon^{(j)}(\Bx)-\text{cap}(\omega_\varepsilon^{(j)})\CH(\Bx, \BO^{(j)})
\}\Big|^2d\Bx
\nonumber\\
&\le &
\text{Const }\int_{B^{(k)}_{3\varepsilon}\backslash {\omega^{(k)}_\varepsilon}}\Big|\sum_{\substack{j \ne k\\ 1\le j \le N}} C_j \Big\{P_\varepsilon^{(j)}(\Bx)-\frac{\text{cap}(\omega_\varepsilon^{(j)})}{4\pi|\BO^{(k)}-\BO^{(j)}|}
\Big\}\Big|^2d\Bx
 \label{ineq_psi_2}
\end{eqnarray}
Lemma  \ref{lem1} can then be applied to obtain the estimate
\begin{eqnarray}
&&\int_{B^{(k)}_{3\varepsilon}\backslash {\omega^{(k)}_\varepsilon}}\Big|\sum_{\substack{j \ne k\\ 1\le j \le N}} C_j \Big\{P_\varepsilon^{(j)}(\Bx)-\frac{\text{cap}(\omega_\varepsilon^{(j)})}{4\pi|\BO^{(k)}-\BO^{(j)}|}
\Big\}\Big|^2d\Bx\nonumber\\
&\le &\text{Const }\varepsilon^2\int_{B^{(k)}_{3\varepsilon}\backslash {\omega^{(k)}_\varepsilon}}\Big|\sum_{\substack{j \ne k\\ 1\le j \le N}} C_j \Big\{\frac{1}{|\Bx-\BO^{(j)}|}-\frac{1}{|\BO^{(k)}-\BO^{(j)}|}
\Big\}\Big|^2d\Bx\nonumber\\
&\le &\text{Const }\varepsilon^7\Big| \sum_{\substack{j \ne k\\ 1\le j \le N}} \frac{C_j}{|\BO^{(k)}-\BO^{(j)}|^2}   \Big|^2\;.
\label{ineq_psi_3}
\end{eqnarray}
Using the Cauchy inequality and Lemma \ref{lemcoeffs} we find the right-hand side is majorised by 
\begin{equation}\label{ineq_psi_4}
\text{Const } \varepsilon^7 d^{-3} \sum_{\substack{j \ne k\\ 1\le j \le N}}\frac{1}{|\BO^{(k)}-\BO^{(j)}|^4}\;.  
\end{equation}
{
Through combining (\ref{ineq_psi_1})--(\ref{ineq_psi_4}), it can then be asserted that 
\begin{eqnarray}
\|\Psi_k\|^2_{L_2(B^{(k)}_{3\varepsilon}\backslash {\omega^{(k)}_\varepsilon})}\le \text{Const } \varepsilon^7 \Big\{{|C_{k}|^2}+ d^{-3} \sum_{\substack{j \ne k\\ 1\le j \le N}}\frac{1}{|\BO^{(k)}-\BO^{(j)}|^4}\Big\}.\nonumber\\ \label{psikest4}
\end{eqnarray}
It then follows
\begin{eqnarray*}
 \sum_{k=1}^N\|\Psi_k\|_{L_2(B^{(k)}_{3\varepsilon}\backslash {\omega_\varepsilon^{(k)}})}^2
&\le&\text{Const }  \varepsilon^7 \Big\{{\sum_{k=1}^N|C_k|^2}+d^{-3}\sum_{k=1}^N \sum_{\substack{j \ne k\\ 1\le j \le N}}\frac{1}{|\BO^{(k)}-\BO^{(j)}|^4}\Big\}\;.
\end{eqnarray*}
Lemma  \ref{lemcoeffs} then gives
\begin{eqnarray}
\sum_{k=1}^N\|\Psi_k\|_{L_2(B^{(k)}_{3\varepsilon}\backslash {\omega_\varepsilon^{(k)}})}^2&\le & \text{Const }  \varepsilon^7 \Big\{{d^{-3}}+d^{-9}\int_{\substack{\omega\times \omega:\\ |\BX-\BY|>d}}\frac{d\BY d\BX}{|\BX-\BY|^4}\Big\}\nonumber\\
\end{eqnarray}
which yields (\ref{PsikestL21}).

\vspace{0.1in}{\bf \emph{
{Proof of inequality $(\ref{DPsikestL21})$}.
}}
Consulting (\ref{1psikh}), we can derive that
\begin{equation}\label{eqnewPsiksum}
 \sum^N_{k=1}\|\nabla \Psi_k\|^2_{L_2(B^{(k)}_{3\varepsilon} \backslash {\omega_\varepsilon^{(k)}})} \le \text{Const }\{\CM +\CN\}\;,
 \end{equation}
with 
\begin{eqnarray}
\CM&=& \sum_{k=1}^N \Big\|  \nabla (C_k \text{cap}(\omega_\varepsilon^{(k)}) \CH(\Bx,\BO^{(k)})) \Big\|_{L_2(B^{(k)}_{3\varepsilon}\backslash {\omega_\varepsilon^{(k)}})}^2\;,\\
\CN&=& \sum_{k=1}^N \Big\| \nabla (\sum_{\substack{j \ne k\\ 1\le j \le N}} C_j\{P_\varepsilon^{(j)}(\Bx)-\text{cap}(\omega_\varepsilon^{(j)})\CH(\Bx, \BO^{(j)})\}) \Big\|_{L_2(B^{(k)}_{3\varepsilon}\backslash {\omega_\varepsilon^{(k)}})}^2\;.\nonumber\\
\end{eqnarray}
 The regular part $\CH(\Bx, \BO^{(j)})$ and its derivatives are bounded for $\Bx \in \omega$. As a consequence, we have
 \begin{eqnarray*}
 \CM &\le & \text{Const }\varepsilon^5
 \sum_{k=1}^N|C_k|^2
 \end{eqnarray*}
Applying  Lemma \ref{lemcoeffs} then gives
  \begin{eqnarray}
 \CM &\le & \text{Const }\varepsilon^5{d^{-3}}
 \;.\label{estM1}
 \end{eqnarray}
For $\CN$, it is appropriate to use Lemma \ref{lem1} where the far-field behaviour of $P^{(j)}_\varepsilon$, $j \ne k$, is given. Thus one obtains the inequality
\begin{eqnarray*}
\CN&\le& \text{Const }\varepsilon^2
\sum_{k=1}^N\int_{B^{(k)}_{3\varepsilon} \backslash {\omega_\varepsilon^{(k)}}}\Big(\sum_{{\substack{j \ne k\\ 1\le j \le N}}}\frac{|C_j|}{|\Bx-\BO^{(j)}|^2} \Big)^2d\Bx \nonumber\\ 
&\le &\text{Const }\varepsilon^5 \sum_{k=1}^N\Big(\sum_{{\substack{j \ne k\\ 1\le j \le N}}}\frac{|C_j|}{|\BO^{(k)}-\BO^{(j)}|^2} \Big)^2
\label{estN1}
\end{eqnarray*}
where the Taylor expansion has been employed about $\Bx=\BO^{(k)}$ in moving to the last line. Next the Cauchy inequality and (\ref{C0jest}) produce
\begin{eqnarray*}
\CN&\le& \text{Const }\varepsilon^5\sum_{m=1}^N |C_m|^2\sum_{k=1}^N \sum_{{\substack{j \ne k\\ 1\le j \le N}}}\frac{1}{|\BO^{(k)}-\BO^{(j)}|^4}\nonumber\\
&\le & \text{Const }\varepsilon^5d^{-3}\sum_{k=1}^N \sum_{{\substack{j \ne k\\ 1\le j \le N}}}\frac{1}{|\BO^{(k)}-\BO^{(j)}|^4}\;.
\end{eqnarray*}
The second sum can be approximated by a double integral  over $\omega$ to give
\begin{eqnarray}
\CN&\le& \text{Const }\varepsilon^5d^{-9} \int\int_{\substack{\omega\times \omega:\\ |\BX-\BY|>d}}\frac{d\BX d\BY}{|\BX-\BY|^4}\le  \text{Const }\varepsilon^5d^{-9}\;.\label{estN12}
\end{eqnarray}

\subsection*{Proof of inequality $(\ref{FNestL2})$}
The characteristic functions $\chi_0$ and $\chi_k^{(j)}$, $1\le j \le N$, are bounded by unity, and  this together with (\ref{estpsi_0}) and (\ref{psikest4}) show that
\begin{eqnarray}
\Big\|\chi_0\Psi_0+\sum_{k=1}^N\chi^{(k)}_\varepsilon \Psi_k\Big\|^2_{L_2(\Omega_N)}&\le &\text{Const }
\Big\{\varepsilon^4d^{-6}+ {\varepsilon^7\sum_{k=1}^N|C_k|^2}
\nonumber\\
&&+\varepsilon^7d^{-3}\sum_{k=1}^N \sum_{\substack{j \ne k\\ 1\le j \le N}}\frac{1}{|\BO^{(k)}-\BO^{(j)}|^4}\Big\}\;.
\end{eqnarray}
The double sum in the right-hand side can be approximated by a double integral over $\omega$. Therefore, with Lemma \ref{lemcoeffs}, one can write the estimate
\begin{eqnarray}
&&\Big\|\chi_0\Psi_0+\sum_{k=1}^N\chi^{(k)}_\varepsilon \Psi_k\Big\|^2_{L_2(\Omega_N)}\nonumber\\&\le &\text{Const }
\Big\{\varepsilon^4d^{-6}+{\varepsilon^7 d^{-3}}
+\varepsilon^7d^{-9}
\int\int_{\substack{\omega\times \omega:\\ |\BX-\BY|>d}}\frac{d\BX d\BY}{|\BX-\BY|^4}\Big\}\nonumber\\
\end{eqnarray}
then we arrive at
\begin{eqnarray}
\Big\|\chi_0\Psi_0+\sum_{k=1}^N\chi^{(k)}_\varepsilon \Psi_k\Big\|^2_{L_2(\Omega_N)}&\le &\text{Const }\varepsilon^4
\Big\{d^{-6}+{\varepsilon^3 d^{-3}}+\varepsilon^3d^{-9}
\Big\}\;\nonumber \\ 
&\le&\text{Const }\varepsilon^4 d^{-6}\;. \label{Sestimate}
\end{eqnarray}


The right-hand side in governing equation (\ref{eqU}) can be estimated in $L_2(\Omega_N)$
by 
considering the term
\[\int_{\Omega_N} \Big| \sum_{j=1}^N
\frac{|C_j| }{ |\Bx-\BO^{(j)}|}\Big|^2\, d\Bx\;.\]
The Cauchy inequality shows this is majorised by 
\[\text{Const }\sum_{j=1}^N |C_j|^2\sum_{k=1}^N \int_{\Omega_N} 
\frac{d\Bx }{ |\Bx-\BO^{(k)}|^2}\;.\]
The above integrals are bounded by a constant, thus we can say owing to Lemma \ref{lemcoeffs} that 
\begin{eqnarray}
\|\Delta U+\Lambda_N U\|_{L_2(\Omega_N)}^2&\le& \text{Const }\varepsilon^4 d^{-9}\Big\{d^{-3}+\sum_{j=1}^N |C_j|^2\Big\}\nonumber\\
&\le & \text{Const }\varepsilon^4 d^{-12}\;.\label{estR1}
\end{eqnarray}
Since $\Lambda_N=O(\varepsilon d^{-3})$, for $\CS$ in (\ref{Sterm1st}), it holds that
\[\CS \le \text{Const }\varepsilon^2d^{-6}\Big\|\chi_0\Psi_0+\sum_{k=1}^N\chi^{(k)}_\varepsilon \Psi_k\Big\|^2_{L_2(\Omega_N)}\;.\]
Using (\ref{Sestimate}) yields
\begin{equation}\label{estSS1}
\CS\le \text{Const }\varepsilon^6 d^{-12}\;.
\end{equation}
The term $\CP$, in (\ref{Pterm1st}), as a result of  $\chi_\varepsilon^{(k)}\in C^\infty_0(B^{(k)}_{3\varepsilon}\backslash \overline{\omega_\varepsilon^{(k)}})$, satisfies
\begin{equation}\label{CPest}
\CP\le \text{Const }\varepsilon^5 d^{-9}\;.
\end{equation}
Combining  
(\ref{FNest}), (\ref{estpsi_0})--(\ref{DPsikestL21}),
  (\ref{estR1})--(\ref{CPest})
yields (\ref{FNestL2}).

\subsection*{{Proof of Lemma \ref{lemremest}}}
 From (\ref{uN1})-(\ref{uN3}), we can then write a boundary value problem for the difference of $\sigma_N$ and $u_N$ as 
\begin{eqnarray}
&&\Delta (\sigma_N-u_N)+\Lambda_N (\sigma_N-u_N)+(\Lambda_N-\lambda_N)u_N=F_N\;, \quad \Bx \in \Omega_N\;,\label{eq1diff}\\
&& \frac{\partial }{\partial n}(\sigma_N-u_N)=0\;, \quad \Bx \in \partial \Omega\;,\\
 &&{\sigma_N-u_N}=0\;, \quad \Bx \in \partial \omega_\varepsilon^{(k)}, 1\le k \le N\;.
\end{eqnarray}
One can then multiply (\ref{eq1diff}) through by the difference $\sigma_N-u_N$ and integrate by parts in $\Omega_N$ to obtain
\begin{eqnarray}
&&-\int_{\Omega_N} |\nabla(\sigma_N-u_N)|^2\,d\Bx+\Lambda_N \int_{\Omega_N}(\sigma_N-u_N)^2\, d\Bx\nonumber\\
&& +(\Lambda_N-\lambda_N)\int_{\Omega_N} u_N(\sigma_N-u_N)\, d\Bx=\int_{\Omega_N} F_N(\sigma_N-u_N)\, d\Bx\;.\label{proofeq1}
\end{eqnarray}
Poincar\'e's inequality implies
\[\int_{\Omega_N} |\nabla(\sigma_N-u_N)|^2\,d\Bx \ge \text{Const} \int_{\Omega_N} |\sigma_N-u_N|^2\,d\Bx\]
which together with (\ref{proofeq1}) shows
\begin{eqnarray}
&&(\Lambda_N-\lambda_N)\int_{\Omega_N} u_N(\sigma_N-u_N)\, d\Bx-\int_{\Omega_N} F_N(\sigma_N-u_N)\, d\Bx\nonumber \\
&\ge& \text{Const } (1-\Lambda_N)\int_{\Omega_N} |\sigma_N-u_N|^2\,d\Bx\;. 
\end{eqnarray}
From this and using the fact $\Lambda_N=O(\varepsilon d^{-3})$ one obtains the inequality
\begin{eqnarray}\nonumber
\text{Const }\| \sigma_N-u_N\|_{L_2(\Omega_N)}&\le&       |\Lambda_N-\lambda_N| 
\|u_N\|_{L_2(\Omega_N)}+\| F_N\|_{L_2(\Omega_N)} 
\\
&=&|\Lambda_N-\lambda_N| 
+\| F_N\|_{L_2(\Omega_N)} \label{proofeq2}
\end{eqnarray}
as $\|u_N\|_{L_2(\Omega_N)}=1$. 

One can obtain an estimate for $\sigma_N-u_N$ in $L_2$ in terms of the small parameters $\varepsilon$ and $d$ from (\ref{proofeq2}). To aid us develop such an estimate we now use (\ref{FNestL2}).

\vspace{0.1in}{\bf\emph{Estimates for the remainders.} \rm}
{Rayleigh's quotient allows one to assert that $\lambda_N=O(\varepsilon d^{-3})$. As a consequence we can say 
\[|\lambda_N-\Lambda_N|\le \text{Const }\varepsilon d^{-3}\;.\]
With (\ref{FNestL2}) and  (\ref{proofeq2}), we derive that
\begin{equation}\label{diffL2}
\| \sigma_N-u_N\|_{L_2(\Omega_N)}\le \text{Const }\varepsilon d^{-3}\;.
\end{equation}}
In addition, using integration by parts, the definitions of $u_N$ in (\ref{uN1})--(\ref{uN3}) and $\sigma_N$ in (\ref{eqsig1})--(\ref{eqsig3}), it is possible to show that
\[(\Lambda_N-\lambda_N)\int_{\Omega_N} \sigma_Nu_N\, d\Bx=\int_{\Omega_N} F_N \sigma_N \, d\Bx+\int_{\Omega_N} F_N (u_N-\sigma_N) \, d\Bx\;.\]
The Cauchy inequality then gives the estimate
\[(\Lambda_N-\lambda_N)\int_{\Omega_N} \sigma_Nu_N\, d\Bx\le \| F_N\|_{L_2(\Omega_N)}(1+\| u_N-\sigma_N\|_{L_2(\Omega_N)})\;.\]
Using (\ref{diffL2}), a lower bound for the left-hand side can be established through the estimate
\begin{equation}\label{eqlamproof1}
\int_{\Omega_N} \sigma_Nu_N\, d\Bx=\int_{\Omega_N}\sigma_N^2\,d\Bx+\int_{\Omega_N}\sigma_N(u_N-\sigma_N)d\Bx\ge 1+O(\varepsilon d^{-3})\;.
\end{equation}
Thus (\ref{FNestL2}), (\ref{diffL2}) and (\ref{eqlamproof1}) prove (\ref{lambda_remainderest}).
It remains to combine this with (\ref{proofeq2}) and deduce that (\ref{remainderest}) holds, completing the proof of Lemma \ref{lemremest}. \hfill $\Box$

\vspace{0.1in}Note that it is possible to write $R_N$ of (\ref{uN_app})
\[R_N=-\chi_0 \Psi_0-\sum_{j=1}^N \chi_\varepsilon^{(j)} \Psi_j 
+Q_N\;,\]
so that with (\ref{Sign})
\[u_N=A^{-1}{\sigma_N}+Q_N\;,\] 
and by Lemma \ref{lemremest} we have
\[\|Q_N\|_{L_2(\Omega_N)}\le \text{Const } \varepsilon^{3/2} \,d^{-9/2}\;.\]
This together with (\ref{Sestimate}) shows 
\[\|R_N\|_{L_2(\Omega_N)}\le \text{Const }\varepsilon^{3/2}d^{-9/2}\;.\]
The remainder estimates of Theorems \ref{th1} and \ref{th1a} follow the same procedure as in Lemma \ref{lemremest}, and require the construction of the higher-order terms in the asymptotic approximations. This relies on the introduction of additional model fields for the inclusions $\omega^{(k)}_\varepsilon$ and an additional algebraic system which removes higher-order discrepancies produced on the small inclusions. 

{\emph{Remark.}} The above estimates are improved further through analysis of  higher-order terms in the next section, as follows
\begin{equation}\label{TH3A30a}
\|R_N\|_{L_2(\Omega_N)}\le \text{{Const} } \varepsilon^{2} d^{-6}\;,
\end{equation}
\begin{equation}
|\lambda_N-\Lambda_N|\le \text{{Const} } \varepsilon^{2} d^{-6}\;.
\end{equation}
\section{Higher-order asymptotics}\label{secHOA}
\subsection{Additional model problem}
To section \ref{modprob}, we now add one more field used to construct the higher-order approximation presented here.
We define a vector function $\BD^{(k)}$ as the solution of a problem posed in the exterior of scaled inclusion 
$\omega^{(k)}:=\{\BGx: \varepsilon \BGx+\BO^{(k)}\in \omega^{(k)}_\varepsilon\}$. This vector function is subject to 
\begin{equation*}
\Delta \BD^{(k)}(\BGx)=\BO\;, \quad \BGx\in  \mathbb{R}^3\backslash \overline{\omega^{(k)}}\;,
\end{equation*}
\begin{equation*}
\BD^{(k)}(\BGx)=\BGx\;, \quad \BGx\in \mathbb{R}^3\backslash \overline{\omega^{(k)}}\;,
\end{equation*}
\begin{equation*}
\BD^{(k)}(\BGx)\to \BO\quad \text{ as } \quad |\BGx|\to \infty\;.
\end{equation*}
The behaviour of this vector field at infinity is summarised in the next lemma (see \cite{MMN_book} for the proof)
\begin{lem} $($\text{\emph{see }} \emph{\cite{MMN_book}}$)$
 \label{lemD}
For $|\BGx|>2$, the vector function $\BD^{(k)}=[D^{(k)}_i]_{i=1}^3$ admits the asymptotic representation:
\begin{equation*}
\BD^{(k)}(\BGx)=\CT^{(k)} \frac{\BGx}{|\BGx|^3}+O(|\BGx|^{-3})
\end{equation*}
where $\BCT^{(k)}=[\CT^{(k)}_{ij}]_{i,j=1}^3$ is a constant matrix whose entries are given by
\begin{equation*}
\CT^{(k)}_{ij}=\text{\emph{meas}}_3(\omega^{(k)})\delta_{ij}+\int_{\mathbb{R}^3\backslash {\omega^{(k)}}} \nabla D^{(k)}_i (\BGx)\cdot  \nabla D^{(k)}_j(\BGx) d\BGx\;.
\end{equation*}
and it is symmetric positive definite.
\end{lem}
We define $\BD^{(k)}_\varepsilon(\Bx) =\varepsilon \BD^{(k)}(\BGx)$ and the matrix $\CT^{(k)}_\varepsilon=\varepsilon^3\CT^{(k)}$ which are quantities associated with the exterior of the small inclusion $\omega^{(k)}_\varepsilon$.

Before moving to the proof of the higher-order approximation, we restate Lemma \ref{lem1}, providing an additional term in the far-field asymptotics of $P^{(j)}_\varepsilon$, $1\le j \le N$:
\begin{lem} $($\text{\emph{see }} \emph{\cite{OPTH1}}$)$
 For $|\Bx-\BO^{(j)}|>2 \varepsilon$, the capacitary potential admits the asymptotic representation
 \begin{equation*}\label{P51}
 P^{(j)}_\varepsilon(\Bx)=\frac{ \text{\emph{cap}}(\omega^{(j)}_\varepsilon)}{4\pi |\Bx-\BO^{(j)}|}+\BGb^{(j)}_\varepsilon\cdot \nabla\Big(\frac{1}{4\pi |\Bx-\BO^{(j)}|}\Big) +O\Big(\frac{\varepsilon^3}{|\Bx-\BO^{(j)}|^3}\Big)\;,
 \end{equation*}
 where $|\BGb_\varepsilon^{(j)}|=O(\varepsilon^2)$.
  \label{lem1a}
 \end{lem}

\subsection{Main result I: Higher-order approximation for the first eigenfunction}
Here we present a theorem concerning a higher-order asymptotic approximation of the first eigenvalue and corresponding eigenfunction of the Laplacian in $\Omega_N$.
Before moving to the theorem regarding this eigenfield, we introduce the new  constant coefficients used in this approximation. We have a vectors $\BB^{(k)}$ that appear as coefficients of the fields $\BD_\varepsilon^{(k)}$, $1\le k \le N$, and are given by:
\begin{equation}
\label{BBk_result}
\BB^{(k)}=C_k  \text{cap}(\omega^{(k)}_\varepsilon)\nabla_\Bx \CH(\BO^{(k)}, \BO^{(k)})-\sum_{\substack{j \ne k\\ 1\le j \le N}} C_j  \text{cap}(\omega^{(j)}_\varepsilon)\nabla_\Bx\CG (\BO^{(k)}, \BO^{(j)})\;.
\end{equation}
Another algebraic system is also used to ensure the asymptotic formulae presented satisfy the boundary conditions to a high accuracy. To this end, we also use the coefficients $A_j$, $1\le j \le N$, which are solutions of 
\begin{eqnarray}
-v^{(k)}&=&A_k(1-\text{{cap}}(\omega^{(k)}_\varepsilon)\{\CH(\BO^{(k)},\BO^{(k)})-\Gamma_\Omega^{(k)} \})\nonumber\\ \nonumber\\
&&+ \sum_{\substack{j\ne k \\ 1\le j \le N}}A_j  \text{{cap}}(\omega^{(j)}_\varepsilon)(\CG(\BO^{(k)}, \BO^{(j)})+\Gamma_\Omega^{(j)})\;.\label{sys2Bj1}
\end{eqnarray}
Here
\begin{eqnarray}
&&v^{(k)}=C_k \BGb^{(k)}_\varepsilon \cdot (\nabla_\Bz \CH(\BO^{(k)}, \Bz)\Big|_{\Bz=\BO^{(k)}}+\BGg^{(k)}_\Omega)\nonumber \\
&&-\sum_{\substack{j \ne k\\ 1\le j \le N}} C_j  \BGb_\varepsilon^{(j)} \cdot (\nabla_\Bz \CG(\BO^{(k)}, \Bz)\Big|_{\Bz=\BO^{(j)}}-\BGg^{(j)}_\Omega)\nonumber \\
&&+\Lambda_1 \sum_{j=1}^N C_j \text{{cap}}(\omega_\varepsilon^{(j)}) \int_\Omega \CG(\By, \BO^{(k)})\CG(\By, \BO^{(j)})\, d\By\;,\label{sys2Bj2}
\end{eqnarray}
where $\Lambda_1$ is given by the right-hand side of (\ref{WN61}) and 
\[\boldsymbol{\gamma}_\Omega^{(j)}=-\int_\Omega \nabla_{\Bz} \left(\frac{1}{4\pi|\Bx-\Bz|}\right)\Big|_{\Bz=\BO^{(j)}}d \Bx\;.\]

We have the theorem:
\begin{thm}\label{th2}
Let the parameters $\varepsilon$ and $d$ satisfy 
\[\varepsilon<c \,d^3\;,\]
where $c$ is a sufficiently small constant. Then the first eigenfunction of the Laplacian in $\Omega_N$ is given by
\begin{eqnarray}\label{UfnAresult}
u_N(\Bx)&=&1+\sum_{j=1}^N (C_j +A_j) \{P_\varepsilon^{(j)}(\Bx)-\text{\emph{cap}}(\omega^{(j)}_\varepsilon) (\CH(\Bx, \BO^{(j)})-\Gamma_\Omega^{(j)})\} \nonumber\\
&& +\sum_{j=1}^N\BB^{(j)}\cdot \BD^{(j)}_\varepsilon(\Bx)+\sum_{j=1}^NC_j \BGb^{(j)}_\varepsilon \cdot [\nabla_\Bz \CH(\Bx, \Bz)\Big|_{\Bz=\BO^{(j)}}+\BGg_\Omega^{(j)}]\nonumber \\
&&+\Lambda_1 \sum_{j=1}^N C_j \text{\emph{cap}}(\omega^{(j)}_\varepsilon) \int_\Omega \CG(\By, \Bx)\CG(\By,\BO^{(j)})d\By+R_N(\Bx)
\end{eqnarray}
where the coefficients $C_j$ and $A_j$, $1\le j \le N$, satisfy the solvable systems $(\emph{\ref{sys_C_thm}})$ and $(\emph{\ref{sys2Bj1}})$--$(\emph{\ref{sys2Bj2}})$, respectively.

The remainder $R_N$ admits the estimate
\begin{equation}\label{TH3A30}
\|R_N\|_{L_2(\Omega_N)}\le \text{\emph{Const} } \varepsilon^{5/2} d^{-15/2}\;.
\end{equation}
\end{thm}

The proof of Theorem \ref{th2} can be found in the Appendix.

\subsection{Main result II: Higher-order approximation for the first eigenvalue}
The next theorem contains the higher-order approximation of $\lambda_N$.
\begin{thm}\label{LHHO}
Let the coefficients $\varepsilon$ and $d$ satisfy 
\[\varepsilon < c \, d^3\]
then the approximation to the first eigenvalue of the Laplacian in $\Omega_N$ has the form
\begin{eqnarray*}
\lambda_N&=&\Lambda_1+\Lambda_2+\lambda_{R,N}\label{lamNHO}
\end{eqnarray*}
where $\Lambda_1$ is the right-hand side of $(\ref{WN61})$, 
\begin{equation}\label{lam2_HOA}
\Lambda_2=-\frac{1}{|\Omega|}\sum_{j=1}^N \text{\emph{cap}}(\omega^{(j)}_\varepsilon)(A_j  +\Lambda_1  C_j \Gamma_\Omega^{(j)} )\;,
\end{equation}
$C_j$, $1\le j \le N$, are the same as in the algebraic system $(\ref{sys_C_thm})$ and $\lambda_{R,N}$ is now the remainder of this approximation with
\begin{equation*}\label{LRNest}
|\lambda_{R,N}|\le \text{\emph{Const} } \varepsilon^{5/2} d^{-15/2}\;.
\end{equation*}
\end{thm}

For the relevant derivation of Theorem \ref{LHHO} we refer to  the Appendix. 

\subsection{Completion of the proofs of Theorems 1 and 2}
Concerning the coefficients $A_j$ and $\BB^{(j)}$, $1\le k \le N$, one can obtain the estimates presented in the next lemma. The detailed proofs are found in the Appendix.
\begin{lem}\label{lemcoeffs2}
Let the small parameters $\varepsilon$ and $d$ satisfy the inequality
\begin{equation*}\label{epsvsd1aa}
\varepsilon < c\, d^3
\end{equation*}
where $c$ is a sufficiently small constant.
Then the system ${\rm(\ref{sys2Bj1})}$--${\rm(\ref{sys2Bj2})}$ is solvable and the estimates
\begin{equation}\label{Bjsumest}
\sum_{j=1}^N A_j^2\le \text{\emph{Const }} \varepsilon^4 d^{-12}\;,
\end{equation}
\begin{equation}\label{BBsumest2}
\sum_{j=1}^N |\BB^{(j)}|^2\le \text{\emph{Const }} \varepsilon^2 d^{-9}\;,
\end{equation}
hold.
\end{lem}

With Lemmas \ref{lemcoeffs} and \ref{lemcoeffs2}, one can show that 
\[|\Lambda_2|\le \text{Const }\varepsilon^2 d^{-6}\;,\]
with $\Lambda_2$ given in (\ref{lam2_HOA}).
This with Theorem \ref{LHHO}, proves Theorem \ref{th1a}.

Note that it is possible to write $R_N$ of (\ref{UfnAresult})
\begin{equation}\label{eqRN_latest}
R_N=-\chi_0 \Psi_0-\sum_{j=1}^N \chi_\varepsilon^{(j)} \Psi_j 
+Q_N\;,
\end{equation}
where for the higher-order approximation presented here
the function $\Psi_0$ is defined as 
\begin{eqnarray}
&&\Psi_0(\Bx)=\sum_{j=1}^N (C_j+A_j)\Big[P^{(j)}_\varepsilon(\Bx)-\frac{\text{cap}(\omega^{(j)}_\varepsilon)}{4\pi |\Bx-\BO^{(j)}|}\Big]\nonumber\\
&&+\sum_{j=1}^N \BB^{(j)}\cdot \BD_\varepsilon^{(j)}(\Bx)+\sum_{j=1}^N C_j \BGb_\varepsilon^{(j)} \cdot \nabla_{\Bz}\left(\frac{1}{4\pi |\Bx-\Bz|}\right)\Big|_{\Bz=\BO^{(j)}}\;. \nonumber \\\label{EqnPsi0_H}
\end{eqnarray} 
For $1\le k \le N$, $\Psi_k$  has the form
\begin{eqnarray}
&&\Psi_k(\Bx)= \sum_{\substack{j\ne k\\ 1\le j \le N}} \Big[ (C_j+A_j)\Big[P^{(j)}_\varepsilon(\Bx)-\frac{\text{cap}(\omega^{(j)}_\varepsilon)}{4\pi |\BO^{(k)}-\BO^{(j)}|}\Big]+\BB^{(j)}\cdot \BD_\varepsilon^{(j)}(\Bx)\Big]\nonumber\\
&&+\sum_{\substack{j\ne k \\ 1\le j \le N}} C_j \BGb_\varepsilon^{(j)} \cdot \nabla_{\Bz}\left(\frac{1}{4\pi |\BO^{(k)}-\Bz|}\right)\Big|_{\Bz=\BO^{(j)}}+\BB^{(k)}\cdot (\Bx-\BO^{(j)})\nonumber\\
&&\nonumber\\
&&-\sum_{j=1}^N  (C_j+A_j) \text{cap}(\omega^{(j)}_\varepsilon) (\CH(\Bx, \BO^{(j)})-\CH(\BO^{(k)}, \BO^{(j)}))\nonumber\\
&&+\sum_{j=1}^N C_j \BGb_\varepsilon^{(j)}\cdot [\nabla_{\Bz}\CH(\Bx, \Bz)\Big|_{\Bz=\BO^{(j)}}-\nabla_{\Bz}\CH(\BO^{(k)}, \Bz)\Big|_{\Bz=\BO^{(j)}}]\nonumber \\
&&+\Lambda_1\sum_{j=1}^N C_j \text{cap}(\omega^{(j)}_\varepsilon) \int_\Omega \CG(\By, \BO^{(j)})(\CG(\By, \Bx)-\CG(\By, \BO^{(k)}))d\By\;.\nonumber\\ \label{EqnPsik_H}
\end{eqnarray}
Here, the functions $\Psi_k$, $0\le k \le N$, are constructed in order to satisfy the properties (\ref{eqDNRN}) and (\ref{eqDNRN_2}), involving $R_N$ defined  in Theorem \ref{th2}, together with the leading order term of the approximation (\ref{UfnAresult}). The latter term we denote by $V$ (see (\ref{UfnA}) in the Appendix)  and this replaces $U$ in (\ref{eqDNRN}) and (\ref{eqDNRN_2}).

In the Appendix, we prove estimates concerning $\Psi_k$, $0\le k \le N$, and their derivatives in $L_2$, which are contained in the next lemma.
\begin{lem} \label{PsiL2ests} 
The function $\Psi_0$ satisfies the $L_2$-estimates
\begin{eqnarray}
&& \|\Psi_0\|_{L_2(\CV)}^2\le \text{\emph{Const }} \varepsilon^6d^{-6}\;,\qquad  \|\nabla \Psi_0\|_{L_2(\CV)}^2\le \text{\emph{Const }} \varepsilon^6d^{-6}\;,\nonumber\\
\label{C1}
\end{eqnarray}
whereas for the $\Psi_k$, $1\le k \le N$,  we have:
\begin{eqnarray}
&&\sum_{k=1}^N \|\Psi_k \|^2_{L_2(B_{3\varepsilon}^{(k)}\backslash {\omega_\varepsilon^{(k)}})}\le \text{\emph{Const }} \varepsilon^9d^{-15}\;,
\end{eqnarray}
and
\begin{eqnarray}&&\sum_{k=1}^N \|\nabla \Psi_k \|^2_{L_2(B_{3\varepsilon}^{(k)}\backslash {\omega_\varepsilon^{(k)}})}\le \text{\emph{Const }} \varepsilon^7d^{-15} \;.\nonumber\\\label{C4}
\end{eqnarray}
\end{lem}

Then, with (\ref{eqRN_latest})
\[u_N=A^{-1}{\sigma_N}+Q_N\;,\] 
with $\sigma_N$ having the form 
\begin{equation}
\label{eqsig2}
\sigma_N=A \,\{V-\chi_0 \Psi_0 -\sum^N_{j=1} \chi^{(j)}_\varepsilon \Psi_j\}\;,
\end{equation}
where (\ref{UfnAresult}) can be used to define $V=u_N-R_N$. From (\ref{TH3A30}) we have
\[\|Q_N\|_{L_2(\Omega_N)}\le \text{Const } \varepsilon^{5/2} \,d^{-15/2}\;.\]
In addition, by Lemma \ref{PsiL2ests} and the definition of the cut-off functions
\[\Big\|\chi_0 \Psi_0+\sum_{j=1}^N \chi_\varepsilon^{(j)} \Psi_j \Big\|_{L_2(\Omega_N)}\le \text{Const }\varepsilon^3 d^{-3}\;.\]
Thus with (\ref{eqRN_latest})
\[\|R_N\|_{L_2(\Omega_N)}\le \text{Const}\{\varepsilon^3 d^{-3}+\varepsilon^{5/2}d^{-15/2}\}\;,\]
proving Theorem \ref{th2}.
Now, using Lemmas \ref{lemD} and \ref{lem1a}, one can show the term
\begin{eqnarray*}\label{UfnA11}
W(\Bx)&=&\sum_{j=1}^N A_j \{P_\varepsilon^{(j)}(\Bx)-\text{cap}(\omega^{(j)}_\varepsilon) (\CH(\Bx, \BO^{(j)})-\Gamma_\Omega^{(j)})\} \nonumber\\
&& +\sum_{j=1}^N\BB^{(j)}\cdot \BD^{(j)}_\varepsilon(\Bx)+\sum_{j=1}^NC_j \BGb^{(j)}_\varepsilon \cdot [\nabla_\Bz \CH(\Bx, \Bz)\Big|_{\Bz=\BO^{(j)}}+\BGg_\Omega^{(j)}]\nonumber \\
&&+\Lambda_1 \sum_{j=1}^N C_j \text{cap}(\omega^{(j)}_\varepsilon) \int_\Omega \CG(\By, \Bx)\CG(\By,\BO^{(j)})d\By\end{eqnarray*}
admits the estimate
\[|W(\Bx)|\le \text{Const }\varepsilon^2 d^{-6}\;, \quad \Bx \in \Omega_N \;,\]
and so
\[\|W\|_{L_2(\Omega_N)}\le \text{Const }\varepsilon^2 d^{-6} \;.\]
This together with Theorem \ref{th2}  completes the proof of Theorem \ref{th1}. \hfill $\Box$

\section{Approximations for dilute clusters versus large clusters of inclusions}\label{finiteVlarge}

We now consult the case of a domain containing a dilute cluster of inclusions, which was considered in \cite{OPTH1}. For this we assume $N$ is finite and we define the domain $\Omega_\varepsilon=\Omega\backslash \cup_{j=1}^N\overline{\omega_\varepsilon^{(j)}}$. We now relax the assumptions of (\ref{assumptions}) and constrain the interior points of the collection of inclusions $\omega^{(j)}_\varepsilon$, $1\le j \le N$, to be separated by a finite distance from each other (so that $d=O(1)$), and we assume these points are sufficiently far away from the exterior boundary $\partial \Omega$.

For this configuration, the first eigenvalue $\lambda_\varepsilon$ and the corresponding eigenfunction $u_\varepsilon$ satisfy:
\begin{equation}\label{uN1_mazya}
\Delta_{\Bx} u_\varepsilon(\Bx)+\lambda_\varepsilon u_\varepsilon(\Bx)=0\;, \quad \Bx \in \Omega_\varepsilon\;,
\end{equation}
\begin{equation}\label{uN2_mazya}
\frac{\partial u_\varepsilon}{\partial n_\Bx}(\Bx)=0\;, \quad \Bx \in\partial  \Omega\;,
\end{equation}
\begin{equation}\label{uN3_mazya}
{u_\varepsilon}(\Bx)=0\;, \quad \Bx \in\partial  \omega_\varepsilon^{(j)}, \quad 1\le j\le N\;.
\end{equation}

According to the method of compound asymptotic expansions presented in \cite{OPTH1} for the dilute cluster of inclusions the first eigenvalue $\lambda_\varepsilon$ and the corresponding eigenfunction $u_\varepsilon$ are approximated as follows:
\begin{thm}\label{th1mazya} 
The asymptotic approximation of the eigenfunction $u_\varepsilon$, which is a solution of $(\ref{uN1_mazya})$--$(\ref{uN3_mazya})$ in $\Omega_\varepsilon$, is given by
\begin{eqnarray*}
u_\varepsilon(\Bx)&=&1-\sum_{j=1}^N
{ \Gamma^{(j)}_\Omega\text{\emph{cap}}(\omega^{(j)}_\varepsilon)}
\nonumber \\
&&-\sum^N_{j=1} \{P^{(j)}_\varepsilon(\Bx)-\text{\emph{cap}}(\omega^{(j)}_\varepsilon)\CH(\Bx, \BO^{(j)})\}+R_\varepsilon(\Bx)\;,
\label{uN_app_mazya}
\end{eqnarray*}
where $R_\varepsilon$ is the remainder term satisfying
\begin{equation*}\label{remainderest1_mazya}
\|R_\varepsilon\|_{L_2(\Omega_\varepsilon)}\le \text{\emph{Const} } \varepsilon^{2}  \;.
\end{equation*}

\end{thm}

\begin{thm}\label{th1a_mazya} 
 The first eigenvalue  $\lambda_\varepsilon$ corresponding to the eigenfunction $u_\varepsilon$ in $\Omega_\varepsilon$ admits the approximation
\begin{equation*}\label{lambda_N_app_mazya}
\lambda_\varepsilon=\frac{1}{|\Omega|}\sum^N_{j=1} {\text{\emph{cap}}(\omega^{(j)}_\varepsilon)}+
O(\varepsilon^{2})
\;.
\end{equation*}
\end{thm}

The results of Theorems \ref{th1mazya} and \ref{th1a_mazya}, for the domain with a finite cluster can be compared with the results of Theorems \ref{th1} and \ref{th1a}. The asymptotic approximations have a similar structure, utilising model problems posed in the domain $\Omega$ and in the exterior of the sets $\omega^{(j)}_\varepsilon$, $1\le j \le N$. One can also obtain the estimates for the remainder  of these approximations by carrying out the approach presented in sections \ref{remainderestsec} and \ref{secHOA}. 

However, we note the uniform approximation for $u_\varepsilon$ does not require the solution  of an algebraic system for unknown coefficients, which are responsible for compensating  the error  produced in the boundary conditions on small inclusions. The approximation for $u_N$ does require the solutions $C_j$, $1\le j \le N$, to system (\ref{sys_C_thm}).
This system contains information about the shape and size of small inclusions, through the presence of the capacity of individual inclusions. In addition, the positions of the inclusions are incorporated in this system, through the arguments of Neumann's function  $G$. 

As a result, it can be concluded from comparing approximations (\ref{lambda_N_app_mazya}) and (\ref{lambda_N_app}) for the first eigenvalue, that the former approximation, to leading order, only takes into account the shape and size of the inclusions and the exterior domain $\Omega$. In addition to this, the leading order term of the approximation in (\ref{lambda_N_app}) incorporates the knowledge of the position of the inclusions through $C_j$, $1\le j \le N$.

It should be noted that the approximations in Theorems \ref{th1} and \ref{th1a_mazya} cannot efficiently serve the case when the inclusions are close together and their number becomes large, whereas
 approximations (\ref{th1}) and (\ref{th1a_mazya}) cover both the case of the domain such as this and the domain with the finite cluster $\Omega_\varepsilon$.

\section{Numerical illustration}\label{numsec}
In this section, we implement the asymptotic formulae of Theorem 1 in numerical schemes and compare with benchmark finite element computations in COMSOL. 

We begin with a general description of the computational geometry, involving a sphere containing small spherical inclusions, in section \ref{sec6p1}. There, we  also present the model fields related to the exterior and interior problems relevant to the asymptotic approximation  (\ref{uN_app}). In section \ref{sec6p2}, the asymptotic formulae of Theorem \ref{th1} and \ref{th1a} are compared with the benchmark finite element computations.
\subsection{Computational geometry and model fields for spherical bodies and inclusions}\label{sec6p1}
Here we consider the domain $\Omega$ to be a sphere $B_R$ of radius $R$, with centre at the origin.
In addition, let the sets $\omega^{(j)}_\varepsilon$, $1\le j\le N$, be small spheres with centres $\BO^{(j)}$ and radii $r^{(j)}_\varepsilon$, respectively.

{\bf Capacitary potential for the spherical inclusion $\omega_{\varepsilon}^{(j)}$. } For the spherical inclusion of radius $r_\varepsilon^{(j)}$ and centre $\BO^{(j)}$ inside in $\mathbb{R}^3$, the capacitary potential is
\[P_\varepsilon^{(j)}(\Bx)=\frac{r^{(j)}_\varepsilon}{|\Bx-\BO^{(j)}|}\;,\]
where the capacity for the cavity is $\text{cap}(\omega_\varepsilon^{(j)})=4\pi r_\varepsilon^{(j)}$.

\vspace{0.1in}\noindent {\bf The Neumann function in $B_R$. } For the sphere $B_R$, the Neumann function $\CG$ is a solution of the problem
\begin{equation*}\label{eN1}
\Delta_{\Bx} \CG(\Bx, \By)+\delta(\Bx-\By)-\frac{3}{4\pi R^3}=0\;, \quad \Bx \in B_R\;,
\end{equation*}
\begin{equation*}\label{eN2}
\frac{\partial \CG}{\partial n_\Bx}(\Bx, \By)=0\;, \quad \Bx \in\partial  B_R\;.
\end{equation*}
The  function $\CG$ is given by 
\[\CG(\Bx, \By)=\frac{1}{4\pi |\Bx-\By|}-\CH(\Bx, \By)\;,\]
where the regular part $\CH$ takes the form
\begin{eqnarray*}
\CH(\Bx,\By)&=&-\frac{|\Bx|^2+|\By|^2}{8\pi R^3} -\frac{R}{4\pi |\By||\Bx-\overline{\By}|}
\nonumber\\
&&-\frac{1}{4\pi R}\log\Big[\frac{2 R^2}{R^2-\Bx \cdot \By +|\By||\Bx-\overline{\By}|}\Big]
\end{eqnarray*}
with $\overline{\By}=R^2\By/|\By|^2$. The above representation can be found through modification of the result in \cite{Namar}, where the last two terms in the above right-hand side can be found. As in \cite{Namar}, we note that logarithmic potentials are characteristic of two dimensional problems, for which they are harmonic. We note that the logarithmic term occurring in the right-hand side is harmonic  and analytic in $B_R$. A detailed proof of these properties are found in \cite{Namar}. The second term is obtained through the classic method of images which yields a harmonic function.

\vspace{0.1in}\noindent {\bf Algebraic system. } In particular if $\Omega=B_R$, we have
\begin{equation}
\label{int0} 
\int_{B_R}\frac{d\Bz}{4\pi |\Bz-\BO^{(j)}|}=\frac{1}{2}\left(R^2-\frac{|\BO^{(j)}|^2}{3}\right)\;,
\end{equation}
which can be computed through Green's formula applied to the kernel of the above integral and the function $|\Bz|^2$ in $\Omega$.

 Then, in combining (\ref{sys_C_thm}), (\ref{eqGAMOM}) and (\ref{int0})
  we receive 
 that for this scenario, the coefficients $C_k$, $1\le k \le N$, can be determined from
 \begin{eqnarray*}\label{sys_C_thm_num_sim}
&&1+C_k+\sum_{\substack{j\ne k\\ 1\le j\le N}} {C_j\text{{cap}}(\omega_\varepsilon^{(j)})}\Big\{\CG(\BO^{(k)}, \BO^{(j)})+ \frac{3}{8\pi R}-\frac{1}{8\pi R^3}{|\BO^{(j)}|^2}
\Big\}=0\;.
\end{eqnarray*}
\subsection{Comparison of asymptotics with the finite element method}\label{sec6p2}
We compute the first eigenvalue, for several configurations of $\Omega_N$, using the approximation (\ref{lambda_N_app}) and compare this with results based on the finite element method in COMSOL. The results are presented in Table \ref{table:lamN}.
 Here, we consider the sphere  $\Omega$, centred at the origin, having radius $R=7$. The spherical inclusions are arranged inside this domain, according to Table \ref{table:lamN}. 
 We note that there is an excellent agreement for values given by the method of finite elements and the asymptotic formula (\ref{lambda_N_app}).
 
First we consider the case when the positions of inclusions form the corners of the cube with centre (0,0,0) and side length 1. In this case, the centres $\BO_{ijk}$ are arranged as follows:
\[\BO_{ijk}=\Big(-\frac{1}{2}+{i-1}, -\frac{1}{2}+{j-1}, -\frac{1}{2}+{k-1}\Big)\;,\]
with $1\le i,j,k\le 2$. We denote this collection of points by the set
\[\BP=\{\BO_{ijk}: 1\le i,j,k\le 2\}\;.\]
In addition, later we use the notations $\BV=(-0.25,0,0)$ and \\$\BW=\{(-0.25,0,0), (0.25,0,0)\}$.
 The radii $r_{ijk}$ corresponding to the inclusion with centre $\BO_{ijk}$, are
 \begin{eqnarray*}
 &r_{111}=0.0125\quad   r_{112}=0.015\quad  r_{121}=0.0075\quad  r_{211}=0.01\\
 &
r_{212}=0.02\quad r_{221}=0.0125
\quad   r_{122}=0.03 \quad  r_{222}=0.01725
    \;,
 \end{eqnarray*} 
 and the set $\BR$ is used to denote the collection of these values.
 
 We define the small parameters as
 \[\varepsilon=R^{-1}{\max_{1\le j\le N}\{r_\varepsilon^{(j)}\}}\quad \text{ and } \quad d={R}^{-1}{\displaystyle{\min_{\substack{k\ne j\\ 1\le k,j\le N}} \text{dist}(\BO^{(j)}, \BO^{(k)})}}\;.\]
For $N=8$, these parameters are $\varepsilon=0.0043$ and $d=0.1428$ for the simulations presented here.
 
 \subsection{Evaluation of the first eigenvalue}
 
 In Table \ref{table:lamN},  we show the first eigenvalue computed in COMSOL and the computations based on the  asymptotic approximation  (\ref{lambda_N_app})
  for various configurations of inclusions. We consider arrangements of inclusions where $N=8,9$ or 10. We begin with the configuration having centres and radii according to $\BP$ and $\BR$. Results are also presented for the $\BP\cup \BV$ and $\BP\cup \BW$, where additional inclusions have been introduced in the simulations. The radii of the additional inclusions are also supplied in Table \ref{table:lamN}.
 
 The computations agree very well with each other.
 The relative error in the computations for $N=8,9,10$ (with $d=0.1428,  0.1072, 0.0714$, respectively) is less than 4\%.
{ This  error between the computations for $\lambda_N$ increases as we increase $N$.  Note that the mesh size for each simulation has the same order. The mesh sizes presented represent those close to the maximum mesh size that the first eigenfield and eigenvalue could be computed with in COMSOL. Therefore, the computations from COMSOL may not be as accurate as one would expect for the case of $N=10$. }
 
\begin{table}[ht]
\caption{Comparison of approximation for $\lambda_N$ with results from COMSOL for arrangements with $N=8,9,10$ inclusions.}
\centering
\begin{tabular}{|c |c |c|c |c |c |c| c| c|}
\hline\hline
Radii & Centres  & Mesh size&$\begin{array}{c}
\text{$\lambda_N$ } \\ \text{ (approx.)}\\ \text{($\times 10^{-3}$)}\end{array}$
& $\begin{array}{c}
\text{$\lambda_N$}\\  \text{(COMSOL)} \\ \text{($\times 10^{-3}$)} \end{array}$ & $\begin{array}{c}\text{Relative}\\ \text{error}\end{array}$
\\ [0.5ex] 
\hline
\BR
& 
\BP
& 1477957 & 0.96588 &  0.98287 &  1.73\%\\
\hline
$\BR \cup \{0.02\}$
& 
$\BP \cup \BV $& 1598887 & 1.08686 &1.11180 & 2.64\%\\
\hline
$\begin{array}{c}
\BR \cup \{0.02\}
\\
\cup \{0.015\}\end{array}$
 & 
$ \begin{array}{c}\BP  \cup \BW
 \end{array}$
 & 1670448 & 1.17062 & 1.21600 & 3.37\%\\
\hline
\end{tabular}
\label{table:lamN}
\end{table}

\subsection{Computations for the first eigenfunction}
 Next, for an arrangement of $N=8$ voids, we compute the first eigenfunction using the asymptotic formula (\ref{uN_app}). The resulting field computed in COMSOL is  shown in Figure \ref{fig:exmeso-0} as a slice plot. Here, the perturbation to the field can be clearly seen near the origin.
 A contour plot of the field along the plane $x_3=-0.5$, in the vicinity of the inclusions, based on the COMSOL computations  is shown in Figure \ref{fig:exmeso-a}. The corresponding computations based on the asymptotic approximation (\ref{uN_app}) are given in Figure \ref{fig:exmeso-b}.  The   computations in Figures \ref{fig:exmeso-a} and \ref{fig:exmeso-b} are visibly indistinguishable. In fact the average absolute error between the results inside this computational window is $2.1 \times 10^{-3}$. The COMSOL computation for  first eigenfield along the plane $x_3=0.5$, near the inclusions, is presented in Figure \ref{fig:exmeso_c}. Once again, the eigenfield computed via (\ref{uN_app}) is shown in Figure \ref{fig:exmeso_c}. 
 There is visibly an excellent agreement between the two computations, with the average absolute error between these results being $3.3 \times 10^{-3}$ inside the computational window. The example here clearly demonstrates the accuracy of the asymptotic approach as this compares well with the benchmark results of COMSOL.
 



 \subsection{The asymptotic coefficients $C_j$, $1\le j \le N$}
 The asymptotic coefficients $C_j$, $1\le j \le N$, contained in the approximation for the first eigenvalue and corresponding eigenfunction of the Laplacian in $\Omega_N$ can be computed by solving the system (\ref{sys_C_thm}). In this section, the cluster inside the spherical body is represented by a collection of many small spherical inclusions positioned close to each other.
 The algebraic system for this case takes the form (\ref{sys_C_thm_num_sim}). For a configuration with $N=1728$ inclusions, with $\varepsilon=1.7369 \times 10^{-6}$ and $d=0.0238$ the quantities $|C_j|$ are plotted as functions of $j$, $1\le j \le N$ in Figure \ref{fig:exmeso-0_coeffs} . The resulting picture shows the coefficients are close to 1 (corresponding to the dilute approximation)
and not comparable with the magnitude  of the $\varepsilon$ and $d$.

 \begin{figure}%
  \centering
       \label{fig:exmeso-0_coeffs}%
\centering
        \includegraphics[width=0.9\textwidth]{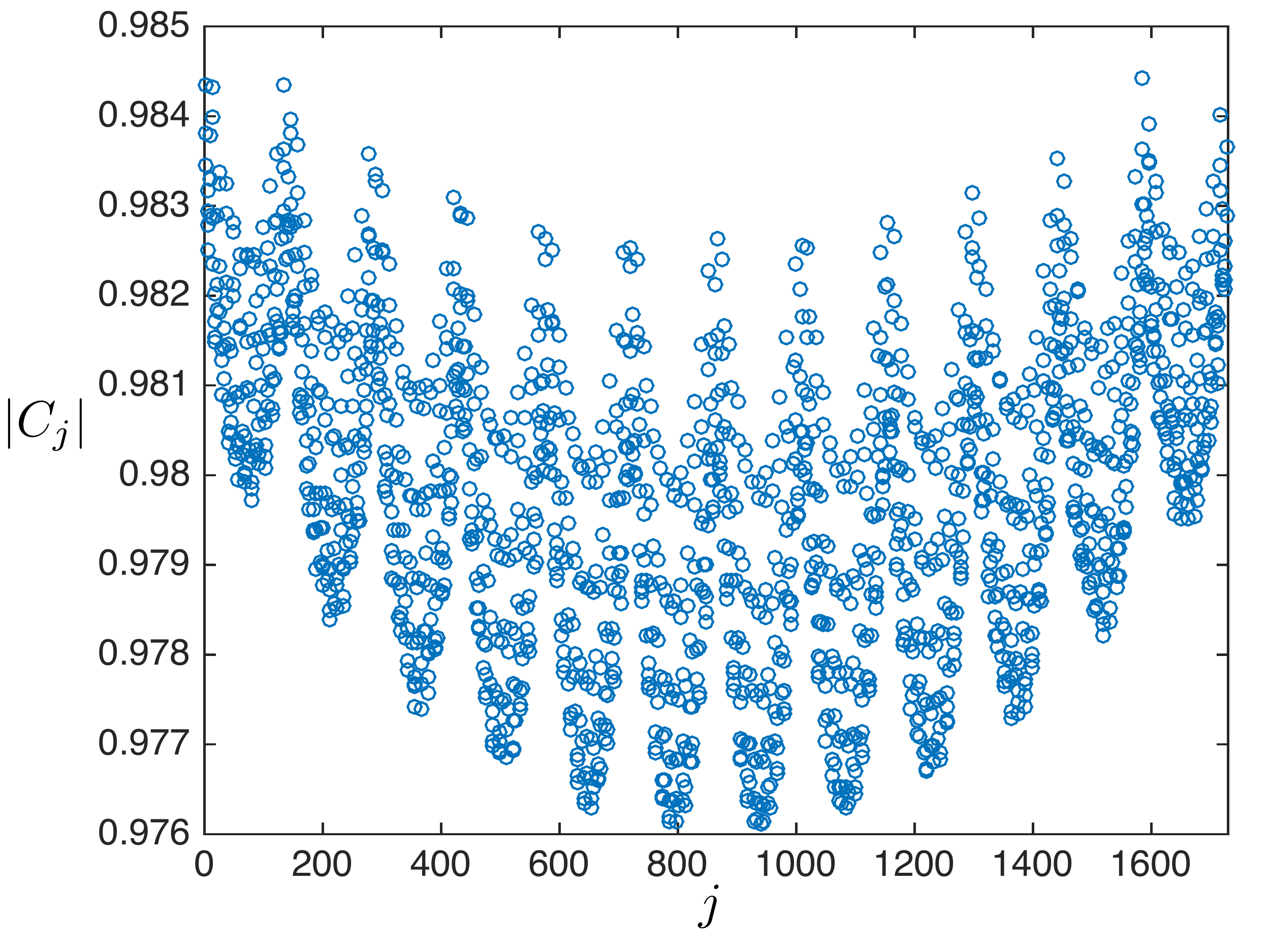}
        \caption{The quantities $|C_j|$, plotted as a function of $j$, $1\le j \le N$, $N=1728$. The coefficients correspond to the case of a  non-periodic  cubic cluster of spherical inclusions (characterised by $\varepsilon=1.7369 \times 10^{-6}$ and $d=0.0238$) contained in a spherical body of radius $7$ with centre at the origin. The index $j$ is assigned in a way that we count the inclusions along the $k^{th}$ plane,  defined by $x_3=(2k-1)/12 $, $1\le k \le 12$, inside the cluster. In each plane there are 144 inclusions, i.e. $C_j$, $1\le j\le 144$ corresponds to the inclusions on the plane $x_3=0$. }
        \end{figure}

 \section{Comparison with the homogenisation approach for a periodic cloud contained in a body}\label{hsolsec}
In this section, we discuss the connection of the algebraic system (\ref{sys_C_thm}) to the homogenised problem obtained in the limit as $N\to \infty$, which we show is a mixed boundary value problem for an inhomogeneous equation. We begin with some underlying assumptions which lead to the homogenised problem.

\subsubsection*{Geometric assumptions}

 We assume the domain $\omega$ is occupied by a periodic distribution of identical inclusions. To describe the cloud $\omega$ inside $\Omega$, we divide the set $\omega$ into $N$ small identical cubes $Q^{(j)}_d= \BO^{(j)}+ Q_d$, with $Q_d=\{\Bx: -d/2<x_j<d/2, 1\le j \le  3\}$, with centres $\BO^{(j)}$ and such that  $\omega^{(j)}_\varepsilon \subset Q^{(j)}_d$, $1\le j \le N$.
 Here, $\varepsilon$  and $d$ are subjected to the constraint (\ref{epstod}).
Each inclusion is defined by $\omega_\varepsilon^{(j)}= \BO^{(j)}+ F_\varepsilon$, for $1\le j \le N$, where $F_\varepsilon$ is a specified set with smooth boundary, containing the origin as an interior point and having a diameter characterised by $\varepsilon$. 
Since the inclusions are identical we have for $1\le j \le N$,  $\text{cap}(\omega^{(j)}_\varepsilon)=\text{cap}(F_\varepsilon)$. 
Here we define
\begin{equation}\label{eqmu}
\mu=\lim_{d\to 0}\frac{\text{cap}(F_\varepsilon)}{d^3}\;.
\end{equation}
In the next section, we consider the case when $N\to \infty$ (and subsequently $d\to 0$, $\varepsilon\to 0$).
In this limit, we will assume the solutions $C_j$, $1\le j \le N$, of the algebraic system (\ref{sys_C_thm})
converge to $\hat{C}_j$, $1\le j \le N$, respectively, and  they are given as
\begin{equation*}
\hat{C}_j=\hat{u}(\BO^{(j)})\;, \quad 1\le j \le N\;,
\end{equation*}
with $\hat{u}$ being the solution of the homogenised problem obtained in the same limit from problem (\ref{uN1})--(\ref{uN3}). 

\subsubsection*{Algebraic system and connection to the auxiliary homogenised  equation}
Let
\[G(\Bx, \By)=\CG(\Bx,\By)+\Gamma_\Omega(\By)\; \]
and 
\[H(\Bx, \By)=\frac{1}{4\pi|\Bx-\By|}-G(\Bx, \By)\;.\]
Here 
\[\Gamma_\Omega(\By)=\frac{1}{4\pi|\Omega|}\int_\Omega \frac{d\Bz}{|\Bz-\By|}\;,\]
and we note $\Gamma_\Omega(\BO^{(j)})=\Gamma_\Omega^{(j)}$, $1\le j \le N$.
In addition, 
\begin{equation}\label{Grel}
\Delta_\Bx G(\Bx, \By)+\delta(\Bx- \By)-\frac{1}{| \Omega|}=0, \quad \Bx\in \Omega\;,
\end{equation}
which follows from the definition of $\CG$ in section \ref{modprob} (see (\ref{N1})).
From (\ref{sys_C_thm}), the algebraic system  may be written as
\begin{eqnarray*}
&&1+C_k(1-\text{{cap}}(F_\varepsilon)H(\BO^{(k)},\BO^{(k)}))\nonumber \\ \nonumber\\
&&+\frac{\text{{cap}}(F_\varepsilon)}{d^3}\sum_{\substack{j\ne k\\ 1\le j\le N}} {C_j}G(\BO^{(k)}, \BO^{(j)}) d^3=0\label{sys_h}
\end{eqnarray*}
for $1\le k \le N.$ 
 By taking the limit $N\to \infty$ (so that $d\to 0$) in the preceding equation, we  replace the Riemann  sum  by an integral over $\omega\backslash\overline{Q_d^{(k)}}$. Simultaneously, as $N\to \infty$, we have  $d\to 0$, $\varepsilon\to 0$ and we retrieve
 the equation 
\begin{eqnarray*}
1+\hat{u}(\Bx)
+\mu\int_\omega G(\Bx, \By) {\hat{u}(\By)} d\By=0,\quad \Bx \in \omega\;, \label{sys_h2}
\end{eqnarray*}
where $\mu$ is defined in (\ref{eqmu}).
 It remains to apply the Laplacian to this equation, to obtain
 \[
 \Delta_\Bx \hat{u}(\Bx)-\mu\Big(  \hat{u}(\Bx)-\frac{1}{|\Omega|}\int_\omega \hat{u}(\Bx)d\Bx \Big)=0\;, \quad \Bx \in \omega\;.
 \]
Here we have used 
  (\ref{Grel}).  In turn, the equation for $\hat{u}$ in $\Omega\backslash \overline{\omega}$ takes the form
\[\Delta_\Bx \hat{u}(\Bx)+\mu =0\;, \quad \Bx \in \Omega\backslash \overline{\omega}\;.
\]
From this, the auxiliary homogenised problem  can now be stated.

\subsection*{Auxiliary homogenised problem }
The function $\hat{u}$, defined  inside the homogenised medium $\Omega$ containing an  effective inclusion $\omega \subset \Omega$, is a solution of the inhomogeneous equation
\begin{equation}\label{heq1}
\Delta \hat{u}(\Bx)-\mu\Big( \chi_\omega(\Bx) \hat{u}(\Bx)-1 \Big)=0\;, \quad \Bx\in \Omega
\end{equation}
with $\chi_\omega$ denoting the characteristic function for the set $\omega$. Together with this, we supply the boundary condition on the exterior of the domain  in the form
\begin{equation}\label{heq2}
\frac{\partial \hat{u}}{\partial n}(\Bx)=0\;,\quad \Bx \in \partial \Omega\;,
\end{equation}
and the transmission conditions across the interface of $\omega$  as:
\begin{equation}\label{heq3}
\Big[{\hat{u}}{}(\Bx)\Big]_{\partial \omega}=0\quad \text{ and }\quad \left[\frac{\partial \hat{u}}{\partial n}(\Bx)\right] _{\partial \omega}=0\;,
\end{equation}
where $[\cdot ]_{\partial \omega}$ indicates the jump across the boundary $\partial \omega$. In addition, we note that  $\hat{u}$ satisfies
\begin{equation*}\label{eqimpose}
\frac{1}{|\Omega|}\int_\omega \hat{u}(\Bx)d\Bx=1\;.
\end{equation*}
One can check that the problem (\ref{heq1})--(\ref{heq3}) is solvable by applying integration parts to $\hat{u}$ inside $\omega\cup\Omega\backslash \overline{\omega} $. 
\subsubsection*{Example: Homogenised problem for a sphere with spherical cluster of inclusions}
We present an example for the case $\Omega=B_R$ and $\omega=B_r$, with $B_\rho:=\{\Bx: |\Bx|<\rho\}$. In this case, the solution of (\ref{heq1})--(\ref{heq3}) can be computed explicitly, and has the form
\begin{equation}
\label{eqe1}\hat{u}(\Bx)=\chi_{\Omega\backslash \overline{\omega}}(\Bx) \hat{u}_O(\Bx)+\chi_{{\omega}} (\Bx)\hat{u}_I(\Bx)\;,
\end{equation}
with 
\begin{eqnarray}
&&\hat{u}_I(\Bx)=\frac{1}{3}\frac{R^3-r^3}{(\sqrt{\mu} r\cosh(\sqrt{\mu}r)-\sinh(\sqrt{\mu}r))}\frac{\sinh(\sqrt{\mu}|\Bx|)}{|\Bx|}+\frac{1}{\mu}\label{eqe2}
\end{eqnarray}
and 
\begin{eqnarray}
\hat{u}_O(\Bx)&=&-\frac{1}{6}|\Bx|^2-\frac{1}{3}\frac{R^3}{|\Bx|}\nonumber \\
&&+\frac{1}{6} \frac{((r^3+2R^3)\mu+6r)\sqrt{\mu} \cosh(\sqrt{\mu}r)-(3r^2\mu +6)\sinh(\sqrt{\mu}r)}{\mu (\sqrt{\mu} r \cosh(\sqrt{\mu}r )-\sinh(\sqrt{\mu}r))}\;.\nonumber\\
\label{eqe3}
\end{eqnarray}
For the case when $R=7$, $r=1$ and $\mu=0.09$, the slice plot of the solution $\hat{u}$ is plotted in Figure \ref{fig:homogenised}. One can see the magnitude of the field inside the effective inclusion $\omega$ drops as $|\Bx| \to 0$.

 \begin{figure}%
  \centering
       
\centering
        \includegraphics[width=0.9\textwidth]{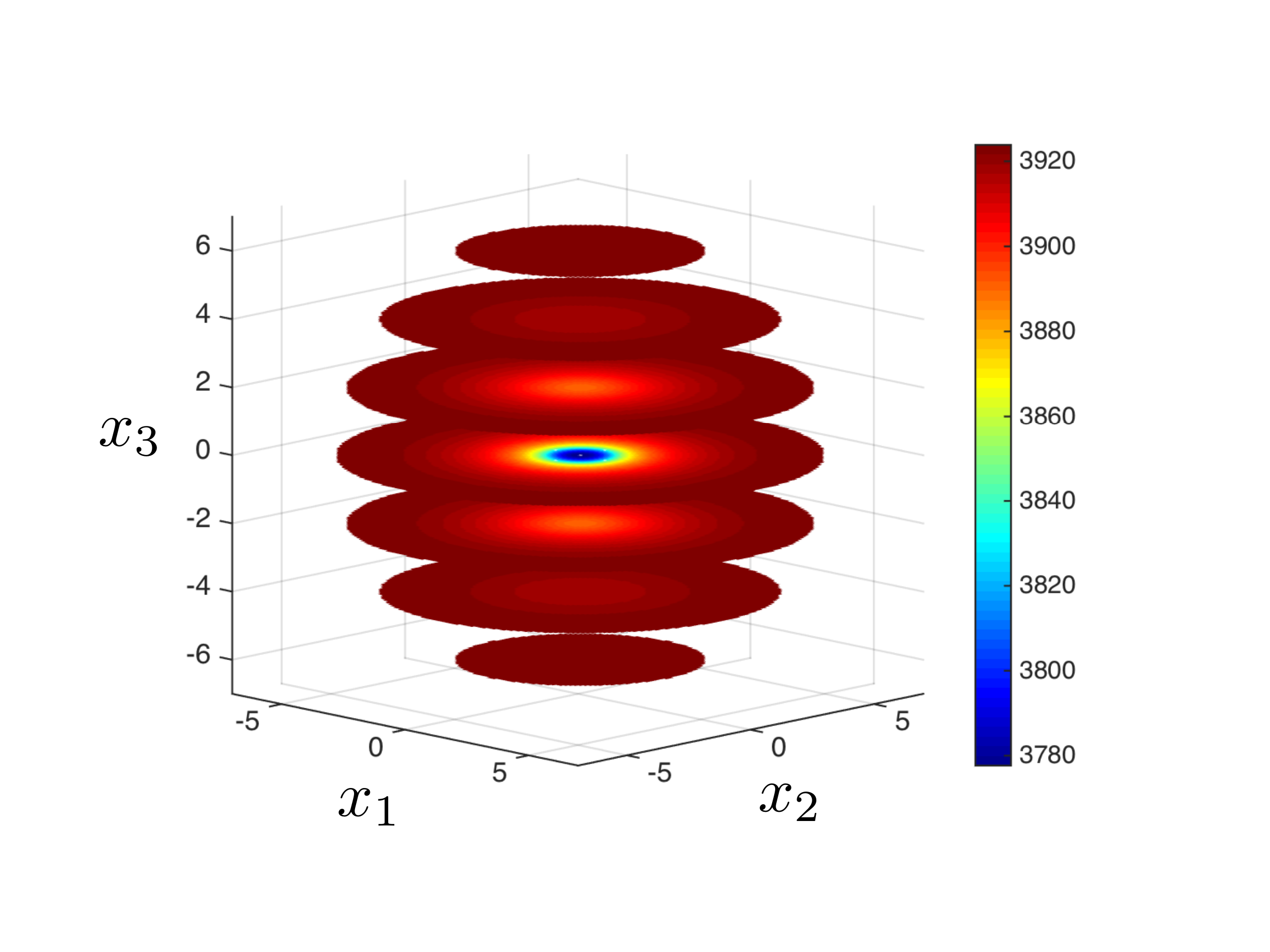}
        \caption{The slice plot of the homogenised solution $\hat{u}$, defined in (\ref{eqe1})--(\ref{eqe3}), satisfying (\ref{heq1})--(\ref{heq2}), for the case then $\Omega=B_7$ and $\omega=B_1$. The computation has been performed using the parameter $\mu=0.09$. }
        \label{fig:homogenised}%
        \end{figure}

\subsubsection*{Comparison with the asymptotic approximation \eq{uN_app}}

Consequently, the homogenisation approach provides the following approximation for the eigenvalue $\lambda_N$ and the coefficients $C_j$ in the representation \eq{uN_app} of the field $u_N$:

\begin{equation*}
\lambda_N \simeq \mu, ~~ C_j \simeq \hat{u}(\BO^{(j)}), ~~ j = 1, \dots, N,
\label{comp_eig}
\end{equation*}
where $\mu$ is defined by \eq{eqmu}, and $\hat{u}$ is the solution of the inhomogeneous transmission problem \eq{heq1}--\eq{heq3}.

 The asymptotic scheme demonstrated in Sections \ref{intro}--\ref{numsec}, has proved to be superior
compared to the homogenisation approximation, as it has delivered 
a uniform approximation of the first eigenfunction over $\GO$ including a disordered cloud $\Go$ of small inclusions.

\section*{Appendix: Higher order approximation}\label{formalalgappendix}
\renewcommand{\theequation}{A.\arabic{equation}}
\setcounter{equation}{0}
\renewcommand{\thesection}{A}
\setcounter{subsection}{0}
\renewcommand{\thesubsection}{A.\arabic{subsection}}
\setcounter{subsection}{0}

We present here more details concerning the proofs associated with the higher order approximations presented in section 
\ref{secHOA}  for the field $u_N$ and the corresponding first eigenvalue $\lambda_N$. Section  \ref{secA} contains the proof of Lemma \ref{lemcoeffs2} and then the proof of Theorems \ref{th2}--\ref{LHHO}, including the proof of the auxiliary estimate Lemma \ref{PsiL2ests}, are presented in Section \ref{secB}.

\section{Proof of Lemma \ref{lemcoeffs2}: Estimates of constant coefficients $A_j$ and $\BB^{(j)}$}\label{secA}

The solvability of $(\ref{sys2Bj1})$--$(\ref{sys2Bj2})$ can be proved in a similar way to steps in the proof of Lemma \ref{lemcoeffs}, as one needs to invert the same matrix to find $C_j$, $1\le j \le N$, as is needed in order to identify $A_j$, $1\le j \le N$. Such a proof yields the inequality
\[\sum_{j=1}^NA_j^2\le \text{Const }\sum_{j=1}^N(v^{(j)})^2\]
where it remains to estimate the right-hand side with (\ref{sys2Bj2}). In fact, from (\ref{sys2Bj2}), we can obtain with Young's  inequality:
\begin{eqnarray*}
&& \sum_{j=1}^N(v^{(j)})^2 \le \text{Const } \varepsilon^4\sum_{k=1}^N\Bigg\{C_k^2+\Big(\sum_{\substack{j\ne k\\ 1\le j \le N}}\frac{C_j}{|\BO^{(k)}-\BO^{(j)}|^2}\Big)^2+ d^{-6}\Big(\sum_{k=1}^N{C_j}\Big)^2\Bigg\}\;.
\end{eqnarray*}
Applying Cauchy's inequality and Lemma \ref{lemcoeffs} it can be deduced:
\begin{eqnarray*}
&& \sum_{j=1}^Nv_j^2 \le \text{Const } \varepsilon^4 d^{-3}\Bigg\{1+d^{-9}+\sum_{k=1}^N\sum_{\substack{j\ne k\\ 1\le j \le N}}\frac{1}{|\BO^{(k)}-\BO^{(j)}|^4}\Bigg\}\\
&\le &\text{Const } \varepsilon^4 d^{-3}\{1+d^{-9}+d^{-6}\}
\end{eqnarray*}
and then (\ref{Bjsumest}) follows.
A similar approach yields the estimate (\ref{BBsumest2}).\hfill $\Box$

\renewcommand{\theequation}{B.\arabic{equation}}
\setcounter{equation}{0}
\renewcommand{\thesection}{B}
\setcounter{subsection}{0}
\renewcommand{\thesubsection}{B.\arabic{subsection}}
\setcounter{subsection}{0}

\section{Proof of Theorems \ref{th2} and \ref{LHHO}}\label{secB}
First we write the formal asymptotic representations for higher order approximations to $u_N$ and $\lambda_N$ in section \ref{secA1}, which also includes the problem for leading order approximation of $u_N$ and associated estimates with proofs. The proof of Lemma \ref{PsiL2ests} is found in section \ref{Plem10}. In section \ref{complete34}, we then complete the proofs of Theorems \ref{th2} and \ref{LHHO}.
\subsection{Formal asymptotic representations}\label{secA1}
The first eigenvalue $\lambda_N$ and corresponding eigenfunction $u_N$ and  are now sought in the form:
\begin{eqnarray}
u_N(\Bx)=V(\Bx)+R_N(\Bx)\;,\label{un_2ordA}
\end{eqnarray}
\begin{eqnarray}\label{lamn_2ordA}
\lambda_N=\Lambda_N +\lambda_{R,N}\;,
\end{eqnarray}
where 
\begin{eqnarray}\label{UfnA}
V(\Bx)&=&1+\sum_{j=1}^N (C_j +A_j) \{P_\varepsilon^{(j)}(\Bx)-\text{cap}(\omega^{(j)}_\varepsilon) (\CH(\Bx, \BO^{(j)})-\Gamma_\Omega^{(j)})\} \nonumber\\
&& +\sum_{j=1}^N\BB^{(j)}\cdot \BD^{(j)}_\varepsilon(\Bx)+\sum_{j=1}^NC_j \BGb^{(j)}_\varepsilon \cdot [\nabla_\Bz \CH(\Bx, \Bz)\Big|_{\Bz=\BO^{(j)}}+\BGg_\Omega^{(j)}]\nonumber \\
&&+\Lambda_1 \sum_{j=1}^N C_j \text{cap}(\omega^{(j)}_\varepsilon) \int_\Omega \CG(\By, \Bx)\CG(\By,\BO^{(j)})d\By\end{eqnarray}
and $\Lambda_N$ is redefined as:
\begin{eqnarray}\label{LpA}
\Lambda_N=\Lambda_1+\Lambda_2\;.
\end{eqnarray}
The term $\Lambda_2$ is defined in (\ref{lam2_HOA}).
In what follows we assume $\Lambda_2=O(\varepsilon^2 d^{-6})$.
\subsection*{Problem for the function $V$}
Before stating the problem that the function $V$ 
satisfies (see (\ref{UfnA})), we first introduce auxiliary functions used in the proof of Theorem \ref{th2}. 
These functions we denote by $\Psi_k$, $0\le k \le N$, and they appear in (\ref{EqnPsi0_H}) and (\ref{EqnPsik_H}). They are constructed in similar way to as in section \ref{remainderestsec}, where 
 $\Psi_0$ is harmonic in $\Omega_N$ and $\Psi_k$, $1\le k \le N$, satisfies 
\[\Delta \Psi_k(\Bx)+\Lambda_N=0\;, \quad \Bx \in \Omega_N\;, \quad 1\le k \le N\;,\] 
with $\Lambda_N$ defined in (\ref{LpA}).

 We have the next lemma, concerning the problem for $V$.
\begin{lem}\label{lemAppRN}
The function $V$ of $(\emph{\ref{UfnA}})$ satisfies the problem
\begin{eqnarray}
&&\Delta V(\Bx)+\Lambda_N V(\Bx)=f(\Bx)\;, \quad \Bx \in \Omega_N\;,\label{NewReq1}\\
&&  \frac{\partial V}{\partial n}(\Bx)=\Psi_0(\Bx)\;, \quad \Bx \in \partial \Omega\;,\label{newRNbc1}\\
&&V(\Bx)=\Psi_k(\Bx)\;, \quad \Bx \in \partial \omega_\varepsilon^{(k)}, 1\le k \le N\;, \label{newRNbc2}
\end{eqnarray}
where $\Lambda_N$ is given in $(\emph{\ref{LpA}})$,
\begin{eqnarray}
&&|f(\Bx)|\le \text{\emph{Const} }\varepsilon^2\, d^{-3}\sum_{j=1}^N\Big\{ \frac{|A_j|}{|\Bx-\BO^{(j)}|}+d^{-3}\frac{|C_j|}{|\Bx-\BO^{(j)}|}+\varepsilon \frac{|C_j|+|A_j|}{|\Bx-\BO^{(j)}|^2}\label{estf11}\nonumber\\
&&\qquad \qquad +\varepsilon^2\frac{|C_j|}{|\Bx-\BO^{(j)}|^3}+\varepsilon\,  d^{-3} {|C_j|}+\varepsilon^2 \frac{|\BB^{(j)}|}{|\Bx-\BO^{(j)}|^2}\Big\}\;,\quad \Bx \in \Omega_N\;,\nonumber\\\label{newf1}
 \end{eqnarray}
 for $\Bx \in \partial \Omega$
 \begin{eqnarray}\label{Psi_0}
|\Psi_0(\Bx)| \le \text{\emph{Const }}\varepsilon^2 \sum_{j=1}^N \Big[\frac{\varepsilon  |C_j|}{|\Bx-\BO^{(j)}|^3}+\frac{|A_j|}{|\Bx-\BO^{(j)}|^2}+\frac{\varepsilon |\BB^{(j)}|}{|\Bx-\BO^{(j)}|^2}\Big]\;,\label{newPsi01}
\end{eqnarray}
and  for $\Bx \in \partial \omega_\varepsilon^{(k)}$, $1\le k \le N$,
\begin{eqnarray}
&& |\Psi_k(\Bx)|\le  \text{\emph{Const} }\varepsilon^2\Big[\sum_{j=1}^N\Big\{|A_j|+ \varepsilon |C_j|+ \varepsilon d^{-3} |C_j|\Big\}
\nonumber \\&&+\varepsilon^2\sum_{\substack{j\ne k\\ 1\le j \le N}}  \Big\{ \frac{\varepsilon |C_j|}{|\BO^{(k)}-\BO^{(j)}|^3}+\frac{|A_j|}{|\BO^{(k)}-\BO^{(j)}|^2}+ \frac{\varepsilon  |\BB^{(j)}|}{|\BO^{(k)}-\BO^{(j)}|^2}\Big\}\Big]\;.\nonumber\\ \label{Psi_k}
\end{eqnarray}

\end{lem}

\subsection*{Proof of (\ref{NewReq1}) and (\ref{newf1})}
Owing to asymptotics of the fields $P_\varepsilon^{(j)}$ and $\BD_\varepsilon^{(j)}$, in Lemmas \ref{lem1} and \ref{lemD}, respectively, from (\ref{UfnA}) it can be shown
\begin{eqnarray*}
&&V(\Bx)=1+\sum_{j=1}^N (C_j +A_j) \text{cap}(\omega^{(j)}_\varepsilon) (\CG(\Bx, \BO^{(j)})+\Gamma_\Omega^{(j)})\nonumber\\ 
&&-\sum_{j=1}^NC_j \BGb^{(j)}_\varepsilon \cdot [\nabla_\Bz \CG(\Bx, \Bz)\Big|_{\Bz=\BO^{(j)}}-\BGg_\Omega^{(j)}]\nonumber \\
&&+\Lambda_1 \sum_{j=1}^N C_j \text{cap}(\omega^{(j)}_\varepsilon) \int_\Omega \CG(\By, \Bx)\CG(\By,\BO^{(j)})d\By\nonumber \\
&&+O\Big(\sum_{j=1}^N\frac{\varepsilon^3 |C_j|}{|\Bx-\BO^{(j)}|^3}\Big)+O\Big(\sum_{j=1}^N\frac{\varepsilon^2 |A_j|}{|\Bx-\BO^{(j)}|^2}\Big)+O\Big(\sum_{j=1}^N\frac{\varepsilon^3 |\BB^{(j)}|}{|\Bx-\BO^{(j)}|^2}\Big)\,.\nonumber\\
 \end{eqnarray*}
Moreover, after multiplication by $\Lambda_N$ in (\ref{UfnA}), one can show
\begin{eqnarray}
&&\Lambda_N V(\Bx)=\Lambda_1+\Lambda_2+\Lambda_1 \sum_{j=1}^N C_j \text{cap}(\omega_\varepsilon^{(j)}) (\CG(\Bx, \BO^{(j)})+\Gamma_\Omega^{(j)})\nonumber\\
&&+O\Big(\varepsilon^2 d^{-3}\sum_{j=1}^N \frac{|A_j|}{|\Bx-\BO^{(j)}|}\Big)+O\Big(\varepsilon^2 d^{-6}\sum_{j=1}^N \frac{|C_j|}{|\Bx-\BO^{(j)}|}\Big)\nonumber\\
&&+O\Big(\varepsilon^3 d^{-3}\sum_{j=1}^N\frac{|C_j|+|A_j|}{|\Bx-\BO^{(j)}|^2}\Big)+O\Big(\varepsilon^4 d^{-3}\sum_{j=1}^N\frac{|C_j|}{|\Bx-\BO^{(j)}|^3}\Big)\nonumber\\ \nonumber
&&+O\Big(\varepsilon^3 d^{-6}\sum_{j=1}^N{|C_j|}\Big)+O\Big(\varepsilon^4 d^{-3}\sum_{j=1}^N\frac{|\BB^{(j)}|}{|\Bx-\BO^{(j)}|^2}\Big)\;.
\nonumber\\\label{LVN}
 \end{eqnarray}
 Using the  model problems in section \ref{modprob}, 
 \begin{equation}\label{eqVN}
 \Delta V(\Bx)=\frac{1}{|\Omega|}\sum_{j=1}^N (C_j +A_j)\text{cap}(\omega^{(j)}_\varepsilon)-\Lambda_1 \sum_{j=1}^N C_j \text{cap}(\omega^{(j)}_\varepsilon) \CG(\Bx, \BO^{(j)}) \;.
 \end{equation}
 Thus, (\ref{LVN}) and (\ref{eqVN}) together with (\ref{WN61}) and  (\ref{lam2_HOA}),  show that $V$ satisfies (\ref{NewReq1}) and (\ref{newf1}).  
 \hfill $\Box$

\subsection*{Proof of (\ref{newRNbc1}) and (\ref{newPsi01})}
The condition (\ref{newRNbc1})  is obtained by using 
(\ref{UfnA}) and the model problems of section \ref{modprob}.
Since $\text{dist}(\omega, \partial \Omega)=O(1)$, for $\Bx \in \partial \Omega$,  Lemmas \ref{lem1} and \ref{lemD} allow one to obtain (\ref{newPsi01}).

\subsection*{Proof of (\ref{newRNbc2}) and (\ref{Psi_k})}
The proof of (\ref{newRNbc2}) again follows from  
(\ref{UfnA}) and the model problems of section \ref{modprob}.
Here we derive estimates for the functions $\Psi_k$,
for $\Bx \in \partial {\omega^{(k)}_\varepsilon}$, $1\le k \le N$.
Lemma \ref{lem1} shows that 
\begin{eqnarray}
&& \sum_{\substack{j\ne k\\ 1\le j \le N}} (C_j+A_j)\Big[P^{(j)}_\varepsilon(\Bx)-\frac{\text{cap}(\omega^{(j)}_\varepsilon)}{4\pi |\BO^{(k)}-\BO^{(j)}|}\Big]\nonumber\\
&&+\sum_{\substack{j\ne k \\ 1\le j \le N}} C_j \BGb_\varepsilon^{(j)} \cdot \nabla_{\Bz}\left(\frac{1}{4\pi |\BO^{(k)}-\Bz|}\right)\Big|_{\Bz=\BO^{(j)}}\nonumber\\
&=  &\sum_{\substack{j\ne k\\ 1\le j \le N}} C_j(\Bx-\BO^{(k)})\cdot \nabla_\Bx \Big(\frac{\text{cap}(\omega^{(j)}_\varepsilon)}{4\pi |\Bx-\BO^{(j)}|}\Big)\Big|_{\Bx=\BO^{(k)}}\nonumber\\
&&+
 O\Big(\varepsilon^2\sum_{\substack{j\ne k\\ 1\le j \le N}}  \Big\{ \frac{\varepsilon |C_j|}{|\BO^{(k)}-\BO^{(j)}|^3}+\frac{|A_j|}{|\BO^{(k)}-\BO^{(j)}|^2}\Big\}\Big)\;,\label{A1}
\end{eqnarray}
where Taylor's expansion about $\Bx=\BO^{(k)}$ has been used. A similar application of this expansion and the use of Lemma \ref{lemD}, provides the estimates
\begin{eqnarray}
&&\sum_{\substack{j\ne k \\1 \le j \le N}} \BB^{(j)}\cdot \BD_\varepsilon^{(j)}(\Bx) \le \text{Const }\sum_{\substack{j\ne k \\1 \le j \le N}}\frac{\varepsilon^3 |\BB^{(j)}|}{|\BO^{(k)}-\BO^{(j)}|^2}\label{A2}\\
&& \sum_{j=1}^N C_j \BGb_\varepsilon^{(j)}\cdot [\nabla_{\Bz}\CH(\Bx, \Bz)\Big|_{\Bz=\BO^{(j)}}-\nabla_{\Bz}\CH(\BO^{(k)}, \Bz)\Big|_{\Bz=\BO^{(j)}}] \le  \text{Const} \sum_{j=1}^N \varepsilon^3 |C_j| \;,\nonumber \\
\end{eqnarray}
and
\begin{eqnarray}
&&\Lambda_1\sum_{j=1}^N C_j \text{cap}(\omega^{(j)}_\varepsilon) \int_\Omega \CG(\By, \BO^{(j)})(\CG(\By, \Bx)-\CG(\By, \BO^{(k)}))d\By\nonumber \\
&\le&  \text{Const }\varepsilon^3 d^{-3}\sum_{j=1}^N  |C_j|\;,
\end{eqnarray}
for $\Bx\in \partial {\omega^{(k)}_\varepsilon}$, $1\le k \le N$.
The Taylor expansion about $\Bx=\BO^{(k)}$ shows that
\begin{eqnarray}
&&\BB^{(k)}\cdot (\Bx-\BO^{(k)})-\sum_{j=1}^N  (C_j+A_j) \text{cap}(\omega^{(j)}_\varepsilon) (\CH(\Bx, \BO^{(j)})-\CH(\BO^{(k)}, \BO^{(j)}))\nonumber\\
&=&(\Bx-\BO^{(k)}) \cdot  \Big( \BB^{(k)}-\sum_{j=1}^N  C_j \text{cap}(\omega^{(j)}_\varepsilon)  \nabla_\Bx \CH(\BO^{(k)}, \BO^{(j)}) \Big)\nonumber\\
 &&+O\Big(\sum_{j=1}^N\varepsilon^2|A_j|\Big)\;,\label{A7}
\end{eqnarray}
for $\Bx\in \partial {\omega^{(k)}_\varepsilon}$, $1\le k \le N$.
The combination of (\ref{EqnPsik_H})
and (\ref{A1})--(\ref{A7}),  yields (\ref{Psi_k}).

\subsection{Proof of Lemma \ref{PsiL2ests}: auxiliary $L_2$-estimates for  $\Psi_k$, $0\le k \le N$ and their derivatives}\label{Plem10}
Here we prove Lemma \ref{PsiL2ests} that concerns the point-wise estimates for the functions $\Psi_k$, $0\le k \le N$. We require  the next auxiliary result.
\begin{lem}\label{lemestPsi}
For $\Bx \in \CV$, where $\CV$ is a neighbourhood of $\partial \Omega$ defined in section $\ref{remainderestsec}$, the function  $\Psi_0$ satisfies
\begin{eqnarray*}\label{Psi_011}
|\Psi_0(\Bx)| \le \text{\emph{Const }}\varepsilon^2 \sum_{j=1}^N \Big[\frac{\varepsilon  |C_j|}{|\Bx-\BO^{(j)}|^3}+\frac{|A_j|}{|\Bx-\BO^{(j)}|^2}+\frac{\varepsilon |\BB^{(j)}|}{|\Bx-\BO^{(j)}|^2}\Big]\;,\nonumber \\
\label{newPsi011}
\end{eqnarray*}
\begin{eqnarray}\label{DPsi0}
|\nabla \Psi_0(\Bx)|\le \text{\emph{Const} }\varepsilon^2\sum_{j=1}^N \Big[\frac{\varepsilon|C_j|}{|\Bx-\BO^{(j)}|^4}+\frac{|A_j|}{|\Bx-\BO^{(j)}|^3}+\frac{\varepsilon |\BB^{(j)}|}{|\Bx-\BO^{(j)}|^3}\Big]\;,\nonumber\\
\end{eqnarray}
whereas for $\Bx \in B^{(k)}_{3\varepsilon}\backslash \overline{\omega^{(k)}_\varepsilon}$, the functions $\Psi_k$, $1\le k \le N$, satisfy the inequalities
\begin{eqnarray}
&& |\Psi_k(\Bx)|\le  \text{\emph{Const} }\varepsilon^2\Big[\sum_{j=1}^N\Big\{|A_j|+ \varepsilon |C_j|+ \varepsilon d^{-3} |C_j|\Big\}
\nonumber \\&&+\varepsilon^2\sum_{\substack{j\ne k\\ 1\le j \le N}}  \Big\{ \frac{\varepsilon |C_j|}{|\BO^{(k)}-\BO^{(j)}|^3}+\frac{ |A_j|}{|\BO^{(k)}-\BO^{(j)}|^2}+ \frac{  \varepsilon |\BB^{(j)}|}{|\BO^{(k)}-\BO^{(j)}|^2}\Big\}\Big]\;,\nonumber\\ \label{Psi_k1}
\end{eqnarray}
and
\begin{eqnarray}
|\nabla \Psi_k(\Bx)|&\le &\text{\emph{Const} }\varepsilon\Big[\sum_{j=1}^N\varepsilon (1+d^{-3}) |C_j|\nonumber \\
&&+\sum_{\substack{j \ne k\\ 1\le j \le N}}\Big\{\frac{\varepsilon  |C_j|}{|\BO^{(k)}-\BO^{(j)}|^3} +\frac{|A_j|}{|\BO^{(k)}-\BO^{(j)}|^2}+\frac{\varepsilon^2 |\BB^{(j)}|}{|\BO^{(k)}-\BO^{(j)}|^3}\Big\} \Big]\;.\nonumber  \\ \label{DPsik1}
\end{eqnarray}
\end{lem}
\emph{Proof. }
Estimates (\ref{Psi_011}) and (\ref{Psi_k1}) are proved in exactly the same way as (\ref{newPsi01}) and (\ref{Psi_k}) of Lemma \ref{lemAppRN} were derived.  

The proof of (\ref{DPsi0}) follows from applying the gradient to (\ref{EqnPsi0_H}) and using the model problems of section \ref{modprob}, Lemmas \ref{lemD} and \ref{lem1a}. It remains to prove (\ref{DPsik1}).
Note that from  (\ref{EqnPsik_H})
\begin{eqnarray}
&&\nabla \Psi_k(\Bx)= \sum_{\substack{j\ne k\\ 1\le j \le N}} (C_j+A_j)\nabla P^{(j)}_\varepsilon(\Bx)+\BB^{(k)}+\sum_{\substack{j\ne k \\1 \le j \le N}} \BB^{(j)}\cdot \nabla \BD_\varepsilon^{(j)}(\Bx)\nonumber\\
&&-\sum_{j=1}^N  (C_j+A_j) \text{cap}(\omega^{(j)}_\varepsilon) \nabla\CH(\Bx, \BO^{(j)})+\sum_{j=1}^N C_j \nabla(\BGb_\varepsilon^{(j)}\cdot \nabla_{\Bz}\CH(\Bx, \Bz)\Big|_{\Bz=\BO^{(j)}})\nonumber \\
&&+\Lambda_1 \sum_{j=1}^N C_j \text{cap}(\omega^{(j)}_\varepsilon) \nabla \int_\Omega \CG(\By, \BO^{(j)})\CG(\By, \Bx)d\By\;.\nonumber\\  \label{B1}
\end{eqnarray}
The last two terms satisfy
\begin{eqnarray}
&&\sum_{j=1}^N C_j \nabla(\BGb_\varepsilon^{(j)}\cdot \nabla_{\Bz}\CH(\Bx, \Bz)\Big|_{\Bz=\BO^{(j)}}) \le \text{Const } \varepsilon^2 \sum_{j=1}^N|C_j |\\
&&\Lambda_1\sum_{j=1}^N C_j \text{cap}(\omega^{(j)}_\varepsilon) \nabla \int_\Omega \CG(\By, \BO^{(j)})\CG(\By, \Bx)d\By\le \text{Const } \varepsilon^2 d^{-3} \sum_{j=1}^N|C_j|\;.\nonumber\\ 
\end{eqnarray}
As with the derivation of (\ref{A2}), we have
\begin{equation}
\sum_{\substack{j\ne k \\1 \le j \le N}} \BB^{(j)}\cdot \nabla \BD_\varepsilon^{(j)}(\Bx) \le \text{Const }\sum_{\substack{j\ne k \\1 \le j \le N}}\frac{\varepsilon^3 |\BB^{(j)}|}{|\BO^{(k)}-\BO^{(j)}|^3}\;.\label{B4}
\end{equation}
The far-field representation of the capacitary potentials gives
\begin{eqnarray}
 &&\sum_{\substack{j\ne k\\ 1\le j \le N}} (C_j+A_j)\nabla P^{(j)}_\varepsilon(\Bx)-\sum_{j=1}^N  (C_j+A_j) \text{cap}(\omega^{(j)}_\varepsilon) \nabla\CH(\Bx, \BO^{(j)})\nonumber \\
 &=&- C_k \text{cap}(\omega^{(k)}_\varepsilon) \nabla_\Bx \CH(\BO^{(k)}, \BO^{(k)})+\sum_{\substack{j\ne k\\ 1\le j \le N}} C_j\text{cap}(\omega_\varepsilon^{(j)})\nabla_\Bx \CG(\BO^{(k)}, \BO^{(j)})\nonumber \\
 &&+O\Big(\varepsilon^2 |C_k|+\sum_{\substack{j \ne k\\ 1\le j \le N}}\Big\{\frac{\varepsilon^2 |C_j|}{|\BO^{(k)}-\BO^{(j)}|^3} +\frac{\varepsilon |A_j|}{|\BO^{(k)}-\BO^{(j)}|^2}\Big\}\Big)\label{B5}
\end{eqnarray}
Now,  gathering   (\ref{B1})--(\ref{B5}) with (\ref{BBk_result}) produces the inequality (\ref{DPsik1}). \hfill $\Box$

\subsection*{Completion of the proof of Lemma \ref{PsiL2ests}}

Using Lemma \ref{lemestPsi} one can obtain the $L_2$-estimates of $\Psi_k$, $k=0, \dots, N$, and their gradients in a similar way to those derived in section \ref{remainderestsec}.
 We use Lemma \ref{lemestPsi} and apply similar estimates, to those employed in section \ref{remainderestsec}, in addition to Lemmas \ref{lemcoeffs} and \ref{lemcoeffs2} to yield the results of Lemma \ref{PsiL2ests}. \hfill $\Box$

\vspace{0.2in} In a equivalent way, one also shows that the function $f$ (in (\ref{NewReq1}) and (\ref{newf1})), satisfies the next estimate.
\begin{lem} \label{lemVopest} 
The following estimate
\begin{equation*}
\|\Delta V+\Lambda_N V\|^2_{L_2(\Omega_N)}\le \text{\emph{Const }}\varepsilon^5  d^{-12}
\end{equation*}
holds.
\end{lem}

\subsection{Completion of the proofs of Theorems 3--4}\label{complete34}
It then follows from Lemmas \ref{PsiL2ests} and \ref{lemVopest}  and the proof of section \ref{remainderestsec}, that the function $\sigma_N$ constructed according to (\ref{eqsig2}), with (\ref{EqnPsi0_H}) and (\ref{EqnPsik_H}),  satisfies the following estimate
\begin{equation*}\label{EQdiffsigun}
\|\sigma_N-u_N\|_{L_2(\Omega_N)}\le \text{Const } \varepsilon^{5/2} \,d^{-15/2}\;.
\end{equation*}
In addition,  for the approximation $\Lambda_N$ (see (\ref{lamn_2ordA}), (\ref{LpA}) and Theorem \ref{LHHO}) to the first eigenvalue $\lambda_N$ admits the estimate
\begin{equation*}
\label{Nlamest}
|\lambda_N-\Lambda_N|\le \text{Const } \varepsilon^{5/2} \,d^{-15/2}\;,
\end{equation*}
holds.
 \hfill $\Box$

\renewcommand{\theequation}{\arabic{equation}}
\setcounter{equation}{10}
\renewcommand{\thesection}{10}
\setcounter{subsection}{10}
\renewcommand{\thesubsection}{\arabic{subsection}}
\setcounter{subsection}{10}



\end{document}